\documentclass[aps,prb,twocolumn,10pt,showkeys,superscriptaddress]{revtex4-2}
\usepackage{times}
\usepackage{amsmath}
\usepackage{amssymb}
\usepackage{amsfonts}
\usepackage{multirow}
\usepackage{mathrsfs}
\usepackage{graphicx}
\usepackage{subfigure}
\usepackage{booktabs}
\usepackage{rotating}
\usepackage{longtable} 
\usepackage{bm}
\usepackage{float}
\usepackage{epstopdf}
\usepackage{comment}
\usepackage{graphicx}
\usepackage{subfigure}
\usepackage{amsmath}
\usepackage{orcidlink}

\makeatletter

\newcommand{\Rmnum}[1]{\expandafter\@slowromancap\romannumeral #1@} 
\makeatother
\makeatletter
\newcommand*{\rom}[1]{\expandafter\@slowromancap\romannumeral #1@}
\makeatother

\renewcommand{\arraystretch}{2}
\makeatletter

\hypersetup{
  colorlinks   = true, 	
  urlcolor     = blue, 	
  linkcolor    = blue, 	
  citecolor   = blue 
}

\begin{document}
\title{Schwarzschild Black Hole Coupled with a Cloud of Strings Immersed in King Dark Matter Halo}

\author{Faizuddin Ahmed\orcidlink{0000-0003-2196-9622}}
\email{faizuddinahmed15@gmail.com}
\affiliation{Department of Physics,  The Assam Royal Global University, Guwahati, 781035, Assam, India}
\author{Edilberto O. Silva\orcidlink{0000-0002-0297-5747}}
\email{edilberto.silva@ufma.br}
\affiliation{Programa de P\'os-Gradua\c c\~ao em F\'{\i}sica \& Coordena\c c\~ao do Curso de F\'{\i}sica -- Bacharelado, Universidade Federal do Maranh\~{a}o, 65085-580 S\~{a}o Lu\'{\i}s, Maranh\~{a}o, Brazil}

\begin{abstract}
In this paper, we examine the geodesic and thermodynamic properties of a Schwarzschild black hole with a cloud of strings (known as the Letelier black hole) immersed in a King dark matter (KDM) halo under an isotropic configuration. The dynamics of both photons and massive particles are analyzed in detail using the effective potential formalism, including particle trajectories, the photon sphere, black hole shadow, and the innermost stable circular orbits (ISCOs). Particular emphasis is placed on how the presence of the KDM halo modifies these geometric and dynamical features. Furthermore, we explore the topological characteristics of photon rings by constructing a normalized vector field, following Duan’s topological current $\phi$-mapping theory, and demonstrate how this field is influenced by both the string cloud and the KDM halo. In the thermodynamic context, we analyze the impact of the KDM halo and the string cloud on the Hawking temperature, Gibbs free energy, thermal stability, and phase transitions of the black hole. Finally, we examine the thermodynamic topology of the system using a theoretical framework that incorporates a generalized Helmholtz free energy and topological current theory. In both the photon sphere and thermodynamic analyses, we show that the string cloud parameter shifts the location of the zero point of the vector field in the equatorial plane. Specifically, in the photon sphere case, the radius of the photon sphere increases with increasing values of the string cloud parameter, while in the thermodynamic topology, the horizon radius decreases as the string cloud parameter increases.
\end{abstract}

\keywords{Black hole; King dark matter; isotropic configuration; accretion disk; topological charge}
\maketitle

\small

\section{Introduction}\label{Sec1}

Black holes have long stood as one of the most fascinating and enigmatic phenomena in astrophysics~\cite{novikov1973astrophysics,novikov2013physics,heckman2014coevolution}. Initially, they were conceived as regions of spacetime devoid of matter, where the gravitational pull is so intense that not even light can escape~\cite{ruffini1971introducing,penrose1972black,rees1974black,novikov1995black}. However, subsequent observations have revealed that BHs are not isolated; instead, they are surrounded by complex environments containing various forms of matter~\cite{chapline1975cosmological,schneider2002first,cyburt2003primordial,volonteri2012black,carr2016primordial}, including gas~\cite{lynden1969galactic,shakura1973black}, dust~\cite{antonucci1993unified,urry1995unified}, and-most intriguingly-dark matter~\cite{zwicky1933rotverschiebung,rubin1980rotational,blumenthal1984formation,trimble1987existence}.

The analysis of galaxy rotation curves and gravitational lensing associated with galaxy clusters strongly indicates the presence of a non-luminous form of matter beyond the observed baryonic components \cite{DM1980, DM1996, DM1998}. Mass profiles derived from stellar dynamics and mass-to-light ratios in spiral galaxies consistently deviate from those inferred through kinematic observations, particularly in the outer regions. This discrepancy is well accounted for by postulating the existence of dark matter (DM), distributed from galactic centers out to their halos. Current cosmological observations suggest that DM constitutes approximately 26.8\% of the total mass-energy content of the universe-significantly exceeding the 5\% contribution from ordinary baryonic matter. DM plays a pivotal role in cosmic structure formation, driving gravitational instabilities that led to the emergence of large-scale structures in the universe. There are several unsolved questions regarding the DM composition, whether it can be described by any fundamental particles, and which interactions are allowed. It plays a pivotal role in explaining numerous astrophysical and cosmological observations and offers a compelling avenue for exploring non-traditional spacetime structures. Strong evidence for DM comes from galactic rotation curves, gravitational lensing, and cluster dynamics, all pointing to an unseen mass component \cite{DM9, DM10, DM11, DM12, DM15}. Despite strong indirect evidence, the true nature of dark matter (DM) remains a fundamental mystery. Proposed candidates span a wide mass range, from ultralight bosons ($\sim 10^{-22}$ eV) to heavy Weakly Interacting Massive Particles (WIMPs). New experimental approaches, including space-based detectors near the Sun, aim to explore local DM distributions with greater sensitivity \cite{DM16, DM17}. Three key halo models have been analyzed in this context: the Navarro-Frenk-White (NFW) profile with a central cusp, the Pseudo-Isothermal (PI) profile with a flat core, and the Perfect Fluid (PF) model representing continuous matter \cite{DM26, DM27}. Each provides insights into how galactic DM structures can influence wormhole geometry and stability in quantum gravity regimes. 

Various BH solutions embedding a dark matter (DM) profile have been studied in the literature. For instance, Li and Yang \cite{LiYang2012} constructed a BH solution surrounded by a DM distribution using a phantom scalar field (without attributing it a cosmological role). That work sparked further interest, and subsequent studies have generalized or extended these ideas in many directions (see, e.g., Refs.\ \cite{Hendi2020, Rizwan2019, Narzilloev2020, Shaymatov2021a, Rayimbaev2021, ShaymatovMalafarina2021, ShaymatovSheoran2022}). These investigations explore variations such as different scalar field coupling, anisotropic DM pressure, rotating spacetimes, or more general metric ansätze. Parallel to that, models of pure DM halo solutions have been developed, in which the BH is treated as embedded in a larger halo environment rather than being the sole gravitational source (see, e.g., Refs. \cite{Cardoso2022, Shen2024, Hou2018}). In these constructions, the halo’s gravity and density profile influence the geometry and observational signatures of the system. More recently, a new DM halo solution has been proposed that employs a Dehnen (1, 4, 0)-type density profile \cite{Dehnen1993, Gohain2024}. In that scenario, a spherically symmetric BH is embedded within a Dehnen-based DM halo, which provides a more flexible and phenomenologically motivated framework for the halo’s inner and outer density slopes. The Dehnen profile is already well established in galactic dynamics and structure studies \cite{Dehnen1993}, and employing it in the context of BH with halo spacetimes allows a more realistic modeling of how galactic-scale DM might affect BH geometries and observational signatures. Some other studies of DM halo were reported in Refs. \cite{B1,B2,B6}.

This revelation has profoundly transformed our understanding of BH dynamics and their role in the cosmos~\cite{hawley2005foundations,volonteri2012black}. Dark matter, which neither emits nor interacts with electromagnetic radiation in the same way as ordinary matter, is believed to constitute a substantial portion of the universe’s total mass~\cite{turner1991dark,overduin2004dark,spergel2015dark,bertone2018history}. As the scientific community seeks to unravel the properties of dark matter, many theoretical models have been developed to describe its behavior, distribution, and interactions-particularly in the vicinity of BHs~\cite{wang2016dark,oks2021brief,shen2024analytical}.

Among the most widely studied models are those that characterize dark matter halos through specific density profiles. Notable examples include the Navarro-Frenk-White (NFW) profile~\cite{navarro1996structure}, the Hernquist profile~\cite{hernquist1990analytical}, the Burkert profile~\cite{burkert1995structure}, the Moore profile~\cite{moore1994evidence}, and the Dehnen profile~\cite{dehnen1993family}. Each of these models offers distinct insights into the influence of dark matter on gravitational dynamics, especially in regions close to BHs. The effects of dark matter on BH systems have been investigated from various perspectives. These include analyses of thermodynamic properties~\cite{xu2019perfect,singh2021thermodynamic,pantig2023black,carvalho2023thermodynamics,gohain2024thermodynamics}, quasinormal modes~\cite{cardoso2016black,jusufi2020quasinormal,bamber2021quasinormal,konoplya2021black,das2023stability,konoplya2025quasinormal}, and optical appearances such as the BH shadow~\cite{konoplya2019shadow,jusufi2019black,saurabh2021imprints,figueiredo2023black,capozziello2023dark,chowdhury2025effect}. Further studies explore broader implications such as Hawking radiation, BH evaporation, and inflationary scenarios~\cite{xu2018black,kavanagh2020detecting,konoplya2019hawking}.

The primary aim of the present work is to investigate the properties of a Letelier BH surrounded by a King-type dark matter (KDM) halo. This setup is of particular interest from both theoretical and observational perspectives. Theoretically, incorporating a realistic dark matter distribution-such as the King model \cite{king1962structure,kar2025diverse}-into the BH spacetime allows for a natural extension of classical BH features. This includes a detailed analysis of geodesic motion for both photons and massive particles, the structure and size of the innermost stable circular orbit (ISCO), modifications to the event horizon, and changes in the thermodynamic behavior, including Hawking radiation. Of particular importance is the study of scalar perturbations in this background, which serve as a powerful probe for assessing the linear stability of the system and for computing the quasinormal mode (QNM) spectrum. These modes encode essential information about the dynamical response of the BH to external perturbations and are key observables in gravitational wave astronomy. From an observational standpoint, the Letelier-King dark matter model provides a framework to predict potentially detectable signatures. These include shifts in the photon sphere radius, modifications to the BH shadow profile, and changes in the topological characteristics, all of which could be constrained or confirmed by current or forthcoming high-resolution observations, such as those from the Event Horizon Telescope. By exploring the impact of a realistic dark matter environment on BH physics, this study aims to bridge the gap between astrophysical modeling and observational signatures in the era of precision BH imaging and gravitational wave detection.

The structure of this paper is organized as follows. In Section~\ref{Sec2}, we introduce the background geometry of the BH solution, which is coupled with a cloud of strings and embedded within a King-type dark matter (KDM) halo. Section~\ref{Sec3} explores the geodesic motion of both massless and massive test particles, identifies the photon sphere, and analyzes the BH shadow, along with observational constraints on the model parameters and the properties of the innermost stable circular orbit (ISCO). Section~\ref{Sec4} investigates the topological features of the spacetime. In Section~\ref{Sec5}, we examine the thermodynamic behavior of the BH, while Section~\ref{Sec6} focuses on thermodynamic topology using BH potential methods. Finally, the main findings and conclusions of the study are summarized in Section~\ref{Sec8}. We adopt the metric signature $(-,+,+,+)$ and work in natural units by setting $\hbar = G = c = 1$. 

\section{Schwarzschild BH with CoS immersed in KDM Halo: background metric}\label{Sec2}

We focus on analyzing the properties of a Schwarzschild BH modified by the presence of King-type dark matter (KDM)~\cite{king1962structure, kar2025diverse} with a cloud of strings proposed by Letelier \cite{Letelier1979}. The King model is particularly well-suited for describing the distribution of matter in galactic halos and has proven effective in characterizing the gravitational structure of galaxies across various regions. Owing to its mathematical tractability and its consistency with astrophysical observations, the King profile has been widely adopted in studies involving BHs and dark matter halos~\cite{konoplya2022solutions}. In Ref.~\cite{kar2025diverse}, the authors construct a spherically symmetric spacetime solution representing a Schwarzschild BH surrounded by a halo of King dark matter, which serves as the geometric foundation for the present work. Subsequently, accretion disk luminosity and topological characteristics of the same BH solution were recently studied in Ref. \cite{SZ2025}. 

The first studies concerning a formalism to treat gravity with a cloud of strings as the source, in the framework of GR, were presented by Letelier \cite{Letelier1979}. He obtained a generalization of the Schwarzschild solution corresponding to a BH surrounded by a spherically symmetric cloud of strings, whose energy-momentum tensor is given by \cite{Letelier1979}:
\begin{equation}
T^{\mu\nu}(\rm CoS)=\rho\,\frac{\Sigma^{\mu\beta}\,\Sigma^{\nu}_{\beta}}{\sqrt{-\gamma}},\label{aa3aa}
\end{equation}
where $\Sigma^{\mu\nu}$ is a bivector that represents the world sheet of the strings and is given by
\begin{equation}
    \Sigma^{\mu\nu}=\epsilon^{ab}\,\frac{\partial x^{\mu}}{d\lambda^{a}}\,\frac{\partial x^{\nu}}{d\lambda^{b}},\label{aa4aa}
\end{equation}
where $\epsilon^{ab}$ is the Levi-Civita bi-dimensional symbol and $\epsilon^{01}=-\epsilon^{10}=1$. By considering only the electric-like component $\Sigma_{01}$ with $\gamma<0$, the non-null components of the stress-energy tensor for the cloud of strings are given by \cite{Letelier1979}:
\begin{equation}
    T^{\mu}_{\nu}(\mbox{CoS})=\frac{\alpha}{8 \pi r^2}\mbox{diag}(-1,1,0,0),\label{aa4}
\end{equation}
where the constant $\alpha$ characterizes the influence of the string cloud on the space-time geometry.

Thereby, a static and spherically symmetric spacetime with a cloud of strings and surrounded by KDM halo is described by the following line element
\begin{eqnarray}
ds^2= -f(r) dt^2+\frac{dr^2}{f(r)}+r^2\, (d\theta^2+\sin^2\theta d\phi^2),\label{aa1}
\end{eqnarray}
where the metric function is given by
\begin{equation}
f(r)=1-\alpha-\frac{2\,m(r)}{r},\label{aa2}
\end{equation}
and $m(r)$ represents the mass function, which is considered, and Einstein's equations are solved, formulating the Einstein field equations as 
\begin{equation}
R_{\mu\nu} - \frac{1}{2}\, R\, g_{\mu\nu} = 8\pi \left(T^\text{CoS}_{\mu\nu}+T^\text{KDM}_{\mu\nu}\right).\label{aa3}
\end{equation}
Here, $g_{\mu\nu}$ is the spacetime metric, $R_{\mu\nu}$ is the Ricci tensor, $R$ is the Ricci scalar, and $T_{\mu\nu}$ is the total energy-momentum tensor defined as $T_{\mu\nu}=\text{diag}[-\rho_\text{eff},\,\tau_\text{eff},\,p,\,p]$, where $\rho_\text{eff}$ represents the effective energy density, $\tau_\text{eff}$ denotes the effective radial pressure, and $p$ is the tangential pressure. 

Through a series of meticulous calculations, the following results are obtained:
\begin{equation}
\rho_\text{KDM} = -\tau_{\rm KDM}=\frac{\partial_r m(r)}{4\pi r^2}\quad\text{and} \quad p_\text{KDM}=-\frac{\partial_r^2 m(r)}{8\pi r}.\label{aa5}
\end{equation}
The parameterized Dekel-Zhao density profile \cite{zhao1996analytical}, given by 
\begin{equation}
\rho_\text{DZ}(r)=\rho_0\frac{\left(\dfrac{r}{R}\right)^{\mu-3}}{\left(1+\left(\dfrac{r}{R}\right)^{\nu}\right)^{\frac{\mu+\alpha}{\nu}}},\label{aa6}
\end{equation}
where $R$ represents the scale radius, $\rho_0$ is the central density, and $(\mu,\, \nu,\, \alpha)$ are dimensionless parameters of the density profile, is then used.
Specifically, the King dark matter density profile is obtained for the values $(3,\,2,\,0)$ in the Dekel-Zhao profile \cite{king1962structure}, such that we have
\begin{equation}
\rho_\text{KDM}(r)=\rho_0\left(1+\left(\frac{r}{R}\right)^2\right)^{-3/2}.\label{aa7}
\end{equation}
This density has also been investigated under the designation of Beta model \cite{sofue2020rotation}.

Next, using Eq.~\eqref{aa7}, the metric function (\ref{aa2}) turns out to be
\begin{equation}
f(r)=1 -\alpha-\frac{2 M}{r}+\frac{8 \pi  \rho_0 R^3}{\sqrt{r^2+R^2}}+\frac{8 \pi  \rho_0 R^3}{r} \ln \left(\sqrt{1+\frac{r^2}{R^2}}-\frac{r}{R}\right).\label{aa8}
\end{equation}
From this expression, it is evident that in the limits of $\rho_0\to 0$ or $R\to 0$, the time-lapse function $f(r)$ approaches the Letelier BH solution.
\begin{figure}[tbhp]
     \centering
    \includegraphics[width=0.9\linewidth]{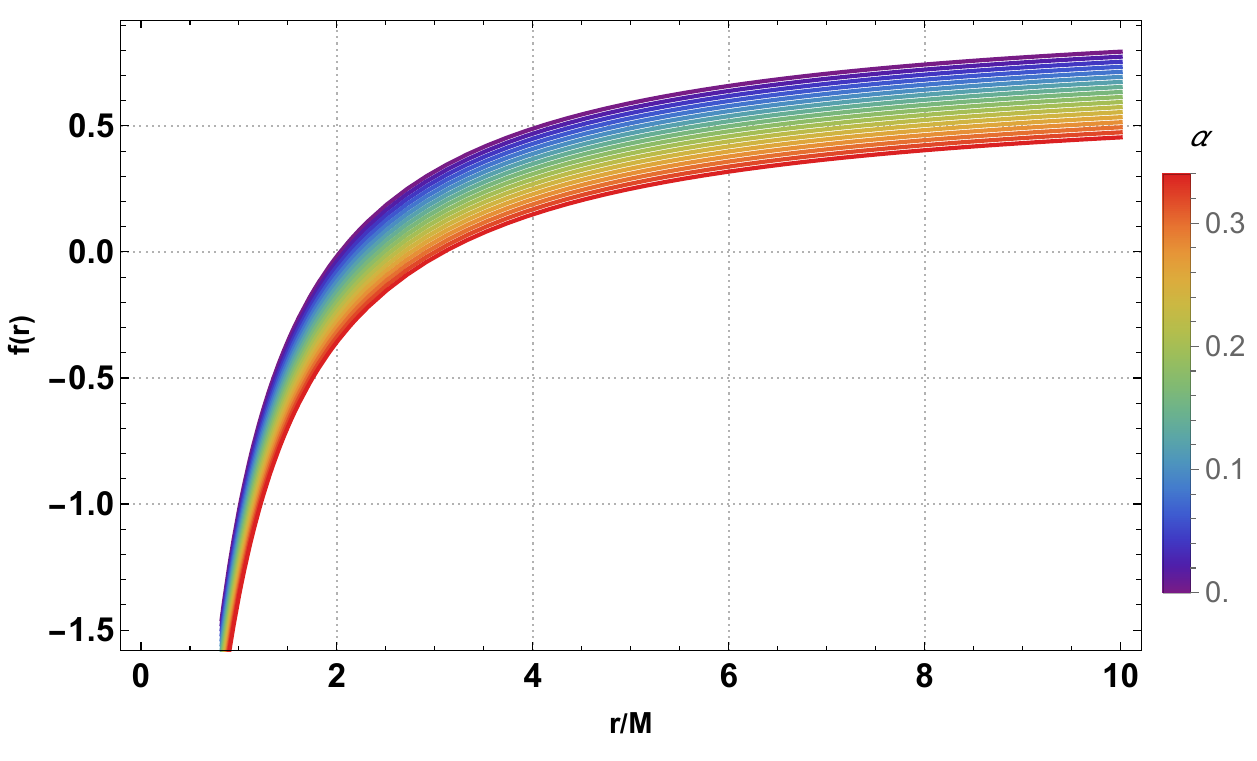}\\
    (i) $R/M=0.1$\\
    \includegraphics[width=0.9\linewidth]{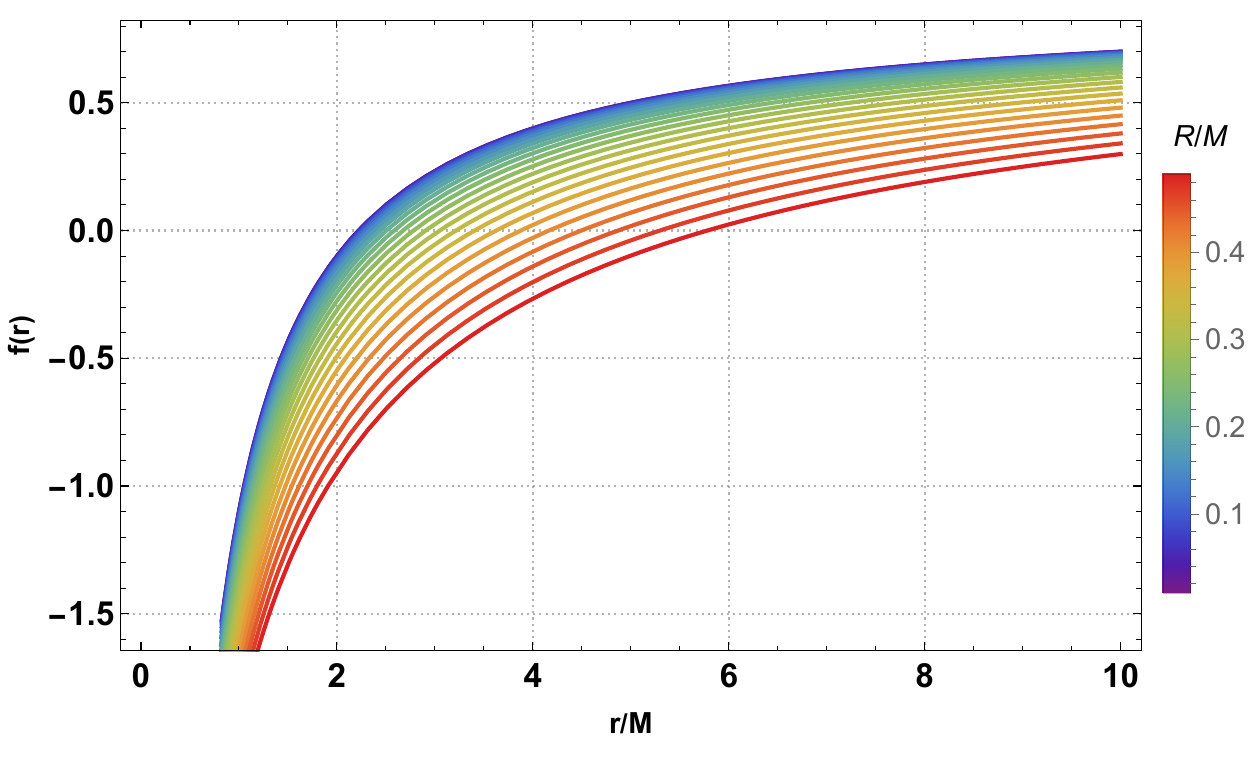}\\
    (ii) $\alpha=0.1$
  \caption{\footnotesize Behavior of the metric function $f(r)$ as radial coordinate $r$ for different values of string parameter $\alpha$ and the scale parameter $R/M$. Here $\rho_0\,M^2=0.5$.}
  \label{fig:metric}
\end{figure}
Figure~\ref{fig:metric} presents the behavior of the metric function $f(r)$ for different values of $\alpha$ and $R/M$. In both panels, increasing either $\alpha$ or $R/M$ led to a larger horizon radius, indicating that both the string cloud and the dark matter halo contribute to the expansion of the BH horizon.

\section{Geodesics Motions of Test Particles around a Letelier BH surrounded by King dark matter}\label{Sec3}

The study of BHs shadows and the paths of light rays near them has gained great importance, particularly following the Event Horizon Telescope (EHT) observations \cite{akiyama2019first,chael2021observing,johnson2023key}, because it provides important insights into the properties and behavior of BHs. To study the shadow, we first examine the BH's photon sphere, where light can orbit the BH in circular paths. As light rays approach this region, they can either escape to infinity or fall into the BH. This boundary creates a disk in the observer's sky, which is interpreted as the BH's shadow  \cite{perlick2022calculating,solanki2022photon}.

To obtain the shadow in the spherically symmetric spacetime introduced into Eq. \eqref{aa1}, for $\theta=\pi/2$, the Lagrangian $\mathcal{L}=\frac{1}{2}g_{\mu\nu}\dot{x}^\mu\dot{x}^\nu$ is written as \cite{chandrasekhar1998mathematical}
\begin{eqnarray}
\mathcal{L}=\frac{1}{2}\left(-f(r)\,\dot{t}^2+\frac{\dot{r}^2}{f(r)}+r^2\, \dot{\phi}^2\right),\label{bb1}
\end{eqnarray}
where $x^\mu=(t,\,r,\,\phi)$, $\dot{x}^{\mu}$ are the derivatives of the coordinates, and $g_{\mu\nu}$ represents the metric tensor of the spacetime. The Euler-Lagrange equations are expressed as
\begin{eqnarray}
\frac{d}{d\lambda}\left(\frac{\partial\mathcal{L}}{\partial\dot{x}^\mu}\right)-\frac{\partial\mathcal{L}}{\partial x^\mu}=0,\label{bb2}
\end{eqnarray}
where $\lambda$ is an affine parameter along the geodesic. 
Using the equations related to the $t$ and $\phi$ components, two constants of motion can be derived as
\begin{eqnarray}
\mathrm{E}=-\frac{\partial\mathcal{L}}{\partial\dot{t}}=f(r)\,\dot{t},\qquad \mathrm{L} =\frac{\partial\mathcal{L}}{\partial\dot{\phi}}=r^2\,\dot{\phi}.\label{bb3}
\end{eqnarray}
Thus, the equation of motion for the radial component is given by
\begin{align}
\dot{r}^2+\left(-\epsilon+\frac{\mathrm{L}^2}{r^2}\right)\,f(r)=\mathrm{E}^2,\label{bb4}
\end{align}
where $\epsilon=0$ for light-like particles  and $\epsilon=-1$ for time-like particles.
\begin{figure*}[tbhp]
  \centering
  \begin{minipage}{0.48\linewidth}
  \centering
      \includegraphics[width=\linewidth]{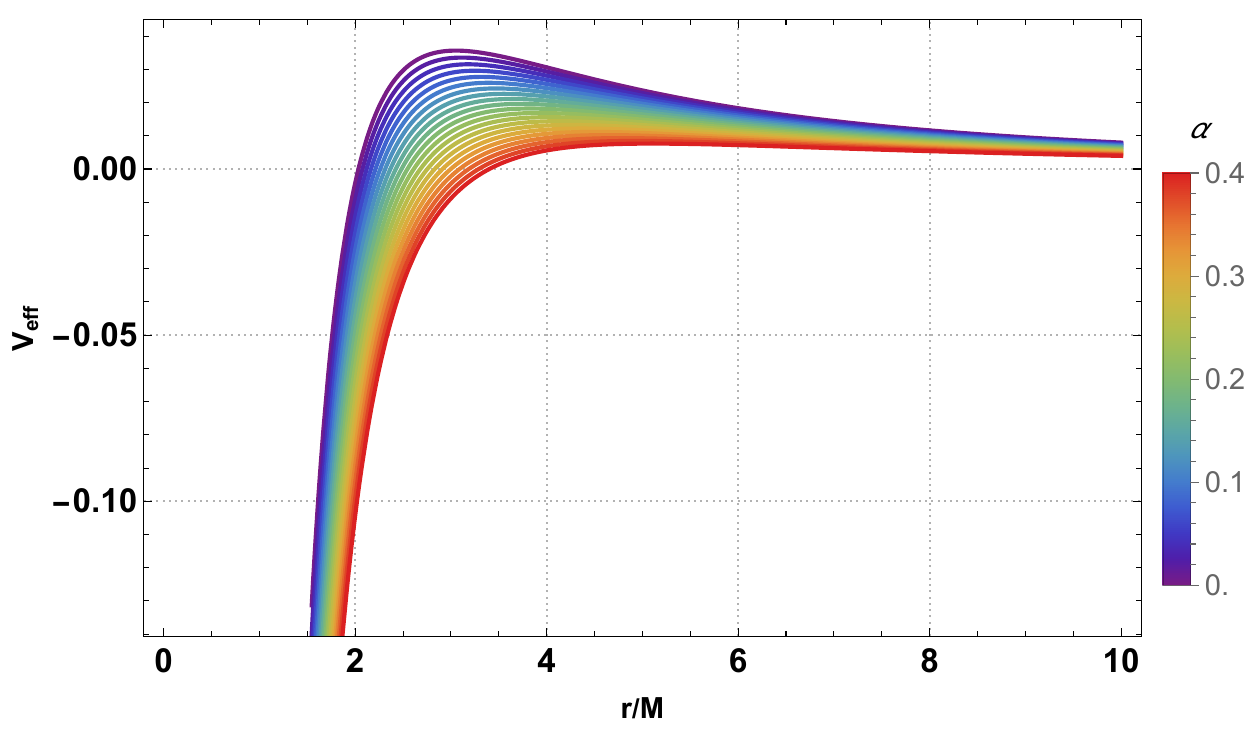}\\[3pt]
    \footnotesize (i) $R/M=0.1$
  \end{minipage}
  \hfill
  \begin{minipage}{0.48\linewidth}
  \centering
    \includegraphics[width=\linewidth]{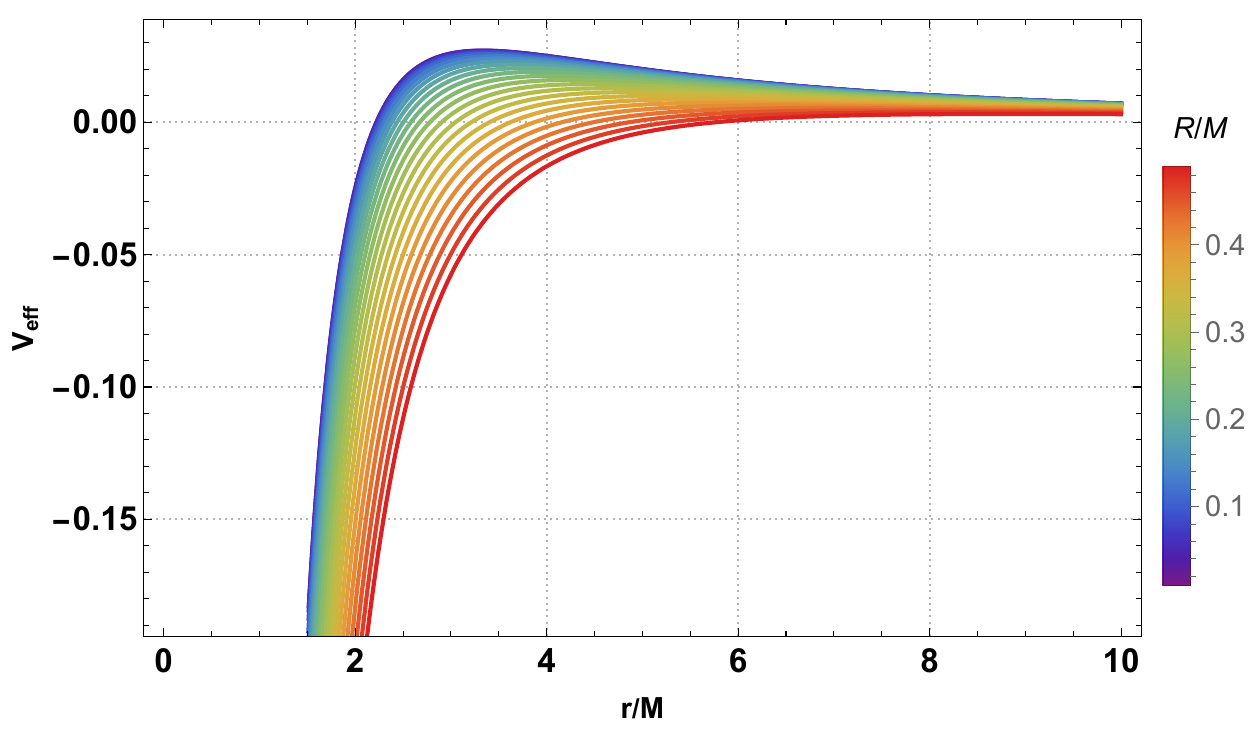}\\[3pt]
    \footnotesize (ii) $\alpha=0.1$
  \end{minipage}
  \caption{\footnotesize Behavior of the effective potential $V_{\rm eff}$ that governs the photon dynamics for different values of the CoS parameter $\alpha$ and the scale parameter $R/M$. Here $\rho_0\,M^2=0.5$ and $\mathrm{L}/M=1$.}
  \label{fig:null-potential}
\end{figure*}

\begin{figure*}[tbhp]
  \centering
  \begin{minipage}{0.48\linewidth}
    \centering
    \includegraphics[width=0.9\linewidth]{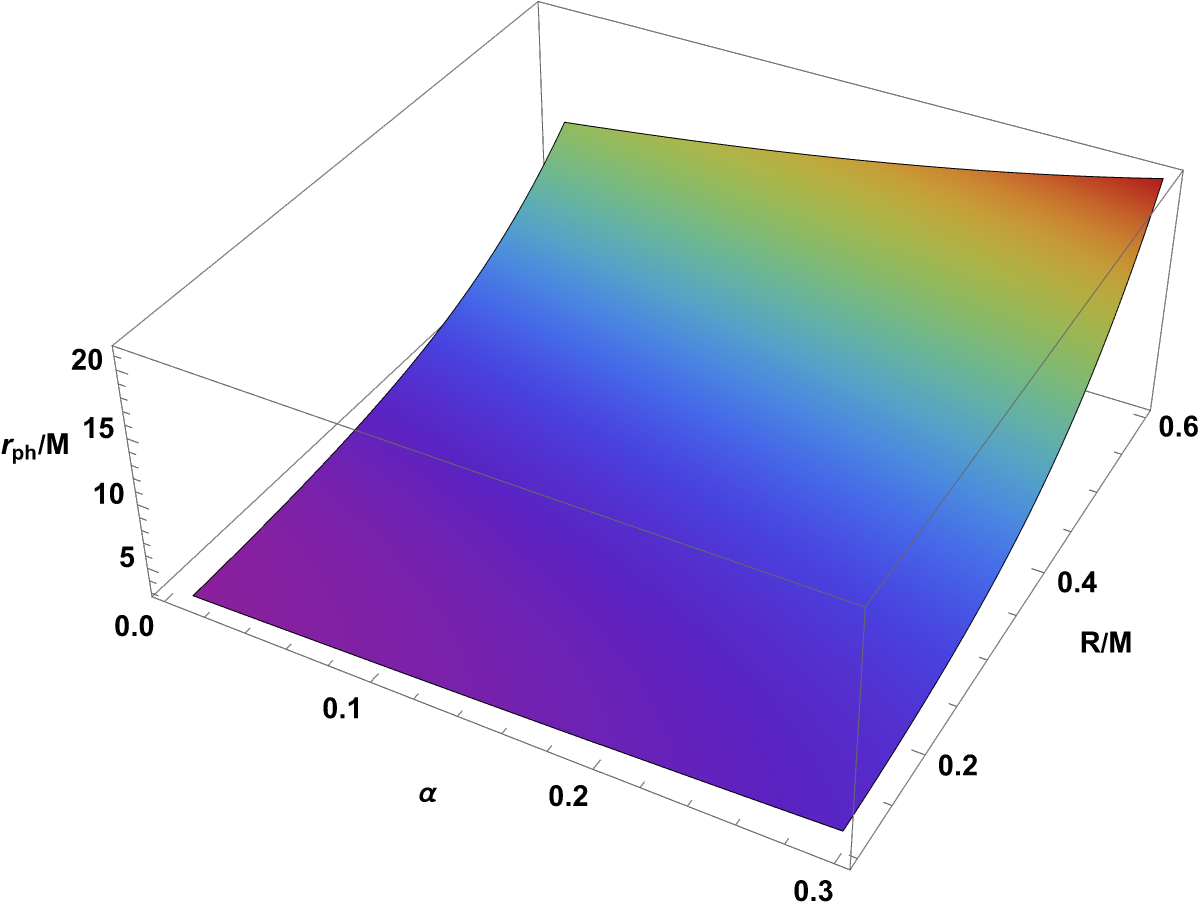}\\[3pt]
    \footnotesize (i) $\rho_0\,M^2=0.5$
  \end{minipage}\hfill
  \begin{minipage}{0.48\linewidth}
    \centering
    \includegraphics[width=0.9\linewidth]{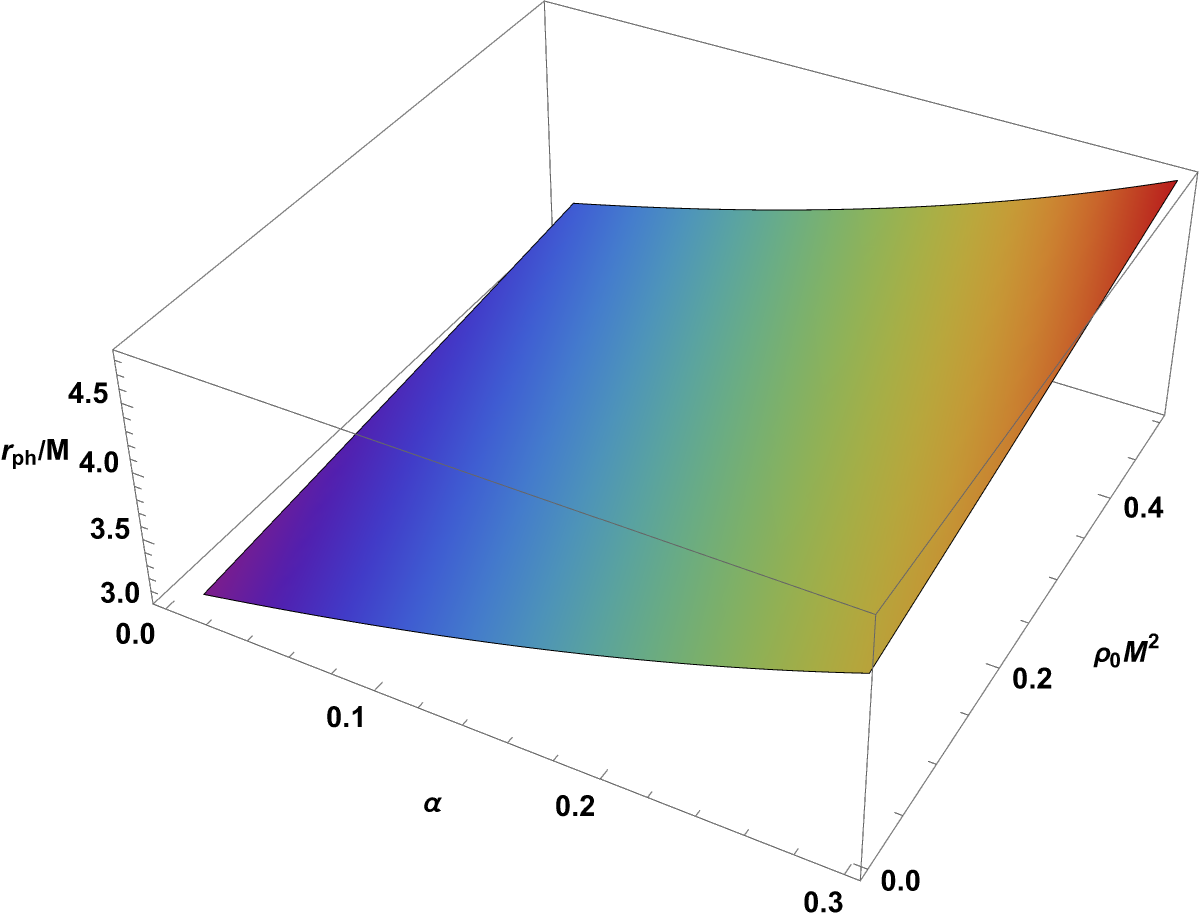}\\[3pt]
    \footnotesize (ii) $R/M=0.2$
  \end{minipage}
  \caption{\footnotesize 3D plot of the photon sphere radius $r_{\rm ph}/M$ as a function of $(\alpha, R/M)$ and $(\alpha,\rho_0\,M^2)$.}
  \label{fig:photon-sphere}
\end{figure*}

Figure~\ref{fig:null-potential} presents the behavior of the effective potential that governs the dynamics of photon particles for different values of $\alpha$ and $R/M$. In both panels, increasing either $\alpha$ or $R/M$ reduces the effective potential, indicating that both the string cloud and the dark matter halo have an impact on photon dynamics.

\vspace{0.2cm}
\begin{center}
    \large{\bf A.\,\,Photon Dynamics}
\end{center}
\vspace{0.2cm}

In this part, we study the dynamics of photons in the gravitational field of the BH solution. We analyze the photon sphere radius, the effective radial force experienced by photons, and the resulting BH shadow size. Particular attention is given to how these optical features are influenced by the key geometric parameters, including the string parameter and the KDM halo profile.

For light-like particles, Eq. \eqref{bb4} can be expressed as
\begin{align}
\dot{r}^2+V_{\rm eff}(r)=\mathrm{E}^2,\label{bb5}
\end{align}
where the effective potential governs the photon dynamics, is given by
\begin{widetext}
\begin{equation}
    V_\text{eff}=\frac{\mathrm{L}^2}{r^2}\,f(r)=\frac{\mathrm{L}^2}{r^2}\,\left[1 -\alpha-\frac{2 M}{r}+\frac{8 \pi  \rho_0 R^3}{\sqrt{r^2+R^2}}+\frac{8 \pi  \rho_0 R^3}{r} \ln \left(\frac{\sqrt{r^2+R^2}-r}{R}\right)\right].\label{bb6}
\end{equation}
\end{widetext}

For the above expression, it becomes clear that the effective potential governs the dynamics of photon particles and is influenced by the KDM halo profile characterized by parameters $(R, \rho_0)$  and the string parameter $\alpha$. Moreover, the angular momentum $\mathrm{L}$ alters this potential too.

Due to the circular orbit at radius $r=r_{\rm ph}$, the following conditions must be satisfied: \cite{perlick2022calculating,virbhadra2000schwarzschild}
\begin{eqnarray}
\dot{r}=0\Longrightarrow \mathrm{L}^2\,\frac{f(r_{\rm ph})}{r^2_{\rm ph}}=\mathrm{E}^2\Longrightarrow \beta_{\rm crit.}=\frac{r_{\rm ph}}{f(r_{\rm ph})}.\label{bb7}
\end{eqnarray}
And
\begin{eqnarray}
\ddot{r}=0\Longrightarrow \partial_{r}\left(\frac{r^2}{f(r)}\right)\Big{|}_{r=r_{\rm ph}}=0.\label{bb8}
\end{eqnarray}

Using the above equation (\ref{bb8}), the BH photon sphere radius, $r=r_{\rm ph}$, would be found. This radius satisfies the following polynomial equation:
\begin{align}
&1 - \alpha - \frac{3M}{r}
+ \frac{12\pi \rho_0 R^3}{\sqrt{r^2 + R^2}}+ \frac{4\pi \rho_0 R^3 r^2}{(r^2 + R^2)^{3/2}}\notag\\&
+ \frac{12\pi \rho_0 R^3}{r} \ln\left( \frac{\sqrt{r^2 + R^2} - r}{R} \right)=0.\label{bb9}
\end{align}
The resulting equation is a transcendental equation in $r$ that can be expressed as an infinite-degree polynomial. Finding an exact analytical solution is highly non-trivial and generally infeasible. However, by choosing suitable values for the parameters, the photon sphere radius can be determined numerically.

For graphical representation of the photon sphere radius $r=r_{\rm ph}$, we employ dimensionless variable $x=r/M$, $y=R/M$ and $\rho_0\,M^2=z$, we can rewrite the above equation (\ref{bb9}) as
\begin{align}
&1 - \alpha - \frac{3}{x}
+ \frac{12 \pi y^3\,z}{\sqrt{x^2 + y^2}}+ \frac{4\pi x^2 y^3 z}{(x^2 + y^2)^{3/2}}
\notag\\&+ \frac{12 \pi z y^3}{x} \ln\left( \frac{\sqrt{x^2 + y^2} - x}{y} \right)=0.\label{bb10}
\end{align}

For the above expression (\ref{bb9}) or (\ref{bb10}), it becomes clear that the radius of the photon sphere is influenced by the KDM halo profile characterized by the parameters $(R, \rho_0)$ and the string parameter $\alpha$. In the limit $\rho_0$ or $R \to 0$, which corresponds to the absence of the KDM halo, one can find the photon sphere radius $r_{\rm ph-Letelier-BH}=\frac{3\,M}{1-\alpha}> r_{\rm ph-Sch}$, which is similar to the Leterlier BH solution \cite{Letelier1979}.

Figure \ref{fig:photon-sphere} shows a 3D plot of the photon sphere radius $r_{\mathrm{ph}}/M$ as a function of parameters $(\alpha, R/M)$ in the left panel and $(\alpha, \rho_{0} M^{2})$ in the right panel. The left panel demonstrates that the photon sphere radius increases with both $\alpha$. This means that the combined effect of the string cloud and scale parameter expands the region where photons can execute unstable circular orbits, which directly impacts the size of the BH shadow.

Now, we examine the shadow cast by the selected BH. Noted that the background space-time geometry described by the line element (\ref{aa1}) with the given metric function (\ref{aa8}) is not asymptotically flat. This is obvious since at $r \to \infty$, the metric function behaves as
\begin{equation}
    \lim_{ r \to \infty} f(r) \sim 1-\alpha-8\pi \rho_0 R^3 \frac{\mbox{ln}r}{r}+\mathcal{O} \left(\frac{1}{r^2}\right).\label{condition}
\end{equation}
A geometry based method determines the angular radius $\vartheta$ of the black
hole’s shadow, discussed in detail in Ref. \cite{perlick2022calculating}. This is given by
\begin{align}
    \sin^2 \vartheta_{\rm sh}&=\frac{h^2(r_{\rm ph})}{h^2(r_{\rm obs})},\nonumber\\
    h(r)&=\frac{r^2}{f(r)},\label{size-1}
\end{align}
where $r_{\rm ph}$ is the photon sphere radius and $r_{\rm obs}$ is the position of an observer.

Due to EHT observations for Sgr A$^*$, it is known that the acceptable range for the shadow radius in the $1\sigma$ and $2\sigma$ regions are $4.55<R_{\rm sh}/M<5.22$ and $4.21<R_{\rm sh}/M<5.56$, respectively \cite{vagnozzi2023horizon}. Similarly, for M 87$^*$ the shadow radius for the $1\sigma$ and $2\sigma$ regions should be within the ranges $4.26<R_{\rm sh}/M<6.03$ and $3.38<R_{\rm sh}/M<6.91$, respectively \cite{akiyama2019first}. Therefore, by comparing the BH shadow calculated from Eq. \eqref{size-1} with the boundaries of the $1\sigma$ and $2\sigma$ regions, we can set upper and lower bounds on the parameters that would allow the shadow to fall within these ranges. In the limit where $R$ or $\rho_0$ approaches zero, the shadow of the BH in the presence of King dark matter corresponds to the shadow of the Letelier BH as $R_{sh}=3\sqrt{3} M (1-\alpha)^{-3/2}$. 


Next, we focus on the photon trajectory and show how geometric and physical parameters alter the photon’s path. The equation of orbit using Eqs.~(\ref{bb3}) and (\ref{bb5})-(\ref{bb6}) is given by
\begin{align}
&\left(\frac{1}{r^2}\,\frac{dr}{d\phi}\right)^2=\frac{1}{\beta^2}-\frac{1}{r^2}\,\Bigg(1 -\alpha-\frac{2 M}{r}+\frac{8 \pi \rho_0 R^3}{\sqrt{r^2+R^2}}\notag\\&+\frac{8 \pi \rho_0 R^3}{r} \ln \left(\frac{\sqrt{r^2+R^2}-r}{R}\right)\Bigg).\label{bb13}
\end{align}    
Transforming to a new variable via $r(\phi)=\frac{1}{u(\phi)}$, we find
\begin{align}
    &\left(\frac{du}{d\phi}\right)^2=\frac{1}{\beta^2}-u^2\,\Bigg[1 -\alpha-2 M u \notag\\&+\frac{8 \pi \rho_0 R^3 u}{\sqrt{1+u^2 R^2}}+8 \pi \rho_0 R^3 u \ln \left(\frac{\sqrt{1+u^2 R^2}-1}{u R}\right)\Bigg].\label{bb14}
\end{align}    
Differentiating both sides w. r. t. $\phi$ and after simplification, results
\begin{align}
    &\frac{d^2u}{d\phi^2}+(1-\alpha)\,u=3\,M\,u^2-\frac{12 \pi \rho_0 R^3 u^2}{\sqrt{1+u^2 R^2}}\notag\\
    &+\frac{4 \pi \rho_0 R^5 u^4}{(1+u^2 R^2)^{3/2}}-12 \pi \rho_0 R^3 u^2 \ln \left(\frac{\sqrt{1+u^2 R^2}-1}{u R}\right)\nonumber\\
    &-\frac{4 \pi \rho_0 R^3 u^3}{\sqrt{1+u^2 R^2}-1}\,\left(\frac{u^2 R^3-R\,(\sqrt{1+u^2 R^2}-1)}{u R}\right).\label{bb15}
\end{align}

The above second-order differential equation describes the photon trajectory in the given gravitational field. It becomes evident that this trajectory is governed by a highly non-linear differential equation. Hence, the deflection of photon particles in the gravitational field is influenced by the KDM halo profile characterized by the parameters $(R, \rho_0)$ and the string parameter $\alpha$. However, in the limit where $R \to 0$ or $\rho_0 \to 0$, the above photon trajectory reduces to the result of the Letelier BH solution.
\begin{figure*}[htbp]
  \centering
  \begin{minipage}{0.48\linewidth}
    \centering
    \includegraphics[width=\textwidth]{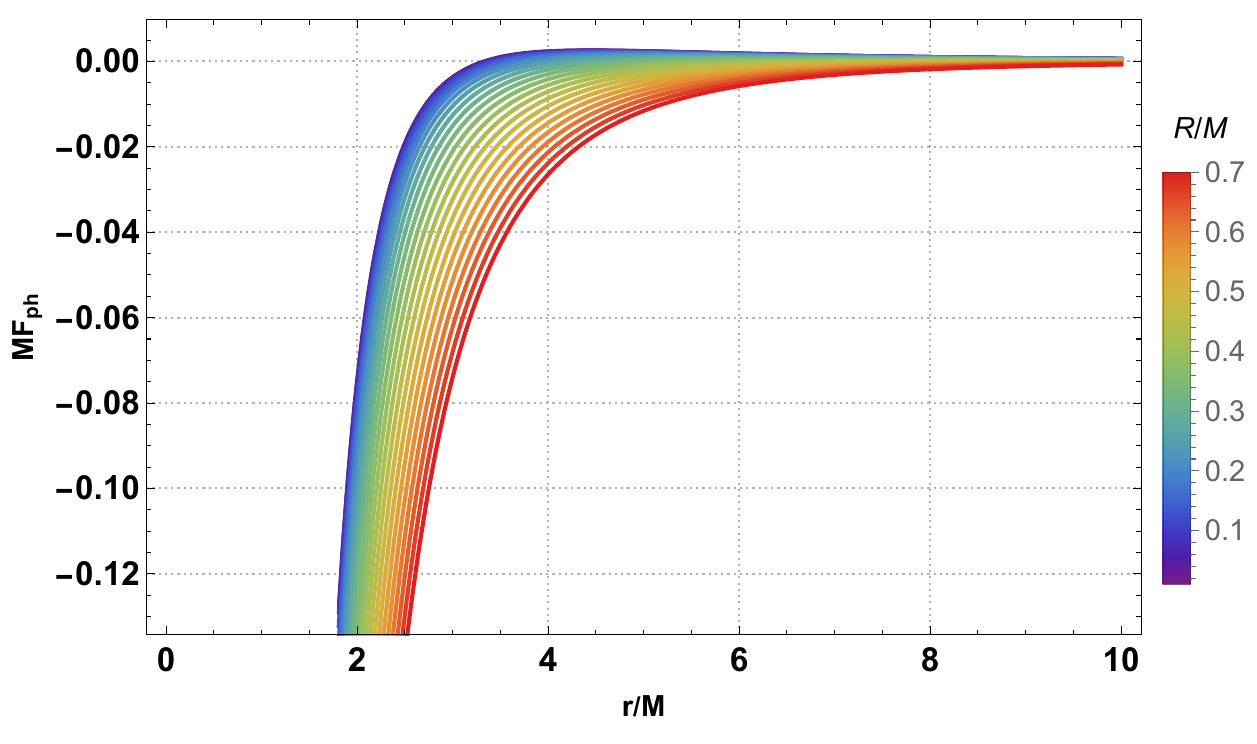}\\[3pt]
    \footnotesize (i) $\rho_0\,M^2=0.5$
  \end{minipage}
  \hfill
  \begin{minipage}{0.48\linewidth}
    \centering
    \includegraphics[width=\textwidth]{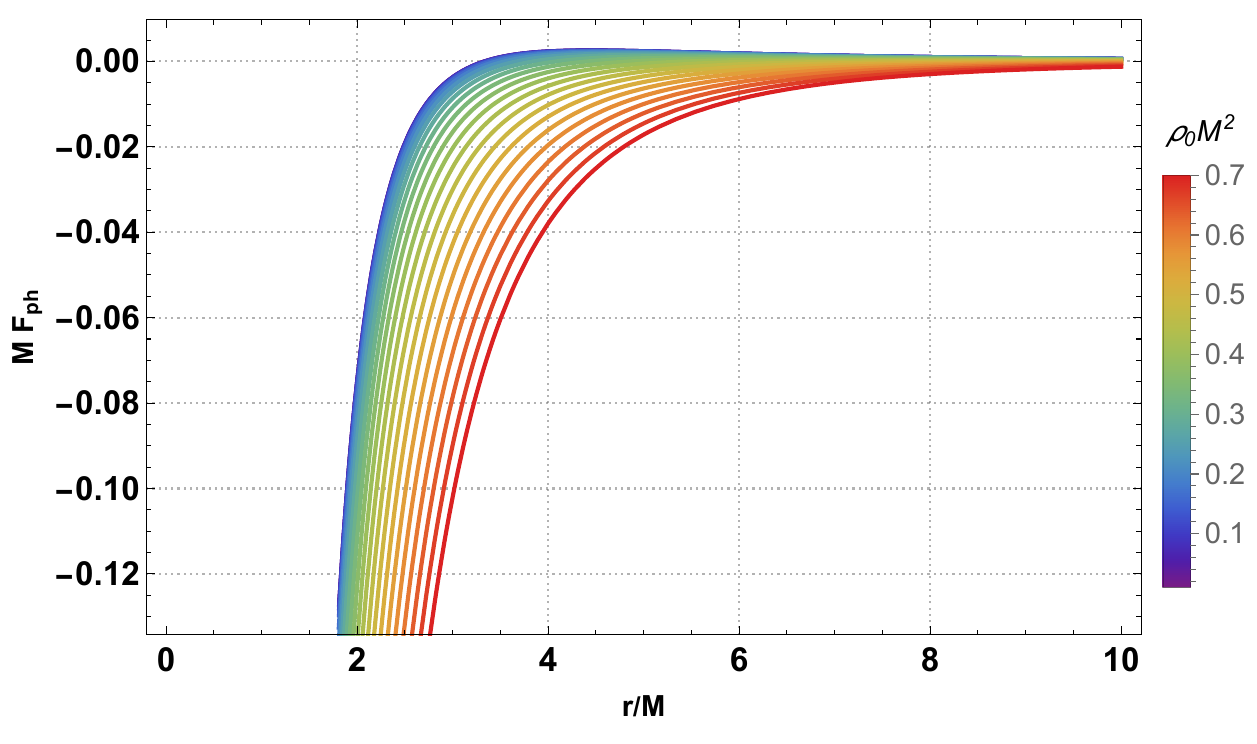}\\[3pt]
    \footnotesize (ii) $R/M=0.2$
  \end{minipage}
  \caption{\footnotesize Behavior of the effective radial force $MF_{\rm eff}$. Here $\alpha=0.1$ and $\mathrm{L}/M=1$.}
  \label{fig:force}
\end{figure*}

Finally, we determine the effective radial force experienced by the photon particles in the gravitational field. This effective force can be defined in terms of the effective potential that governs the photon dynamics as the negative gradient given by $F_{\rm ph}=-\frac{1}{2}\,\frac{dV}{dr}$. In our case, the expression of effective radial force is given by
\begin{align}\label{force}
&F_{\rm ph}=\frac{\mathrm{L}^2}{r^3}\,\Bigg[1 - \alpha - \frac{3M}{r}
+ \frac{12\pi \rho_0 R^3}{\sqrt{r^2 + R^2}}+ \frac{4\pi \rho_0 R^3 r^2}{(r^2 + R^2)^{3/2}}\notag\\&
+ \frac{12\pi \rho_0 R^3}{r} \ln\left( \frac{\sqrt{r^2 + R^2} - r}{R} \right)\Bigg]
\end{align}

By employing dimensionless variables $x=r/M$, $y=R/M$ and $\rho_0\,M^2=z$, we can rewrite the above equation (\ref{force}) as
\begin{align}\label{force2}
    &\frac{M^3 F_{\rm ph}}{\mathrm{L}^2}=\frac{1}{x^3}\,\Bigg[1 - \alpha - \frac{3}{x}
+ \frac{12\pi y^3 z}{\sqrt{x^2 + y^2}}\notag\\&+ \frac{4\pi x^2 y^3 z}{(x^2 + y^2)^{3/2}}
+ \frac{12\pi y^3 z}{x} \ln\left( \frac{\sqrt{x^2 + y^2} - x}{y} \right)\Bigg]
\end{align}

For the above expression (\ref{force}), it becomes clear that the effective radial force experienced by the photon particles is influenced by the KDM halo profile characterized by the parameters $(R, \rho_0)$ and the string parameter $\alpha$. In the limit $\rho_0$ or $R \to 0$ corresponds to the absence of KDM halo, one can obtain this radial force $F_{\rm ph-Letelier-BH}=\frac{\mathrm{L}^2}{r^3}\,\left(1-\alpha-\frac{3\,M}{r}\right)$ which is similar to that result obtained for the Leterlier BH solution \cite{Letelier1979}.

Figure \ref{fig:force} illustrates the behavior of the effective radial force for different values of the halo parameters $(R, \rho_0)$. We observed that as the values of $R/M$ and $\rho_0\,M^2$ increase, the effective radial force decreases, indicating lesser attraction by the gravitational field when halo parameters increase. 

\begin{center}
    \large{\bf B.\,\,Particle Dynamics}
\end{center}

The motion of test particles around BHs shows key aspects of gravitational interactions and spacetime structure. A fundamental concept is the innermost stable circular orbit (ISCO), which marks the transition between stable and plunging trajectories. For a Schwarzschild BH, the ISCO lies at $r_{\rm ISCO-Sch}=6M$. Modifications in the surrounding matter content or in the gravity theory itself, such as the inclusion of string clouds \cite{Letelier1979} or DM halo, can alter the ISCO location and particle dynamics. The effective potential method is commonly used to study such motion.

To study the dynamics of massive particles in the gravitational field, we employ a model similar to the one in the previous section. According to Eq. \eqref{bb4}, the effective potential for timelike particles is given by
\begin{eqnarray}\label{cc1}
V_{\text{eff}}(r)=f(r)\,\left(1+\frac{\mathrm{L}^2}{r^2}\right),
\end{eqnarray}
and the equation of motion of the radial components can be expressed as
\begin{eqnarray}
\dot{r}^2+V_{\rm eff}(r)=\mathrm{E}^2.\label{cc2}
\end{eqnarray}
\begin{figure*}[htbp]
  \centering
  \begin{minipage}{0.48\linewidth}
    \centering
    \includegraphics[width=\linewidth]{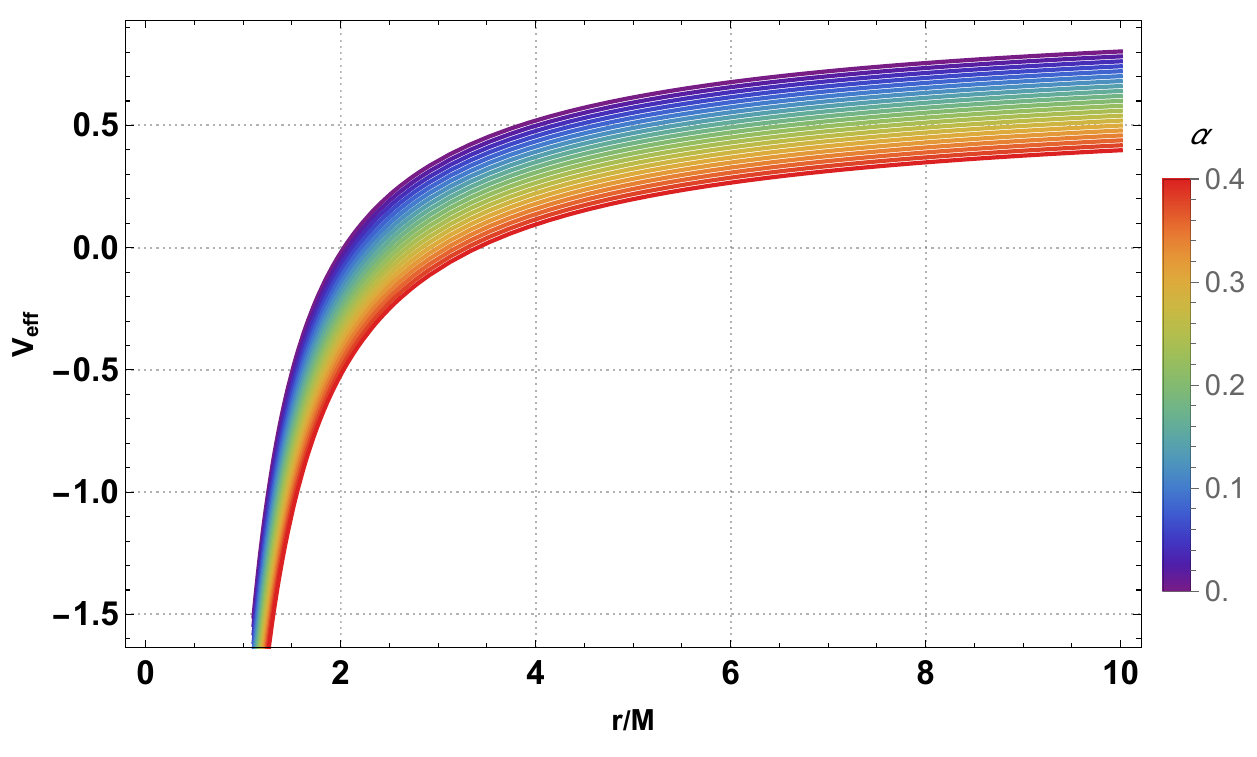}\\[3pt]
    \footnotesize (i) $R/M=0.1$
  \end{minipage}\hfill
  \begin{minipage}{0.48\linewidth}
    \centering
    \includegraphics[width=\linewidth]{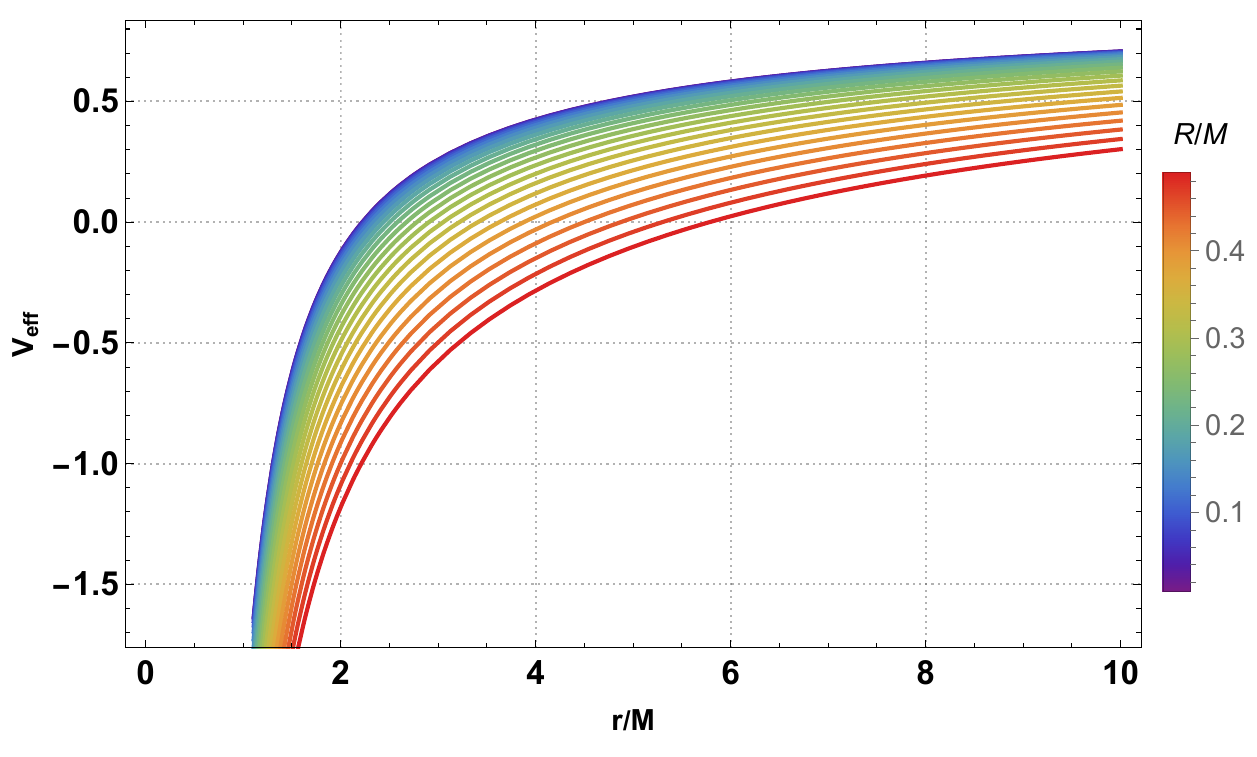}\\[3pt]
    \footnotesize (ii) $\alpha=0.1$
  \end{minipage}
  \caption{\footnotesize Behavior of the effective potential $V_{\rm eff}$ that governs the dynamics of test particles for different values of the CoS parameter $\alpha$ and the scale parameter $R/M$. Here $\rho_0\,M^2=0.5$ and $\mathrm{L}/M=1$.}
  \label{fig:timelike-potential}
\end{figure*}

Figure~\ref{fig:timelike-potential} presents the behavior of the effective potential that governs the dynamics of massive test particles for different values of $\alpha$ and $R/M$. In both panels, increasing either $\alpha$ or $R/M$ reduces the effective potential, indicating that both the string cloud and the dark matter halo have an impact on the dynamics of massive particles.

The study of the circular orbits of particles in the accretion disk is crucial for calculating the luminosity.
For a marginally stable circular trajectory denoting the ISCO radius, which is the closest stable circular orbit that massive particles can maintain around a BH, the following conditions must be satisfied:
\begin{align}
\dot{r}&=0\longrightarrow V_{\text{eff}}(r)=\mathrm{E}^2,\label{cc3a}\\
\ddot{r}&=0\longrightarrow \partial_r V_{\text{eff}}(r)=0,\label{cc3b}\\
\dddot{r}& \geq 0\longrightarrow \partial^2_r V_{\text{eff}}(r) \geq 0.\label{cc3c}
\end{align}
Using Eq. \eqref{cc3c}, one can determine the ISCO radius, which is calculated from the following relation:
\begin{equation}\label{cc4}
\left(r\,f(r)\,f''(r)-2\,r\,(f'(r))^2+3\,f(r)\,f'(r)\right)\Big{|}_{r=r_{\rm ISCO}}=0.
\end{equation}

\squeezetable 
\begin{table}[tbhp]
\centering
\scriptsize
\setlength{\tabcolsep}{2pt}
\renewcommand{\arraystretch}{1.05}
\resizebox{\columnwidth}{!}{
\begin{tabular}{|c|c|c|c|c|c|c|}
\hline
\multirow{2}{*}{$\alpha$ $\downarrow$} & \multirow{2}{*}{$R/M$ $\downarrow$} &
\multicolumn{5}{c|}{$r_{\rm ISCO}/M$} \\
\cline{3-7}
 & & $\rho_0 M^2=0.1$ & & $\rho_0 M^2=0.2$ & & $\rho_0 M^2=0.3$ \\
\hline
0.05 & 0.05 & 6.31896 & & 6.32214 & & 6.32532 \\
     & 0.10 & 6.33572 & & 6.35570 & & 6.37576 \\
     & 0.15 & 6.37242 & & 6.42968 & & 6.48755 \\
     & 0.20 & 6.43272 & & 6.55275 & & 6.67583 \\
\hline
0.10 & 0.05 & 6.67007 & & 6.67348 & & 6.67689 \\
     & 0.10 & 6.68815 & & 6.70971 & & 6.73133 \\
     & 0.15 & 6.72798 & & 6.78995 & & 6.85259 \\
     & 0.20 & 6.79372 & & 6.92412 & & 7.05781 \\
\hline
0.15 & 0.05 & 7.06249 & & 7.06617 & & 7.06984 \\
     & 0.10 & 7.08208 & & 7.10541 & & 7.12881 \\
     & 0.15 & 7.12545 & & 7.19280 & & 7.26086 \\
     & 0.20 & 7.19742 & & 7.33963 & & 7.48540 \\
\hline
0.20 & 0.05 & 7.50397 & & 7.50794 & & 7.51192 \\
     & 0.10 & 7.52528 & & 7.55064 & & 7.57608 \\
     & 0.15 & 7.57272 & & 7.64622 & & 7.72050 \\
     & 0.20 & 7.65184 & & 7.80762 & & 7.96726 \\
\hline
0.25 & 0.05 & 8.00432 & & 8.00864 & & 8.01296 \\
     & 0.10 & 8.02761 & & 8.05531 & & 8.08310 \\
     & 0.15 & 8.07976 & & 8.16037 & & 8.24183 \\
     & 0.20 & 8.16718 & & 8.33865 & & 8.51434 \\
\hline
0.30 & 0.05 & 8.57615 & & 8.58087 & & 8.58559 \\
     & 0.10 & 8.60176 & & 8.63218 & & 8.66270 \\
     & 0.15 & 8.65939 & & 8.74830 & & 8.83812 \\
     & 0.20 & 8.75652 & & 8.94632 & & 9.14076 \\
\hline
\end{tabular}}
\caption{\footnotesize Numerical results for the ISCO radius $r_{\rm ISCO}/M$ as a function of $\alpha$, $R/M$, and $\rho_0 M^2$ of the Letelier BH with KDM halo.}
\label{tab:1}
\end{table}

The ISCO radius can, in principle, be obtained by solving the above polynomial relation in $r$ analytically. However, finding an exact closed-form expression is quite challenging. Instead, one can determine the ISCO radius numerically by fixing suitable values for the parameters involved in the polynomial. 

Table~\ref{tab:1} presents numerical values of the ISCO radius for varying values of the string parameter $\alpha$, and the halo parameters $R/M$ and $\rho_0 M^2$. From this table, it is evident that the ISCO radius $r = r_{\rm ISCO}$ is significantly influenced by the KDM halo profile characterized by $(R, \rho_0)$, as well as by the string parameter $\alpha$. In the limiting case where $\rho_0 \to 0$ or $R \to 0$, corresponding to the absence of the KDM halo, the ISCO radius reduces to $ r_{\rm ISCO-Letelier-BH} = \frac{6M}{1-\alpha}, $ which matches the known result for the Letelier black hole solution \cite{Letelier1979}. Increasing the halo parameters $R/M$ and $\rho_0 M^2$ leads to an expansion of the ISCO radius. Similarly, increasing the string parameter $\alpha$ also enlarges the ISCO radius. Due to modifications in the gravitational field, the impact of increasing $R/M$ or $\rho_0 M^2$ on the ISCO radius becomes more pronounced at larger radii compared to smaller ones.

On the other hand, the specific angular velocity of the test particles in the accretion disk using the equation \eqref{bb3} and (\ref{cc1}) is derived from
\begin{widetext}
\begin{align}\label{cc5}
\Omega_{\phi}(r)&=\sqrt{\frac{\partial_r f(r)}{2\,r}}=\frac{1}{r}\sqrt{\frac{M}{r}-\frac{4 \pi  \rho_0 R^3}{r}\,\ln \left(\frac{\sqrt{r^2+R^2}-r}{R}\right)-\frac{4 \pi  \rho_0 R^3 \left(2 r^2+R^2\right)}{\left(r^2+R^2\right)^{3/2}}}.
\end{align}    
\end{widetext}
\begin{figure}[htbp]
    \centering
    \includegraphics[width=0.48\linewidth]{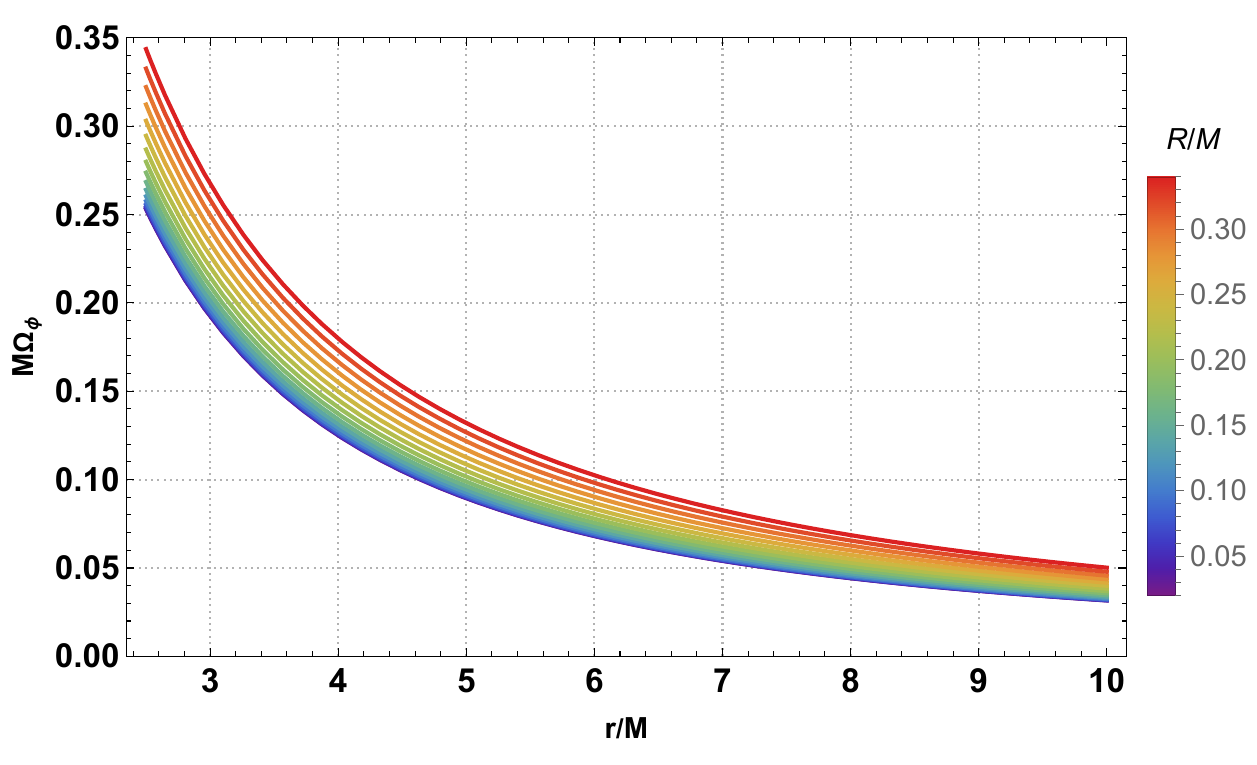}\qquad
    \includegraphics[width=0.48\linewidth]{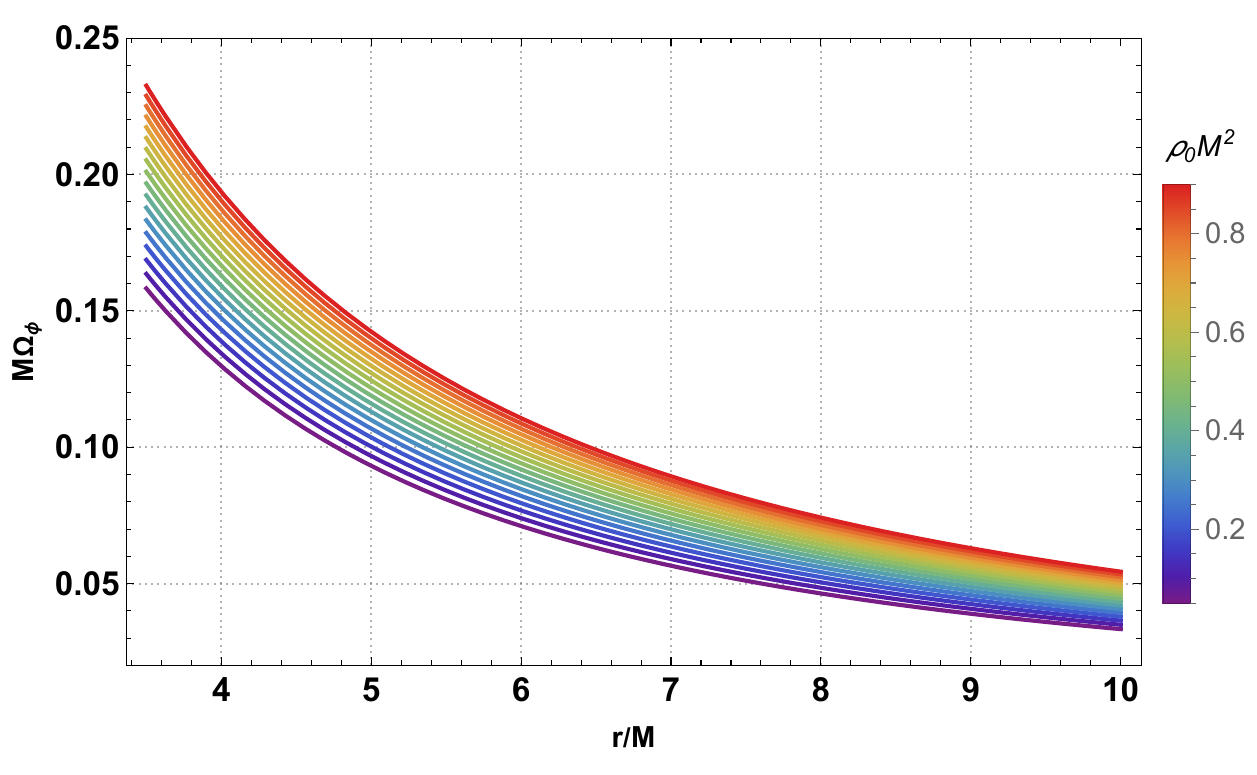}\\
    (i) $\rho_0=0.5/M^2$ \hspace{8cm} (ii) $R=0.3M$
    \caption{\footnotesize Behavior of the angular velocity term $M\Omega_{\phi}$ of test particles.}
    \label{fig:angular-velocity}
\end{figure}

Furthermore, using the equations (\ref{cc3a}) and (\ref{cc3b}), one can determine the specific energy and the specific angular momentum of a test particle, and these are given by
\begin{widetext}
\begin{equation}\label{cc6}
\mathrm{E}_\text{sp}(r)=\pm\,\sqrt{\frac{2}{2\,f(r)-r\,f'(r)}}\,f(r)=
\pm\,\dfrac{\left(1 -\alpha-\frac{2 M}{r}+\frac{8 \pi  \rho_0 R^3}{\sqrt{r^2+R^2}}+\frac{8 \pi  \rho_0 R^3}{r} \ln \left(\frac{\sqrt{r^2+R^2}-r}{R}\right)\right)}{\sqrt{1 - \alpha - \frac{3M}{r}
+ \frac{12\pi \rho_0 R^3}{\sqrt{r^2 + R^2}}+ \frac{4\pi \rho_0 R^3 r^2}{(r^2 + R^2)^{3/2}}
+ \frac{12\pi \rho_0 R^3}{r} \ln\left( \frac{\sqrt{r^2 + R^2} - r}{R} \right)}}
,
\end{equation}
and
\begin{equation}\label{cc7}
\mathrm{L}_\text{sp}(r)=r\,\sqrt{\frac{r\,f'(r)}{2\,f(r)-r\,f'(r)}}=
\frac{r \sqrt{\frac{M}{r}
- \frac{4\pi \rho_0 R^3 r^2}{(r^2 + R^2)^{3/2}}+ 4\pi \rho_0 R^3 \left[
-\frac{1}{r}\ln\left( \frac{\sqrt{r^2 + R^2} - r}{R} \right)
- \frac{1}{\sqrt{r^2 + R^2}}\right]}}{\sqrt{1 - \alpha - \frac{3M}{r}
+ \frac{12\pi \rho_0 R^3}{\sqrt{r^2 + R^2}}+ \frac{4\pi \rho_0 R^3 r^2}{(r^2 + R^2)^{3/2}}
+ \frac{12\pi \rho_0 R^3}{r} \ln\left( \frac{\sqrt{r^2 + R^2} - r}{R} \right)}}.
\end{equation}    
\end{widetext}
\begin{figure*}[tbhp]
  \centering
  \begin{minipage}{0.48\linewidth}
    \centering
    \includegraphics[width=\linewidth]{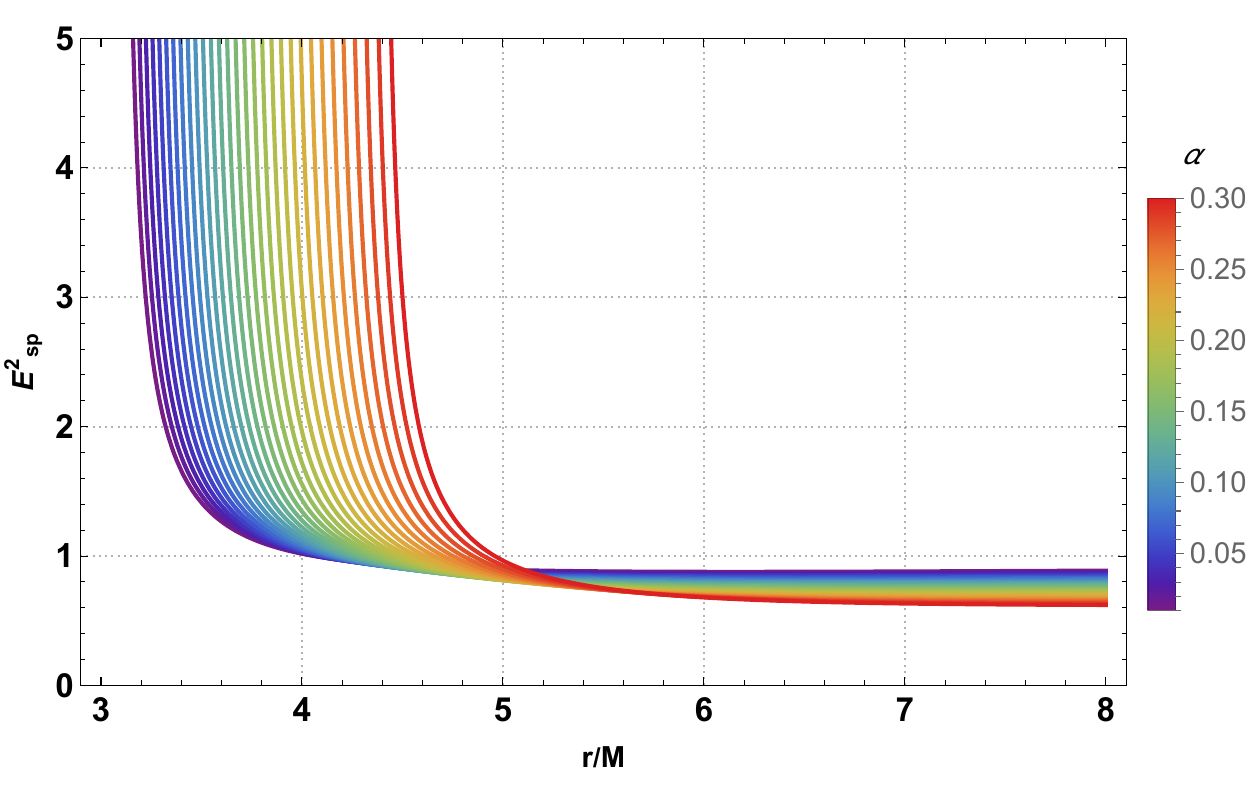}\\[3pt]
    \footnotesize (i) $R/M=0.1$
  \end{minipage}\hfill
  \begin{minipage}{0.48\linewidth}
    \centering
    \includegraphics[width=\linewidth]{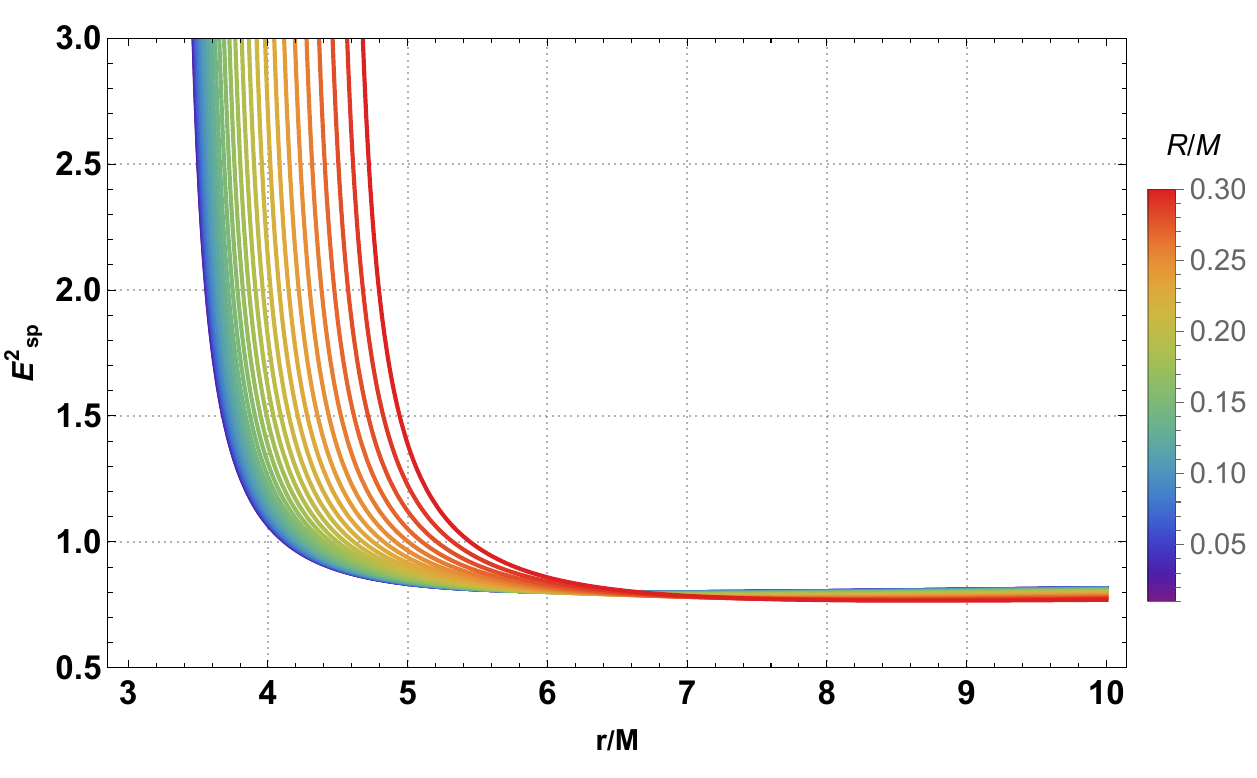}\\[3pt]
    \footnotesize (ii) $\alpha=0.1$
  \end{minipage}
  \caption{\footnotesize Behavior of the squared specific energy $E_{\rm sp}$ of test particles for different values of the CoS parameter $\alpha$ and of the scale parameter $R/M$. Here $\rho_0\,M^2=0.5$.}
  \label{fig:energy}
\end{figure*}

\begin{figure*}[tbhp]
  \centering
  \begin{minipage}{0.48\linewidth}
    \centering
    \includegraphics[width=\linewidth]{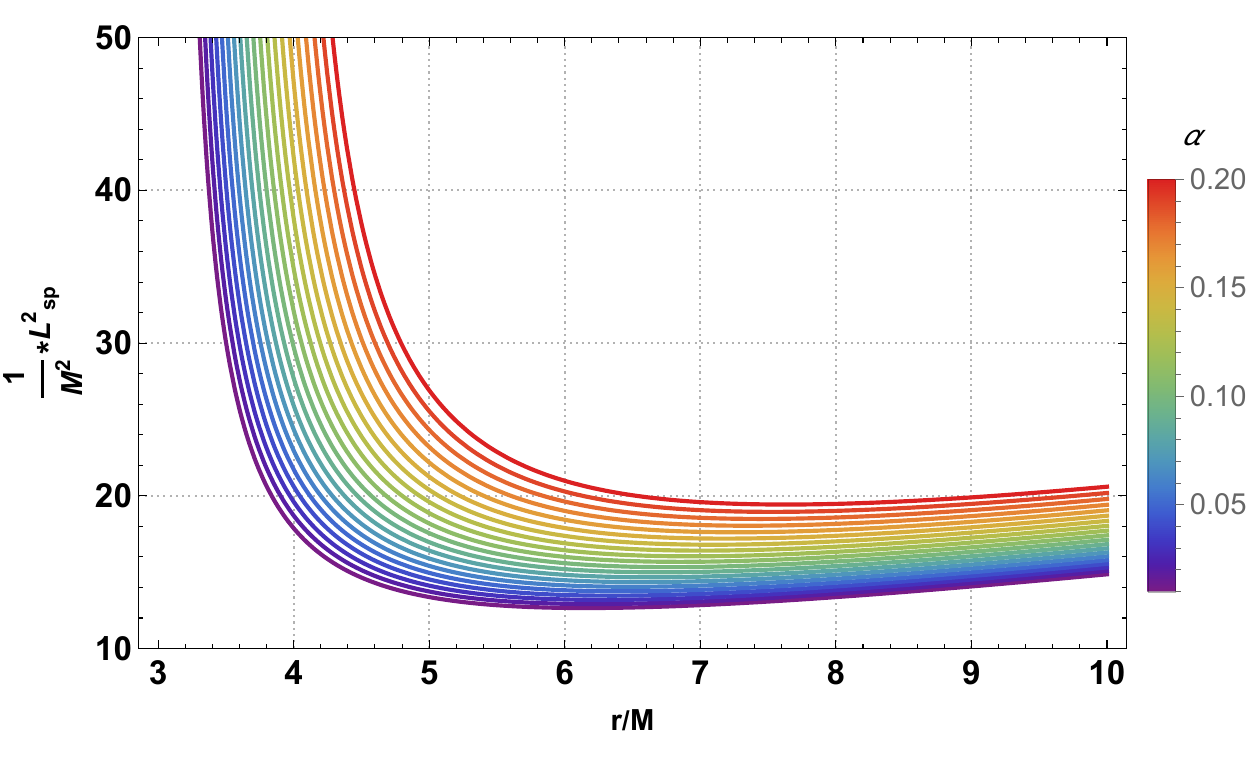}\\[3pt]
    \footnotesize (i) $R/M=0.1$
  \end{minipage}\hfill
  \begin{minipage}{0.48\linewidth}
    \centering
    \includegraphics[width=\linewidth]{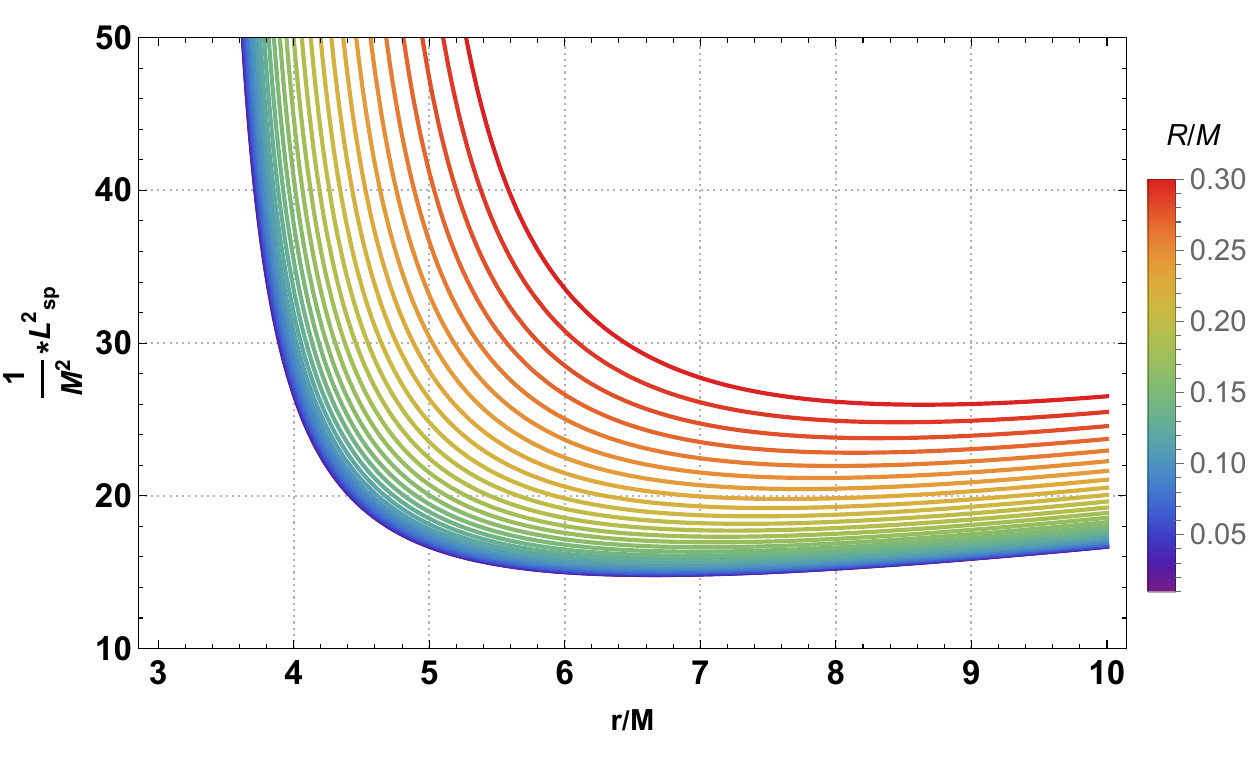}\\[3pt]
    \footnotesize (ii) $\alpha=0.1$
  \end{minipage}
  \caption{\footnotesize Behavior of the squared specific angular momentum per unit mass $L_{\rm sp}/M$ of test particles for different values of the CoS parameter $\alpha$ and of the scale parameter $R/M$. Here $\rho_0\,M^2=0.5$.}
  \label{fig:momentum}
\end{figure*}

From the above expressions (\ref{cc6}) and (\ref{cc7}), we observe that both quantities $(\mathrm{E}_\text{sp}, \mathrm{L}_\text{sp})$ of test particles are influenced by the KDM halo profile characterized by the parameters $(R, \rho_0)$ and the string parameter $\alpha$. However, in the limit $\rho_0$ or $R \to 0$ corresponding to the absence of KDM halo, these are 
\begin{align}
&\mathrm{E}_\text{sp}(r)=
\pm\,\dfrac{\left(1 -\alpha-\frac{2 M}{r}\right)}{\sqrt{1 - \alpha - \frac{3M}{r}}},\\&
\mathrm{L}_\text{sp}(r)=
\sqrt{\frac{M\,r}{1 - \alpha - \frac{3M}{r}}},\label{special}
\end{align}
which is similar to those results obtained for the Letelier BH solution.

Figure \ref{fig:angular-velocity} illustrates the behavior of the angular velocity of the azimuthal component for different values of the halo parameters $(R, \rho_0)$. We observed that as the values of $R/M$ and $\rho_0\,M^2$ increase, the azimuthal angular velocity increases, indicating test particles orbiting with higher speed. Furthermore, the impact of parameters $R/M$ and $\alpha$ on the physical quantities $\mathrm{E}_{\rm sp}$ and $\mathrm{L}_{\rm sp}$ of test particles is illustrated in Figures \ref{fig:energy} and \ref{fig:momentum}, respectively. It is observed that, in the presence of a King dark matter halo and a string cloud, the test particles exhibit higher angular momentum and energy. Specifically, it can be stated that at small values of $r/M$, the presence of King dark matter leads to an increase in mass at a fixed radius, resulting in a stronger gravitational field experienced by test particles. Consequently, the value of angular velocity becomes larger in the presence of King dark matter. However, as $r/M$ increases, $\Omega(r)$ approaches zero due to the reduction in gravity at greater distances. Also, angular momentum is greater with King dark matter than without at the same $r/M$, because dark matter increases gravitational attraction, which requires particles to have more angular momentum to balance this force and avoid falling into the BH. Moreover, the string cloud also impacts the gravitational field, and consequently, the test particles have higher angular momentum when $\alpha$ increases. Thus, in the presence of King dark matter as well as the string cloud, particles must move faster to maintain their orbits. On the other hand, in relativistic mechanics, $\mathrm{E}(r)$ represents the energy per unit mass of a particle. At infinity, specific energy reaches a maximum of $(1-\alpha)$, less than unity. A smaller $\mathrm{E}(r)$ indicates a more strongly bound particle, and in the presence of a King dark matter halo and string cloud, particles are more tightly bound. 

\section{Topological Characteristics of Photon Rings}\label{Sec4}

In Section~\ref{Sec3}, we examined the photon sphere of a BH in the presence of King dark matter and confirmed the existence of such a sphere. However, we did not analyze the \emph{stability} or \emph{instability} of the photon sphere, an aspect that has garnered significant interest in recent studies~\cite{qiao2025existence,qiao2022curvatures,koga2019stability}. Our previous focus was limited to computing the BH shadow. A stable photon sphere refers to a region where small perturbations in a light ray's trajectory do not lead to escape or capture, allowing the light to remain in orbit. Conversely, an unstable photon sphere is highly sensitive to perturbations, resulting in the light either falling into the BH or escaping, thereby contributing to shadow formation~\cite{qiao2022curvatures,koga2019stability,shoom2017metamorphoses}. 

It is well established that every BH possesses at least one unstable photon sphere responsible for the creation of its shadow~\cite{cvetivc2016photon,cunha2017fundamental}. Given that the distinction between stable and unstable photon spheres plays a critical role in interpreting BH observations, we now turn our attention to investigating the (in)stability of the photon sphere using a topological approach. For this purpose, we define a potential as outlined in~\cite{wei2020topological,sadeghi2024role,sadeghi2024thermodynamic,shahzad2025topological} given by
\begin{widetext}
\begin{eqnarray}\label{dd1}
H(r,\theta)=\frac{\sqrt{f(r)}}{r\,\sin\theta}=\frac{1}{r\,\sin \theta}\,\sqrt{1 -\alpha-\frac{2 M}{r}+\frac{8 \pi \rho_0 R^3}{\sqrt{r^2+R^2}}+\frac{8 \pi \rho_0 R^3}{r} \ln \left(\frac{\sqrt{r^2+R^2}-r}{R}\right)}.
\end{eqnarray}    
\end{widetext}
\begin{figure*}[tbhp]
    \centering
    \begin{minipage}{0.48\linewidth}
    \centering
        \includegraphics[width=\linewidth]{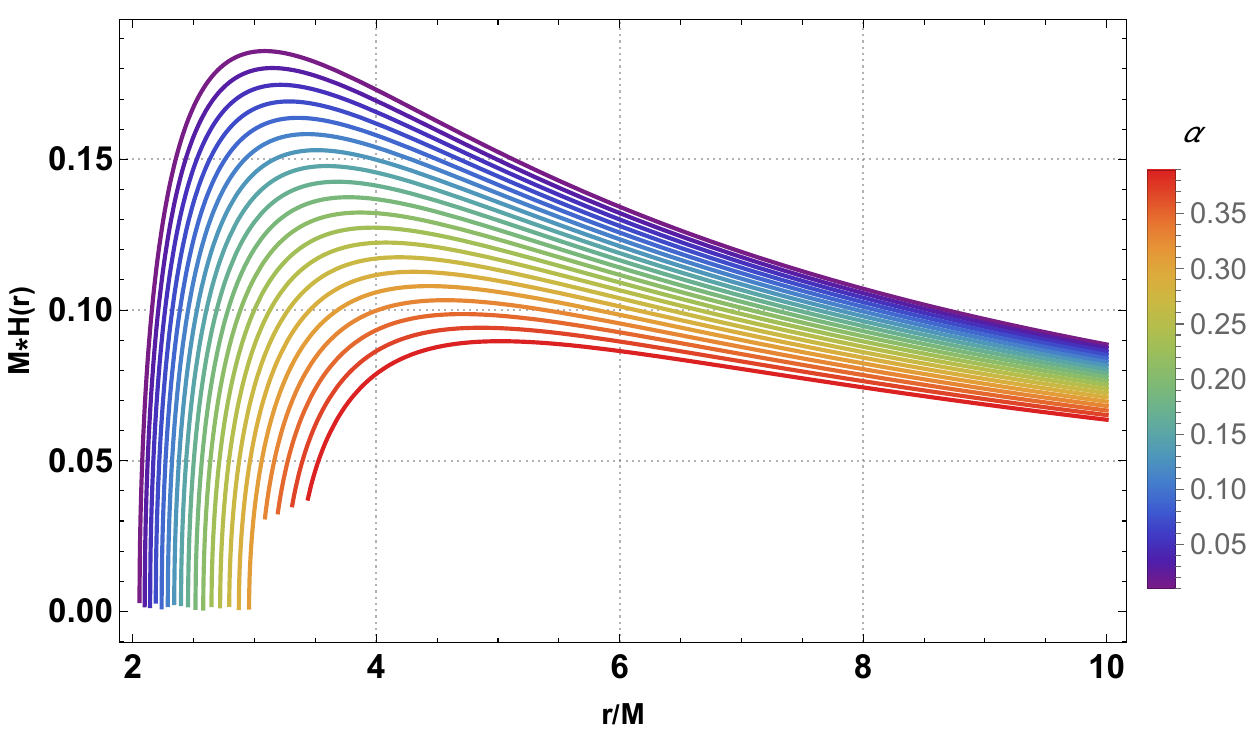}\\[3pt]
        \footnotesize (i) $R/M=0.1$ 
    \end{minipage}
    \hfill
    \begin{minipage}{0.48\linewidth}
    \centering
        \includegraphics[width=\linewidth]{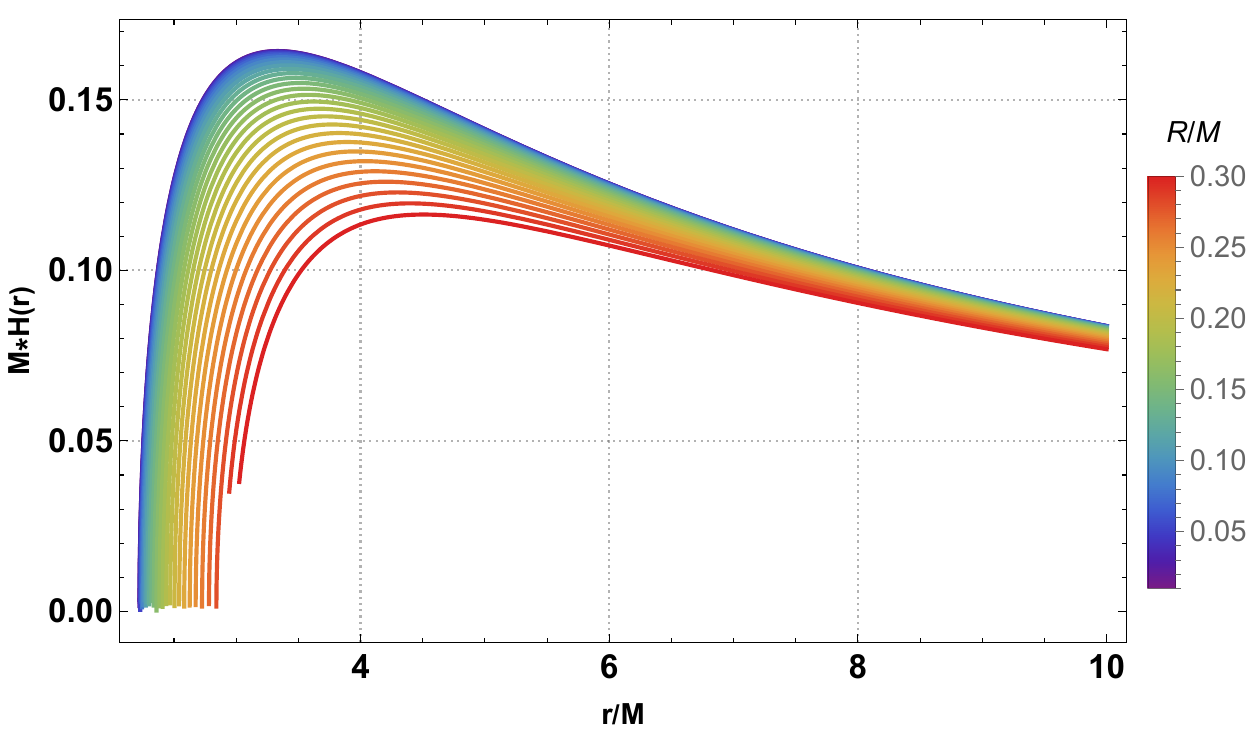}\\
    \footnotesize (ii) $\alpha=0.1$
    \end{minipage}
    \caption{\footnotesize Behavior of the potential function terms ``$MH(r,\pi/2)$" as a function of $r/M$ for different values of $\alpha$ and $R/M$. Here $\rho_0\,M^2=0.5$.}
    \label{fig:potential-function}
\end{figure*}

The potential function $H(r, \pi/2)$ as a function of the dimensionless radius $r/M$ is plotted in Fig.~\ref{fig:potential-function} for varying values of the string parameter $\alpha$ and the halo parameter $R/M$, while keeping $\rho_0 M^2 = 0.5$ fixed. In both panels, we observe that the potential function decreases as either $\alpha$ or $R/M$ increases. This behavior indicates that a stronger KDM halo or a higher string cloud parameter leads to a less stable photon sphere profile within this topological framework.

Considering a vector field $\phi_H$ of the potential $H(r, \theta)$, which can be visualized on a plane using coordinates, which are essentially the components of its gradient vector. Expressing the vector field $\phi_H=(\phi^r_H\,,\,\phi^{\theta}_H)$, where the components of this vector field in terms of potential $H(r, \theta)$ are defined as
\begin{widetext}
\begin{align}
\phi^{r}_H&=\sqrt{f(r)} \partial_r H(r,\theta)=-\frac{1}{r^2\,\sin \theta}\,\left[1 - \alpha - \frac{3M}{r}
+ \frac{12\pi \rho_0 R^3}{\sqrt{r^2 + R^2}}+ \frac{4\pi \rho_0 R^3 r^2}{(r^2 + R^2)^{3/2}}
+ \frac{12\pi \rho_0 R^3}{r} \ln\left( \frac{\sqrt{r^2 + R^2} - r}{R} \right)\right],\label{dd2}\\
\phi^{\theta}_H&=\frac{1}{r^2}\partial_{\theta} H(r,\theta)=-\frac{\sqrt{1 -\alpha-\frac{2 M}{r}+\frac{8 \pi \rho_0 R^3}{\sqrt{r^2+R^2}}+\frac{8 \pi \rho_0 R^3}{r} \ln \left(\frac{\sqrt{r^2+R^2}-r}{R}\right)}}{r^2}\,\frac{\cot \theta}{\sin  \theta},\label{dd3}
\end{align}
normalized via \cite{wei2020topological}
\begin{eqnarray}\label{dd4}
n^r_H=\frac{\phi^{r}_H}{||\phi||}\qquad,\qquad
n^{\theta}_H=\frac{\phi^{\theta}_H}{||\phi||},\qquad\text{where}\qquad
||\phi||=\sqrt{(\phi^r_{H})^2+(\phi^{\theta}_H)^2}.
\end{eqnarray}    
\end{widetext}
\begin{figure*}[tbhp]
  \centering
  \begin{minipage}{0.32\linewidth}
    \centering
    \includegraphics[width=\linewidth]{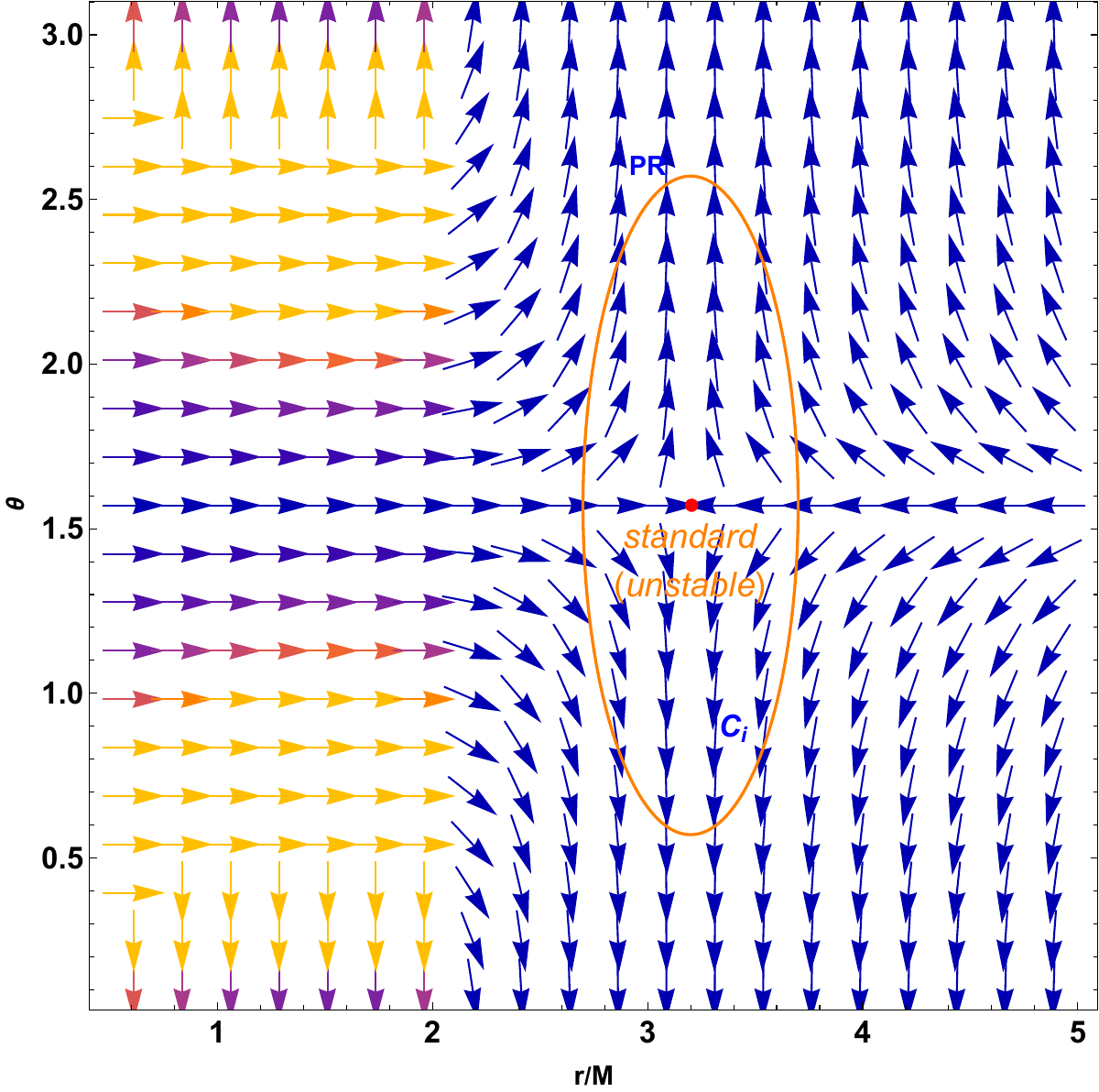}\\[3pt]
    \footnotesize (i) $\alpha=0.05$
  \end{minipage}\hfill
  \begin{minipage}{0.32\linewidth}
    \centering
    \includegraphics[width=\linewidth]{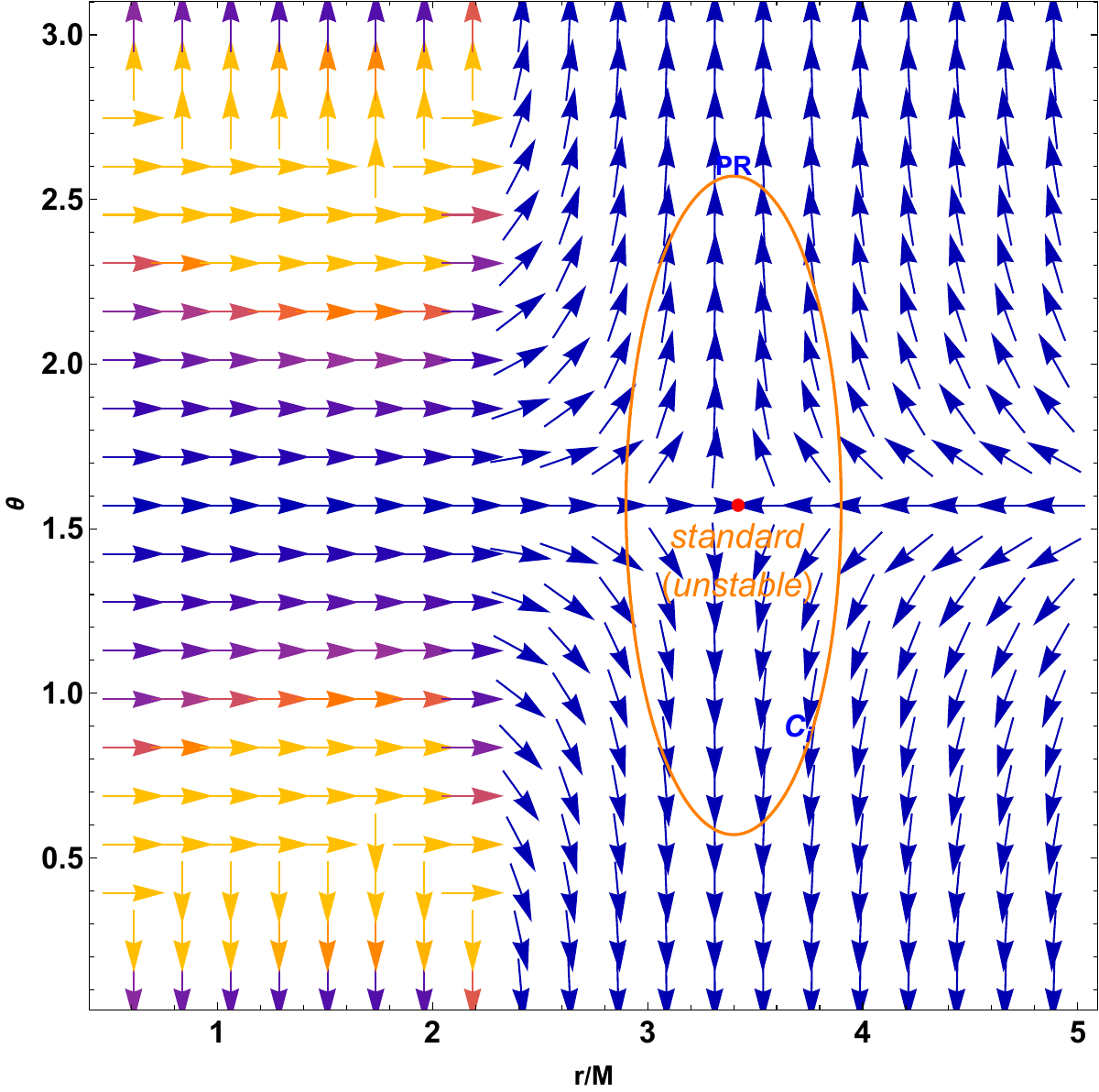}\\[3pt]
    \footnotesize (ii) $\alpha=0.1$
  \end{minipage}\hfill
  \begin{minipage}{0.32\linewidth}
    \centering
    \includegraphics[width=\linewidth]{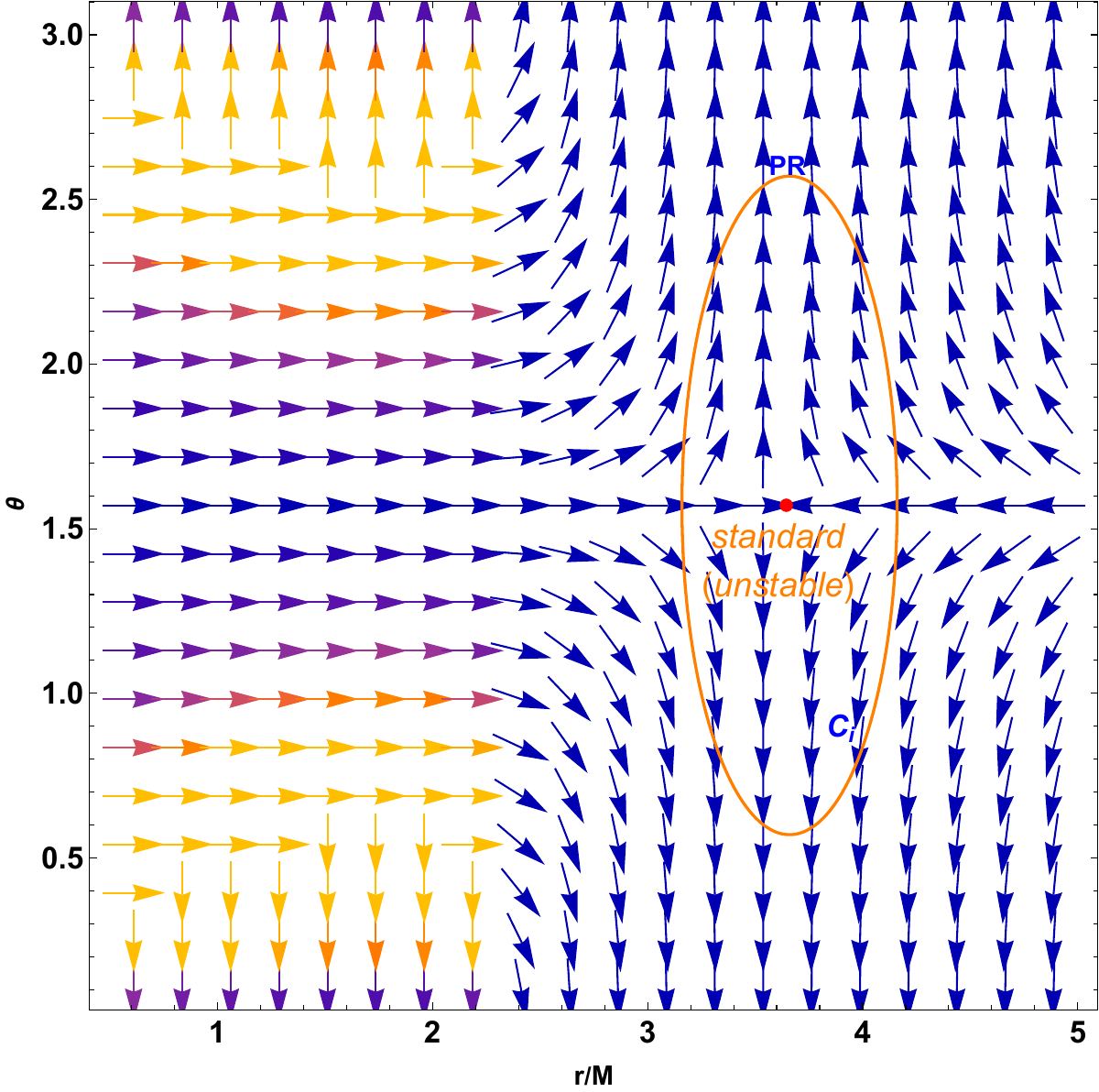}\\[3pt]
    \footnotesize (iii) $\alpha=0.15$
  \end{minipage}
  \caption{\footnotesize The arrows represent the unit vector field $\vec{n}_H$ on a portion of the $r$-$\Theta$ plane for the Letelier black holes surrounded by KDM halo with $R/M = 0.1$ and $\rho_0=0.1/M^2$. The photon ring (PR), marked with a red dot, is at $(r, \theta) =(3.205, \pi/2)$ for $\alpha=0.05$; $(r, \theta) =(3.421,\pi/2)$ for $\alpha=0.10$; and $(r, \theta)=(3.645, \pi/2)$ at $\alpha=0.15$. The orange contour $\mathcal{C}_i$ is a closed loop enclosing the photon ring. Obviously, the topological charge of the photon ring is $Q = -1$.}
  \label{fig:unit-vector-1}
\end{figure*}
\begin{figure*}[tbhp]
  \centering
  \begin{minipage}{0.32\linewidth}
    \centering
    \includegraphics[width=\linewidth]{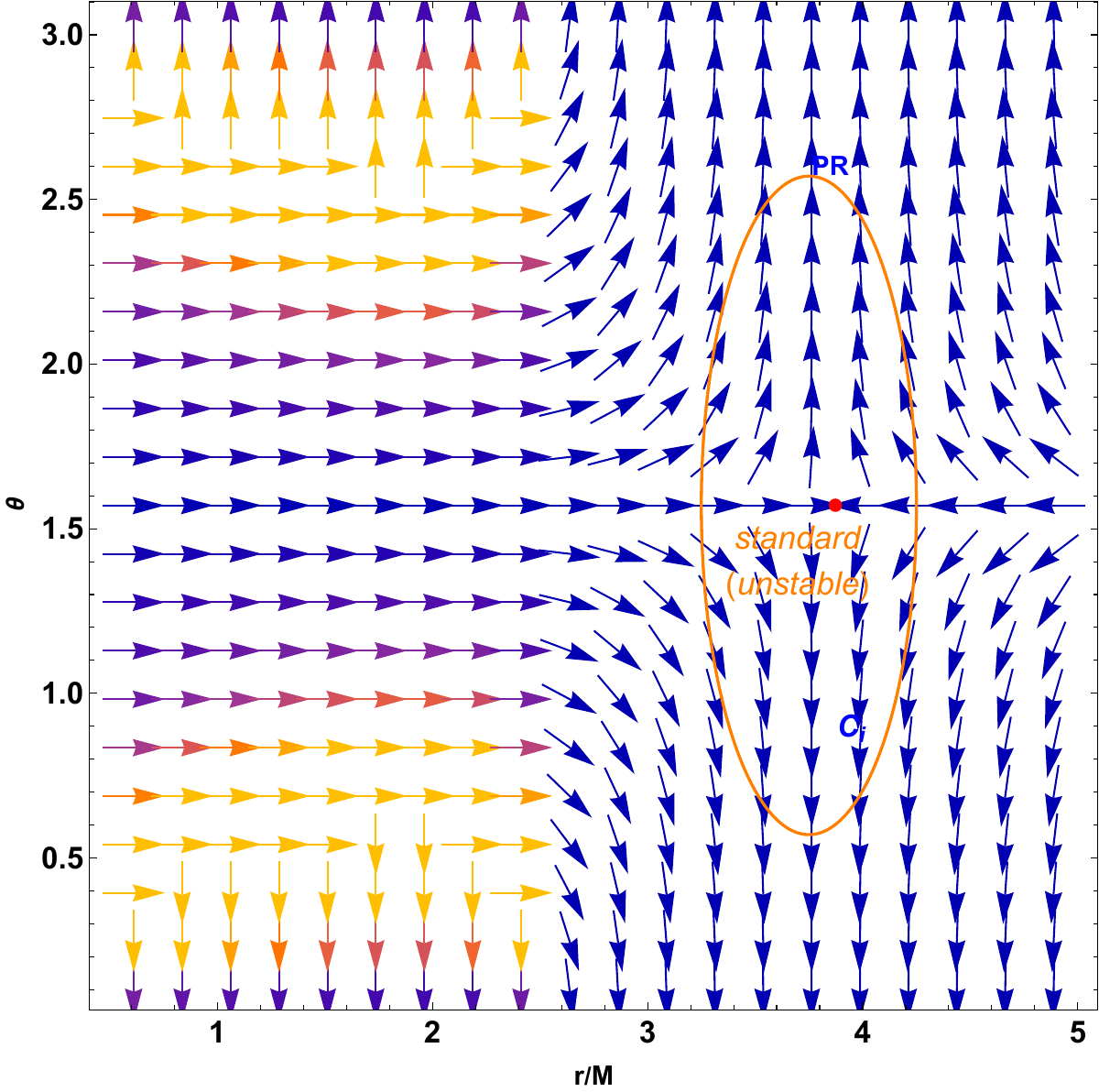}\\[3pt]
    \footnotesize (i) $\rho_0=0.6/M^2$
  \end{minipage}\hfill
  \begin{minipage}{0.32\linewidth}
    \centering
    \includegraphics[width=\linewidth]{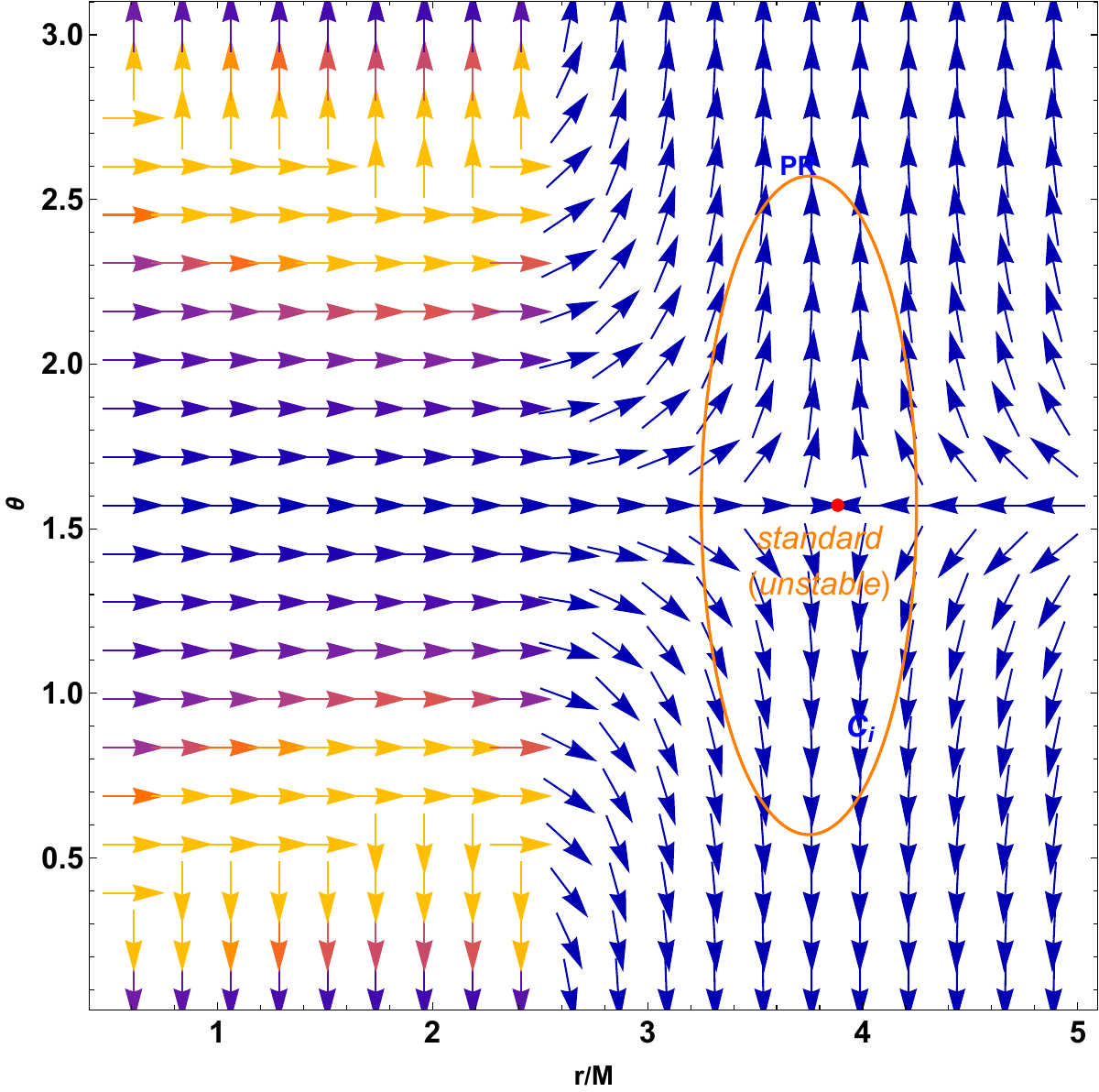}\\[3pt]
    \footnotesize (ii) $\rho_0=0.7/M^2$
  \end{minipage}\hfill
  \begin{minipage}{0.32\linewidth}
    \centering
    \includegraphics[width=\linewidth]{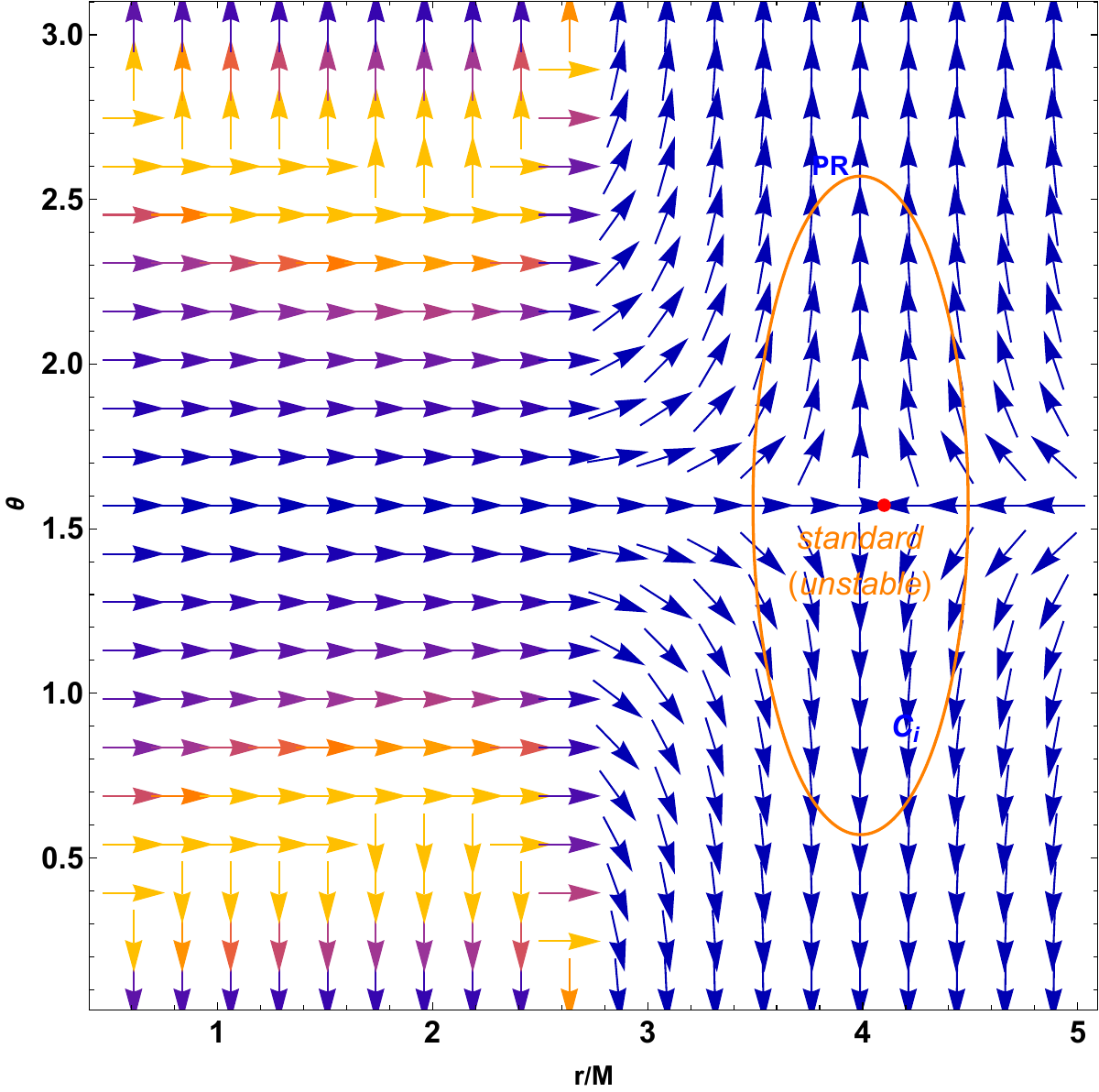}\\[3pt]
    \footnotesize (iii) $\rho_0=0.85/M^2$
  \end{minipage}
  \caption{\footnotesize The arrows represent the unit vector field $\vec{n}_H$ on a portion of the $r$-$\Theta$ plane for the Letelier black holes surrounded by KDM halo with $R/M = 0.1$ and $\alpha=0.1$. The photon ring (PR), marked with a red dot, is at $(r, \theta) =(3.873, \pi/2)$ for $\rho_0\,M^2=0.6$; $(r, \theta) = (3.884,\pi/2)$ for $\rho_0\,M^2=0.7$; and $(r, \theta)=(4.11, \pi/2)$ for $\rho_0\,M^2=0.85$. Analogous to Fig.~\ref{fig:unit-vector-1}, the orange contour $\mathcal{C}_i$ is a closed loop enclosing the photon ring with topological charge $Q = -1$.}
  \label{fig:unit-vector-2}
\end{figure*}

It is evident that at points with coordinates $\theta = \pi/2$, the angular component of the vector field $\phi^{\theta}_H$ vanishes, and the radial component satisfies $\partial_r H(r,\theta)\big|_{r=r_c} = 0$. Here, the value $r_c$ corresponds to the solution of Eq.~\eqref{bb9}, which essentially defines the location of the BH's photon sphere. These points are referred to as zero points in the vector field, where vectors either converge or diverge, and can be interpreted as topological defects. Each of these defects is characterized by a topological number, which describes the rotation behavior of vectors around the zero point. For such points, the topological number takes values of $+1$ or $-1$, depending on whether the vector field rotates counterclockwise or clockwise, respectively, while it is zero at non-defect points. To determine this number, one can examine the closed curve formed by $\phi^r_H - \phi^\theta_H$. If the direction of the vector field along this curve is clockwise at a zero point, the topological number is $-1$; if counterclockwise, it is $+1$~\cite{zhu2025universal,dong2025thermodynamic}.

An alternative approach involves drawing a closed contour around an arbitrary point and analyzing the rotation of the vectors within that loop. This allows the computation of the topological number as proposed in~\cite{wei2020topological} given by
\begin{eqnarray}\label{dd5}
W=\sum w_i=\frac{1}{2\pi}\oint_{c_i} d(\arctan\frac{n_\theta}{n_r}),
\end{eqnarray}
determine the topological number of each point in the vector space. To do this, we employ a parameter change in the form  \cite{wei2020topological}
\begin{eqnarray}
r=a\, \cos\vartheta +r_i,\qquad
\theta=b\, \sin\vartheta+\frac{\pi}{2},\label{dd6}
\end{eqnarray}
where $r_i$ is the radial coordinate of the point under consideration. 

It is evident that the selected BH has only one photon sphere located at $r_{c_i}$, which depends on both the string parameter $\alpha$ and KDM halo profile parameters $(R,\rho_0)$. In the topological analysis of the photon sphere, a charge of $-1$ means its instability \cite{wei2020topological}, which leads to the formation of the BH shadow, a result consistent with our findings in the previous section.

The unit vector field $n_H$ on a portion of the $r-\theta$ plane is plotted in Figure \ref{fig:unit-vector-1} with parameters $R/M=0.1$ and $\rho_0\,M^2=0.1$ for three different values of $\alpha$. It is easy to observe that there is a photon ring, marked with a red dot, is located at $(r, \theta)=(3.205, \pi/2)$ for $\alpha=0.05$; at $(r, \theta)=(3.421, \pi/2)$ for $\alpha=0.1$; and at $(r, \theta)=(3.645, \pi/2)$ for $\alpha=0.15$. Furthermore, the winding number $W$ associated with the orange contours $C_i$ characterizes the behavior of the vector field. Thereby, the topological charge of the photon ring is $Q = -1$. Based on the classification of photon rings, this configuration corresponds to a standard and unstable photon ring~\cite{Cunha2020, Wei2020}.

Similarly, the unit vector field $n_H$ on a portion of the $r-\theta$ plane is plotted in Figure \ref{fig:unit-vector-2} with $R/M=0.1$ and $\alpha=0.1$ for three different values of $\rho_0$. Here also, we observe a photon ring located at $(r, \theta)=(3.873, \pi/2)$ for $\rho_0\,M^2=0.6$; at $(r, \theta)=(3.884, \pi/2)$ for $\rho_0\,M^2=0.7$; and at $(r, \theta)=(4.11, \pi/2)$ for $\rho_0\,M^2=0.85$. Analog to the previous Figure \ref{fig:unit-vector-1}, the topological charge of the photon ring is $Q = -1$, which is unstable. 

\section{Thermodynamic Properties of BH}\label{Sec5}

BH thermodynamics, grounded in the Bekenstein-Hawking area law and Hawking’s derivation of thermal radiation, offers a robust framework to study stability and phase transitions in classical and semiclassical gravity~\cite{Bekenstein1973,Hawking1975,Wald2001}. More recently, topological ideas have been incorporated into thermodynamic analyses, either by classifying critical points via Duan’s topological current~\cite{Wei2022a} or by using off-shell free energies to identify topological defects in parameter space~\cite{Wei2022b,wu2023topological}. In this section, we apply the standard (non-extended) formalism to the Letelier black hole immersed in a King (KDM) halo, keeping $(\alpha,\rho_0,R)$ fixed. We show that the halo lowers the Hawking temperature at fixed $r_h$, modifies the free energy, and introduces critical radii where the specific heat diverges, fully consistent with the topological criteria discussed in the next section. We work in natural units $\hbar=G=c=1$.

\paragraph*{ADM mass from the horizon condition.}

The event horizon $r=r_h$ is determined by the condition $f(r_h)=0$. With
\begin{align}
f(r)=1-\alpha&-\frac{2M}{r}+\frac{8\pi\rho_0 R^3}{\sqrt{r^2+R^2}}
\notag\\&+\frac{8\pi\rho_0 R^3}{r}\,\ln\!\left(\frac{\sqrt{r^2+R^2}-r}{R}\right),
\end{align}
from the horizon condition, we find the ADM mass $M$ as
\begin{align}
M(r_h)=\frac{r_h}{2}&\!\Bigg[\,1-\alpha+\frac{8 \pi \rho_0 R^3}{\sqrt{r_h^2+R^2}}
\notag\\&+\frac{8 \pi \rho_0 R^3}{r_h} \ln \!\left(\frac{\sqrt{r_h^2+R^2}-r_h}{R}\right)\Bigg].
\label{thermo1}
\end{align}

\paragraph*{Hawking temperature.} The Hawking temperature follows from the surface gravity,
\begin{equation}
    T_H=\frac{f'(r)}{4\pi}\Bigg|_{r=r_h}
    =\frac{1}{4\pi r_h}\,\left[1 - \alpha-\frac{ 8 \pi \rho_0 R^3 r_h^{2}}{(r_h^2 + R^2)^{3/2}}\right].
    \label{thermo2}
\end{equation}

In the halo-free limit $\rho_0\to 0$ or $R\to 0$, one recovers
\begin{equation}
T_{H}^{\rm Letelier}=\dfrac{1-\alpha}{4\pi r_h}
=\dfrac{(1-\alpha)^2}{8\pi M}<T_{H}^{\rm Sch.},    
\end{equation}
with $r_h=2M/(1-\alpha)$.

The radial derivative (used below) is
\begin{equation}
\frac{dT_H}{dr_h}
=\frac{1}{4\pi}\!\left[
-\frac{1-\alpha}{r_h^{2}}
+\frac{8\pi\rho_0 R^3\,(2r_h^{2}-R^{2})}{(r_h^{2}+R^{2})^{5/2}}
\right].
\label{dTdr}
\end{equation}

\paragraph*{Entropy and Gibbs free energy.} 

The entropy of the system is determined by
\begin{equation}
    S=\int \frac{dM(r_h)}{T_H}=\pi r_h^2,\label{ee3}
\end{equation}
which is the Bekenstein-Hawking area law. At fixed $(\alpha,\rho_0,R)$, the first law holds in the non-extended form $dM=T_H\,dS$.

The canonical free energy is
\begin{equation}
F(r_h)=M(r_h)-T_H(r_h)\,S(r_h).
\end{equation}
Substituting Eqs.~\eqref{thermo1}, \eqref{thermo2} and \eqref{ee3}, we obtain
\begin{align}
F(r_h)&=\frac{r_h}{4}\left[\,(1-\alpha)
+\frac{8\pi\rho_0 R^3}{(r_h^2+R^2)^{3/2}}\,(3r_h^2+2R^2)\right]\notag\\&
+4\pi\rho_0 R^3\,
\ln\!\left(\frac{\sqrt{r_h^2+R^2}-r_h}{R}\right).
\label{F_final}
\end{align}
In the halo-free limit, $F=\tfrac{1-\alpha}{4}\,r_h$, which further reduces to the Schwarzschild value as $\alpha\to 0$.

\paragraph*{Specific heat and stability.} The specific heat at fixed $(\alpha,\rho_0,R)$ is
\begin{align}
&\mathbb{C}\equiv T_H\,\frac{\partial S}{\partial T_H}
= T_H\;\frac{dS/dr_h}{dT_H/dr_h}\notag\\&
= \frac{\displaystyle \frac{1}{4\pi r_h}\!\left[1-\alpha-\frac{8\pi\rho_0 R^3 r_h^{2}}{(r_h^2+R^2)^{3/2}}\right]\,(2\pi r_h)}
{\displaystyle \frac{1}{4\pi}\!\left[-\frac{1-\alpha}{r_h^{2}}+\frac{8\pi\rho_0 R^3(2r_h^{2}-R^{2})}{(r_h^{2}+R^{2})^{5/2}}\right]},
\end{align}
which can be written compactly as
\begin{equation}
\mathbb{C}(r_h)=
2\pi\;
\frac{\displaystyle 1-\alpha-\dfrac{8\pi\rho_0 R^3 r_h^{2}}{(r_h^2+R^2)^{3/2}}}
{\displaystyle -\dfrac{1-\alpha}{r_h^{2}}+\dfrac{8\pi\rho_0 R^3(2r_h^{2}-R^{2})}{(r_h^{2}+R^{2})^{5/2}}}.
\label{C_compact}
\end{equation}
Hence, $\mathbb{C}$ diverges (second-order phase transition) when the denominator vanishes,
\begin{equation}
\frac{1-\alpha}{r_h^{2}}
=\frac{8\pi\rho_0 R^3\,(2r_h^{2}-R^{2})}{(r_h^{2}+R^{2})^{5/2}}.
\label{C_divergence_condition}
\end{equation}
The zeros of $\mathbb{C}$ coincide with the zeros of $T_H$, delimiting the physical ($T_H>0$) branches. In the halo-free limit, one recovers $\mathbb{C}\to -2\pi r_h^2$ (Schwarzschild with Letelier rescaling). Equations \eqref{thermo1}, \eqref{thermo2}, \eqref{F_final} and \eqref{C_compact} fully determine the thermodynamics of the Letelier-King black hole for fixed $(\alpha,\rho_0,R)$, while the condition~\eqref{C_divergence_condition} specifies the critical radii across which local thermal stability changes.
\begin{figure*}[tbhp]
\centering
\includegraphics[width=0.45\linewidth]{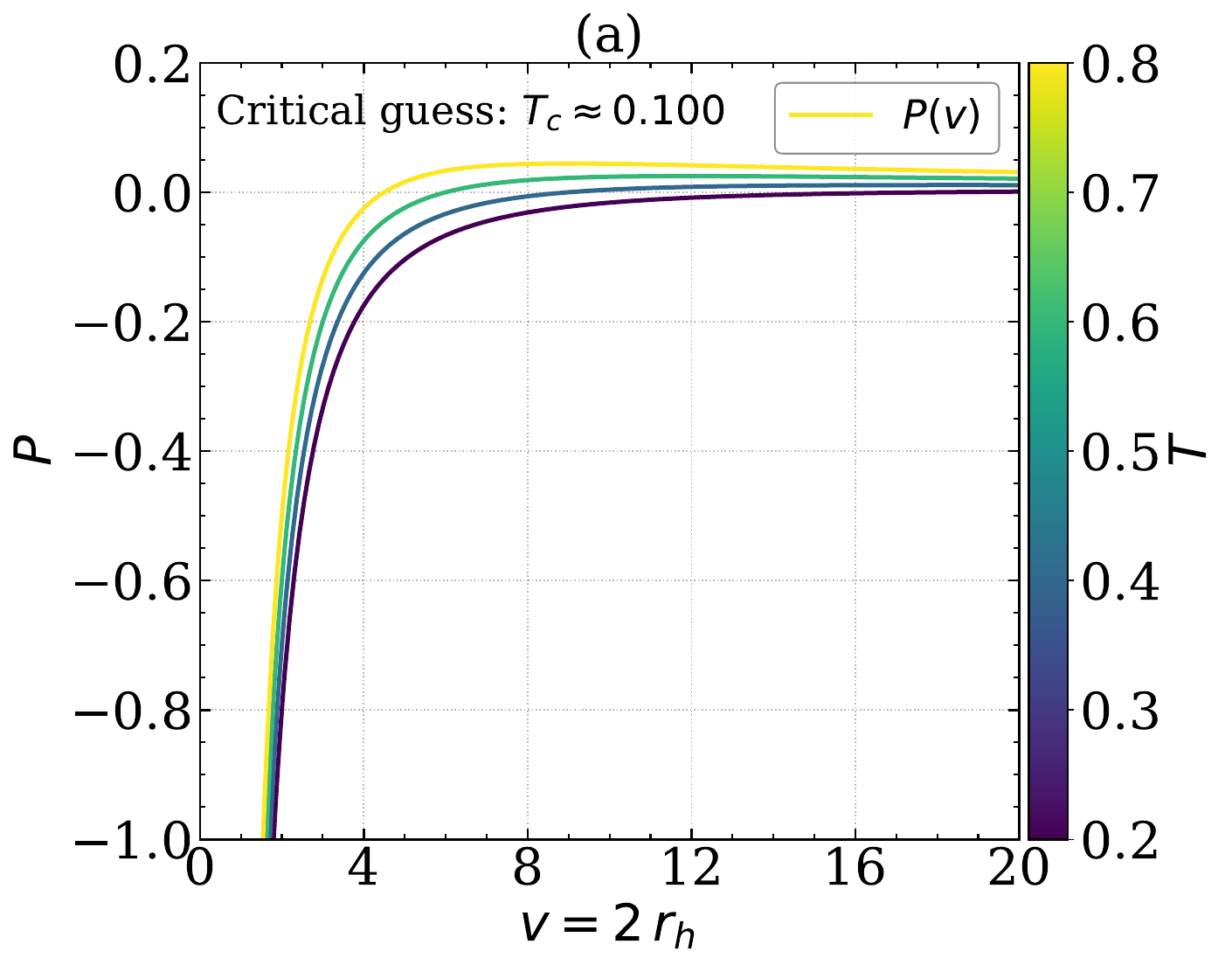}\qquad
\includegraphics[width=0.45\linewidth]{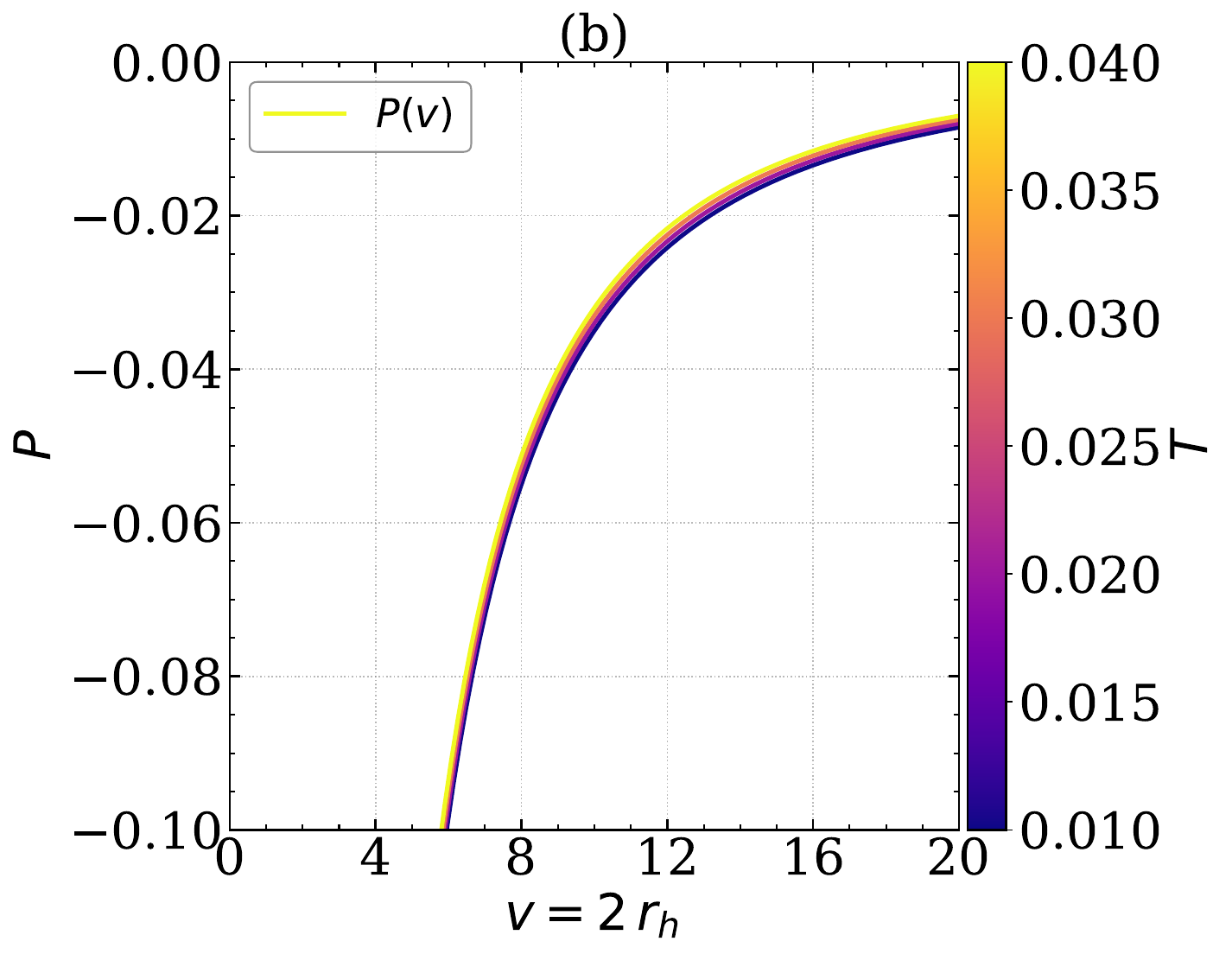}
\caption{\footnotesize Equation-of-state isotherms $P(v)$ for the Letelier BH in KDM halo, with $v=2r_h$ and units $8\pi=1$. Parameters are fixed to $\alpha=0.1$, $\rho_0=0.05$, $R=0.5$. Figure (a) shows higher temperatures: for each $T$, $P$ decreases monotonically with $v$, and larger $T$ shifts the isotherm upward. Figure (b) shows lower temperatures with a tighter vertical range to reveal the small pressure differences; the curves remain monotonic and lie at lower $P$ than in (a). These baselines will be used to calibrate the full KDM contribution when the complete bracket structure of the EoS is reinstated.}
  \label{fig:eos-isotherms-two}
\end{figure*}

Figure~\ref{fig:eos-isotherms-two} displays the equation-of-state (EoS) isotherms $P(v)$ with the specific volume $v=2r_h$ for the Letelier black hole in a KDM halo, working in units $8\pi=1$ and fixing $\alpha=0.1$, $\rho_0=0.05$, $R=0.5$. In Fi.~\ref{fig:eos-isotherms-two}(a), for the higher temperatures $T=\{0.20,\,0.40,\,0.60,\,0.80\}$, each isotherm is strictly monotonic in $v$, and increasing $T$ shifts the entire curve upward, as expected from the analytic form $P(T,r_h)=\tfrac{T}{2r_h}-\tfrac{1-\alpha}{r_h^2}$. The vertical scale makes the relative separation of the isotherms evident over the range $v\in[0,20]$. In Fig.~\ref{fig:eos-isotherms-two}(b) zooms into the low-temperature regime $T=\{0.010,\,0.020,\,0.030,\,0.040\}$. Because the pressure differences are much smaller, a tighter vertical window is chosen to resolve the curves cleanly; the isotherms remain monotonic and lie at lower pressures than those in panel (a). These two baselines (high-$T$ and low-$T$) will be used to calibrate axis ranges and legend settings before reinstating the full KDM contribution in the EoS bracket, which can introduce additional structure (e.g., inflection points or near-critical behavior) at suitably chosen parameters.

\begin{figure*}[tbhp]
\centering
\includegraphics[width=0.42\linewidth]{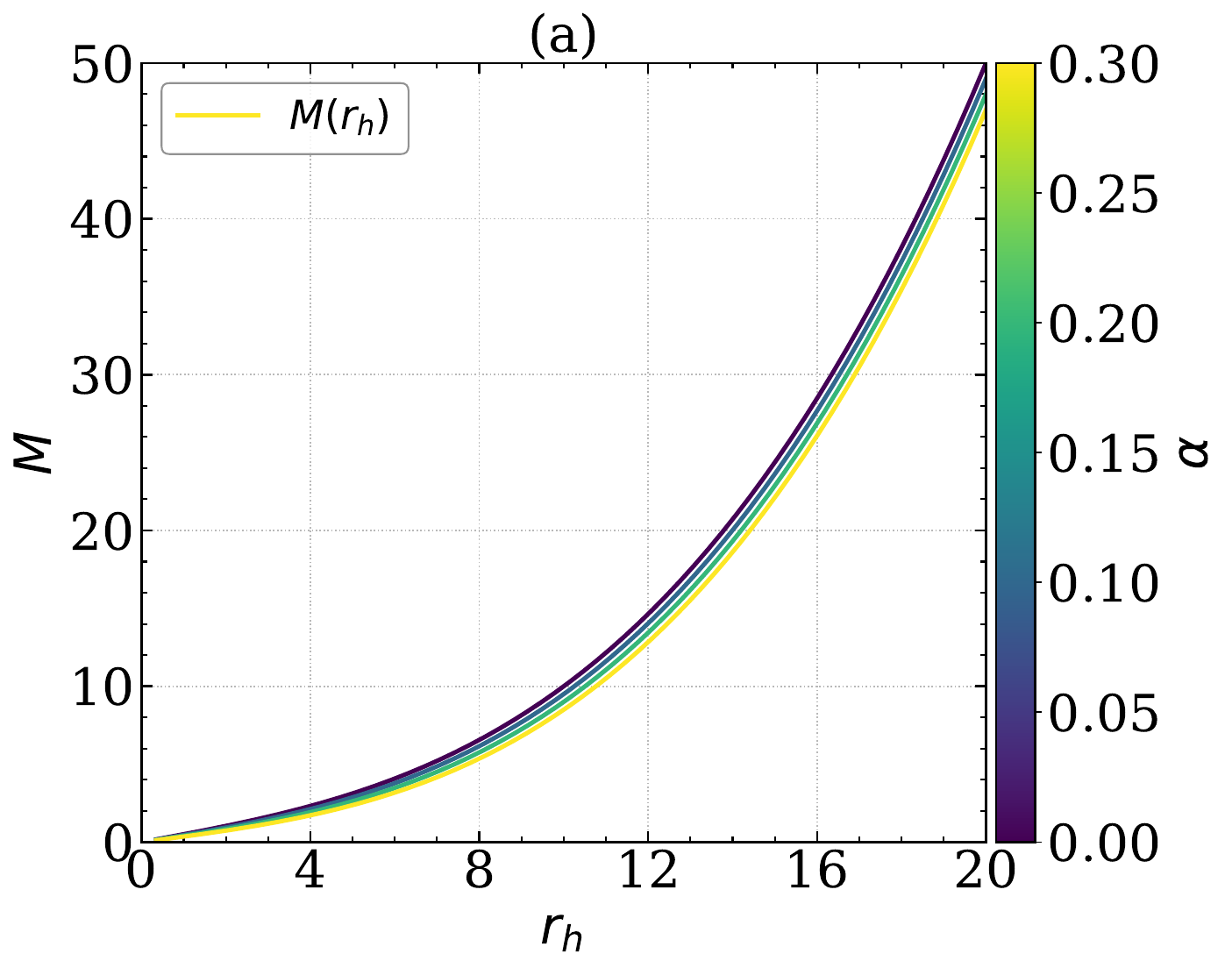}\qquad
\includegraphics[width=0.42\linewidth]{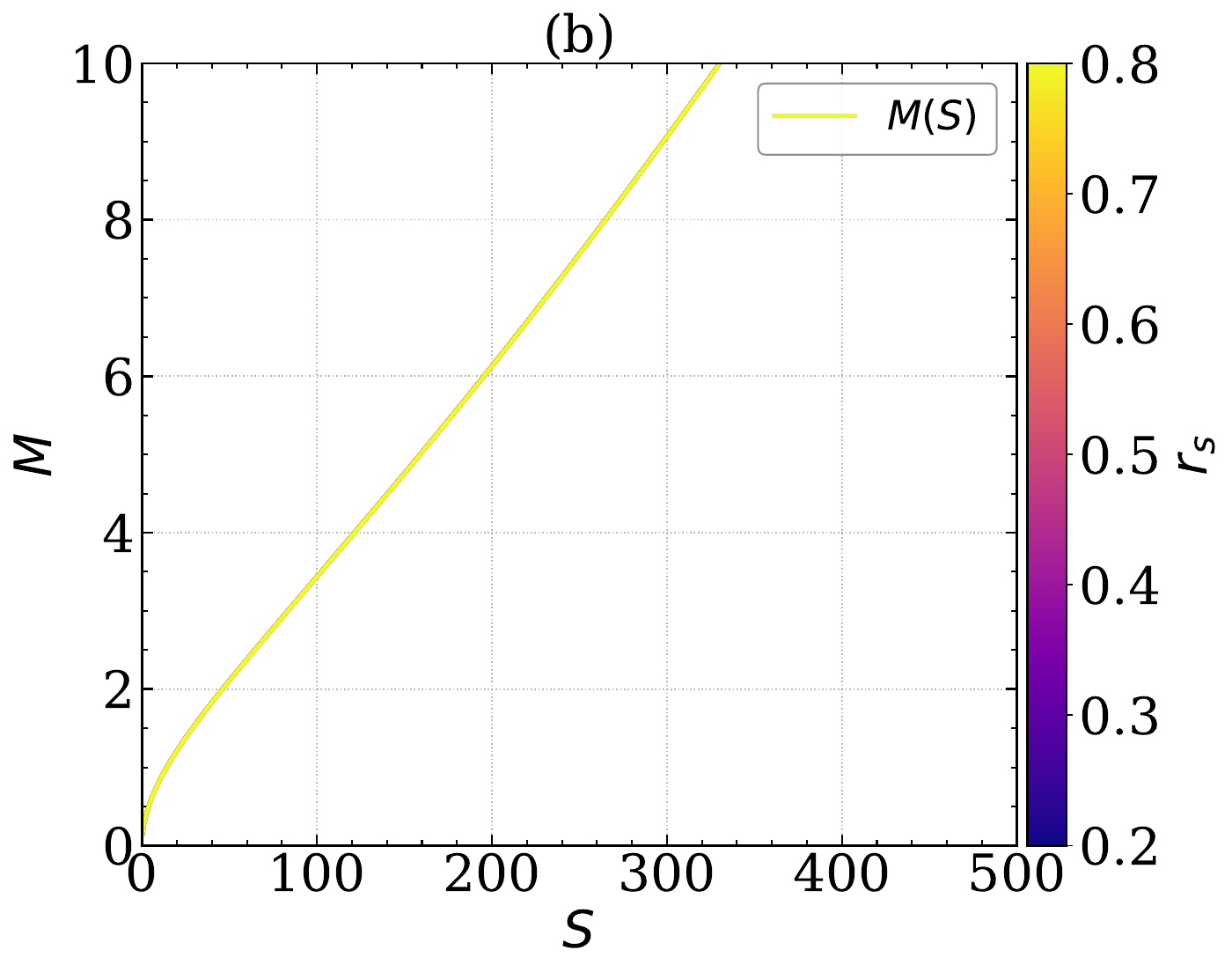}
\includegraphics[width=0.42\linewidth]{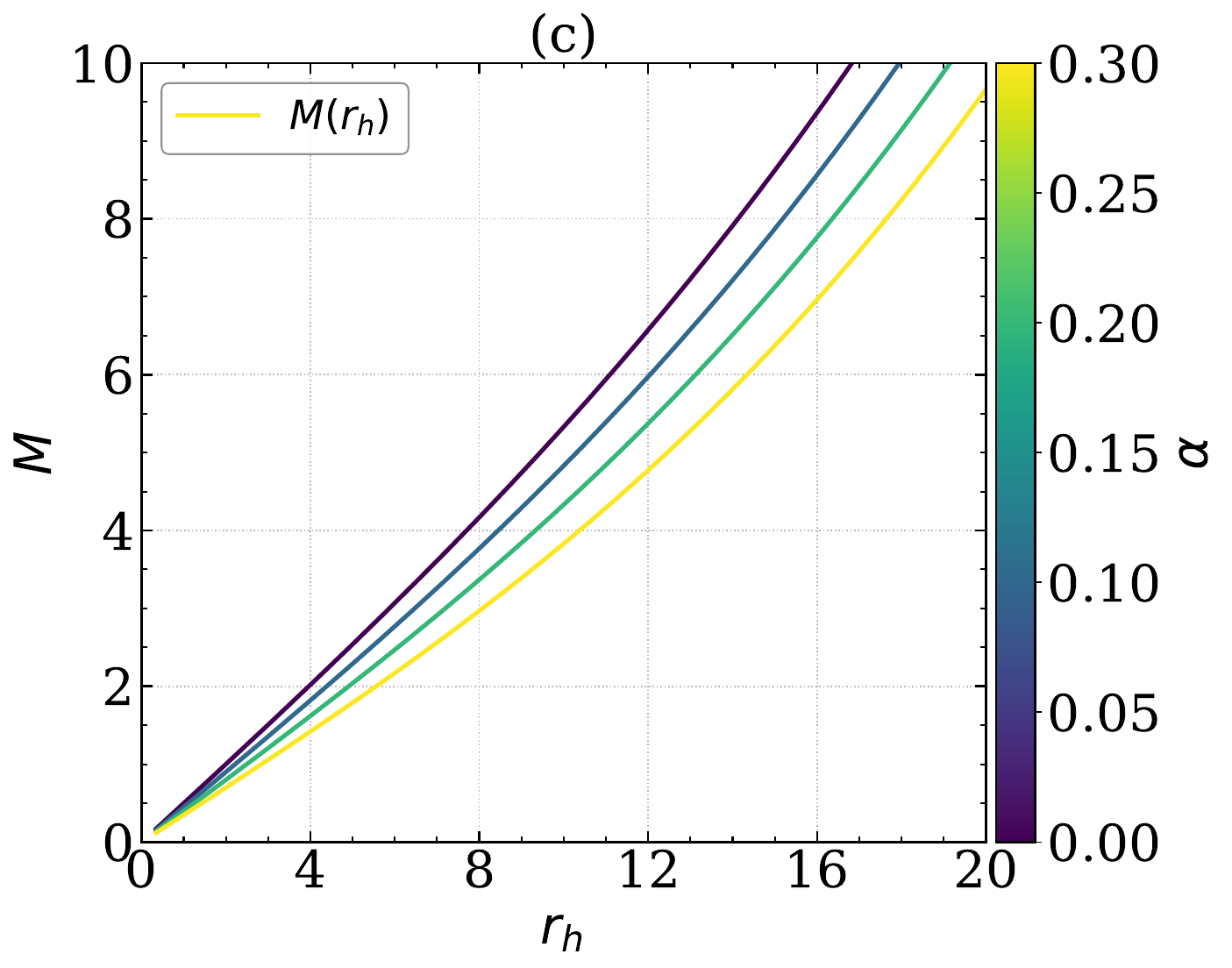}\qquad
\includegraphics[width=0.42\linewidth]{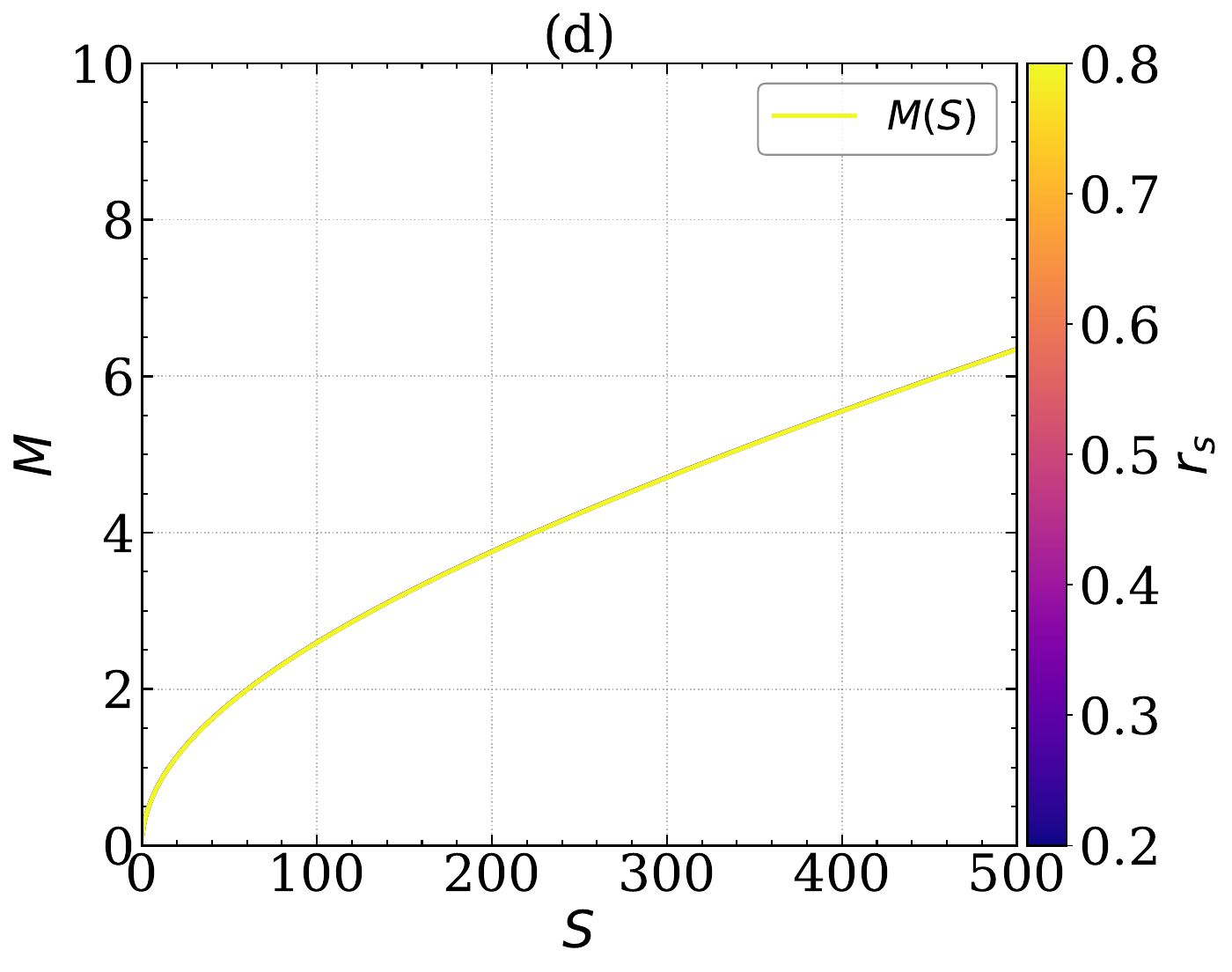}
\caption{\footnotesize Mass function panels for the string-cloud SAdS model used in our plotting routine (units with $8\pi=1$). Figures (a)-(b) keep the AdS length $\ell_p$ fixed, while (c)-(d) keep the thermodynamic pressure $P=3/\ell_p^2$ fixed. In (a) and (c), we scan the string parameter $\alpha$; in (b) and (d), we scan the scale $r_s$ that controls the matter correction in the mass formula. In all cases, $M$ grows with the horizon size, with slopes and curvature modulated by the matter parameters. At fixed background scale (top row), increasing $\alpha$ shifts the curves downward and reduces $dM/dr_h$ at a given $r_h$, reflecting the effective deficit produced by the cloud of strings. At fixed pressure (bottom row), the same trend persists, but the AdS contribution is locked through $P$, which slightly reshapes the large-$r_h$ behavior (where the AdS term dominates). Expressing the mass as $M(S)$ via $S=\pi r_h^2$ (right column) makes the growth manifestly superlinear at large $S$ because of the AdS term $\propto r_h^3/\ell_p^2$, and highlights how larger $r_s$ enhances the matter-induced shift. These plots will be used below to correlate geometric mass growth with the thermal features extracted from $T_H(r_h)$ and the specific heat.}  \label{fig:mass-panels}
\end{figure*}

Figure~\ref{fig:mass-panels}(a) shows $M(r_h)$ at fixed $\ell_p$ for several $\alpha$: the curves are monotonic and increasingly suppressed as $\alpha$ grows, consistent with the factor $(1-\alpha)$ that multiplies the leading Schwarzschild-like contribution. Figure~\ref{fig:mass-panels}(b) displays $M(S)$ at fixed $\ell_p$ for different $r_s$; since $S=\pi r_h^2$, changing $r_s$ mainly shifts the overall level of $M$ and slightly alters the curvature at intermediate $S$, while the AdS-controlled large-$S$ growth is preserved. Figures~\ref{fig:mass-panels}(c) and (d) repeat the analyses at fixed $P$ (thus fixed $\ell_p$ through $P=3/\ell_p^2$ in our convention). The qualitative influence of $\alpha$ and $r_s$ matches the fixed-$\ell_p$ case, but the high-$r_h$ (or high-$S$) tails are more strictly organized because the AdS contribution is tied to $P$. Taken together, the four panels make explicit how the string cloud and the matter scale deform the mass function without spoiling its monotonicity, setting the stage for the thermodynamic behavior discussed in the next sections.
\begin{figure*}[tbhp]
\centering
\includegraphics[width=0.42\linewidth]{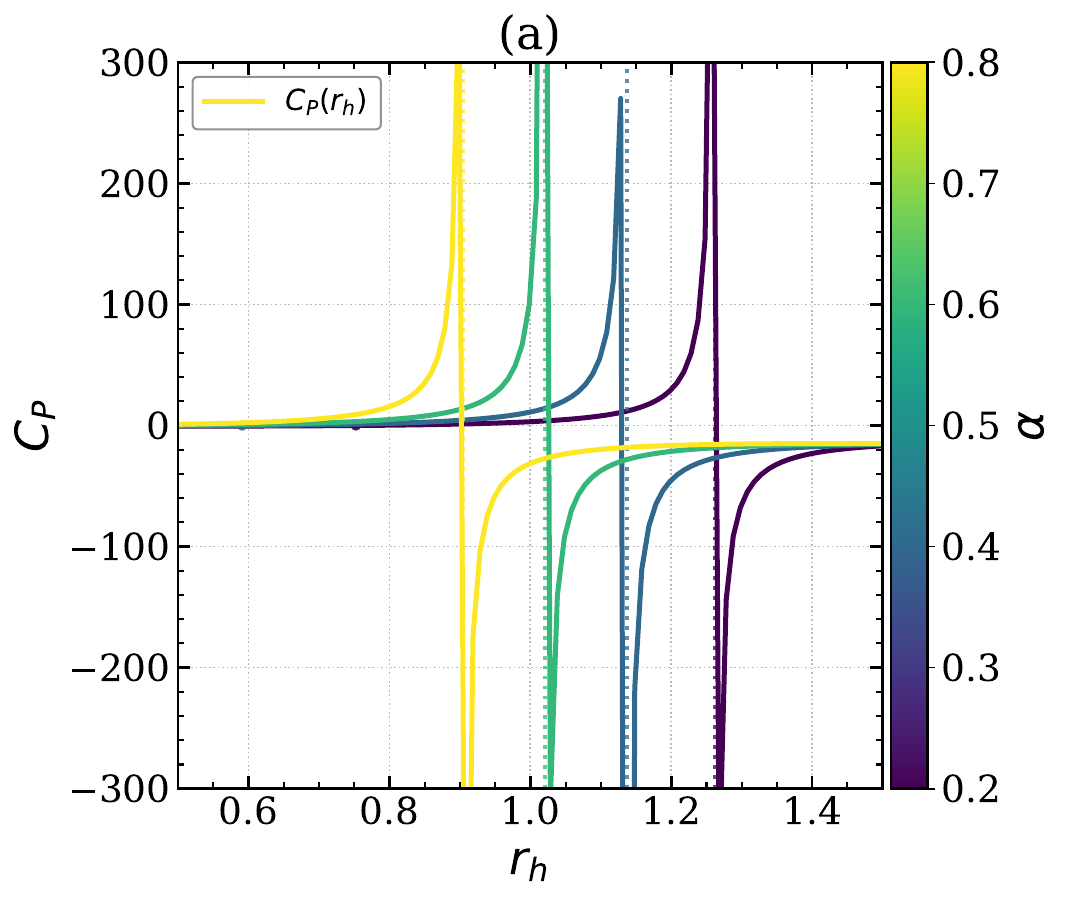}\qquad
\includegraphics[width=0.42\linewidth]{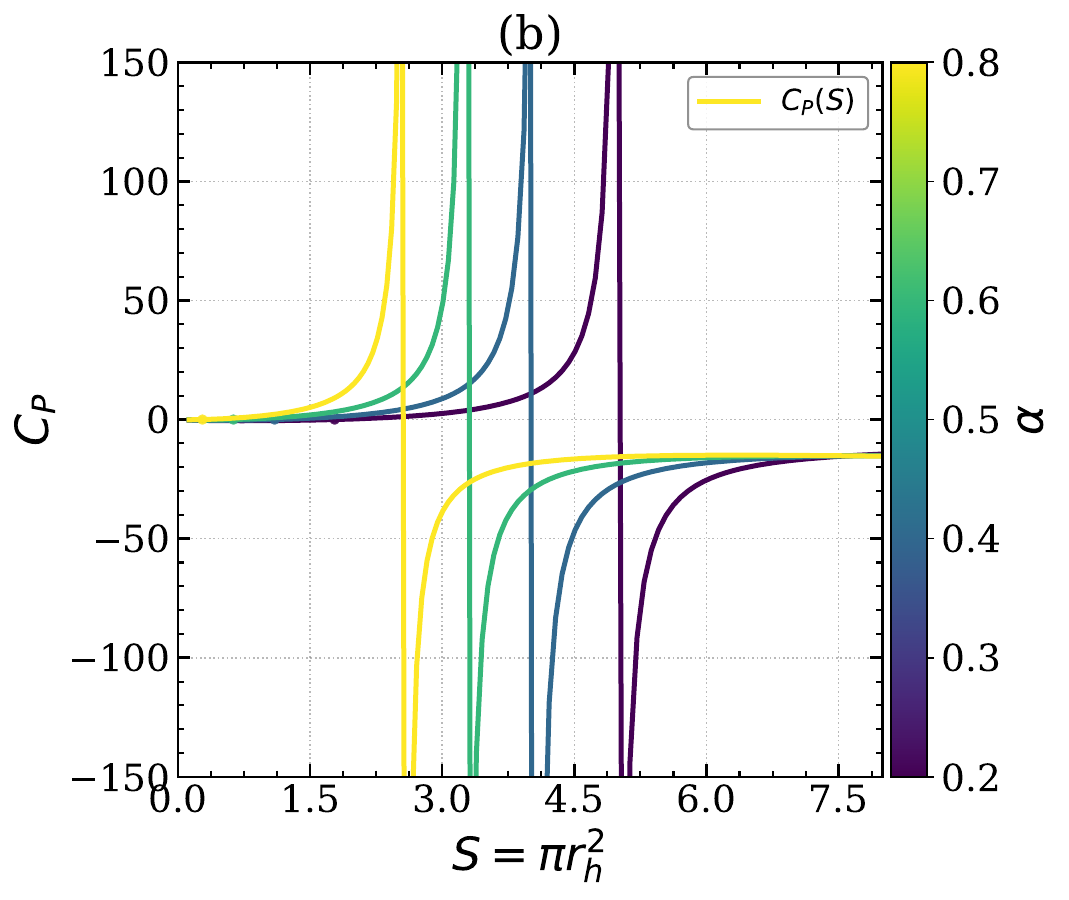}
\caption{\footnotesize Specific heat at fixed $(\alpha,\rho_0,R)$ for the Letelier BH in a KDM halo. We use $\rho_0=0.10$, $R=1.1$, and vary the string-cloud parameter $\alpha=\{0.2,0.4,0.6,0.8\}$. Panel (a): $C_P(r_h)$ on the interval shown; Panel (b): the same data in the entropy variable $S=\pi r_h^2$. The divergence(s) of $C_P$ occur exactly where $\partial_{r_h}T_H=0$ (second-order phase transition), as predicted by Eq.~(\ref{dTdr}), while zeros of $C_P$ coincide with $T_H=0$. Increasing $\alpha$ shifts the critical radius to a smaller $r_h$ and enlarges the domain where $C_P<0$, indicating a reduction of local thermal stability due to the string cloud.}
\label{fig:CP-panels}
\end{figure*}
\begin{figure*}[tbhp]
  \centering
  \includegraphics[width=0.9\linewidth]{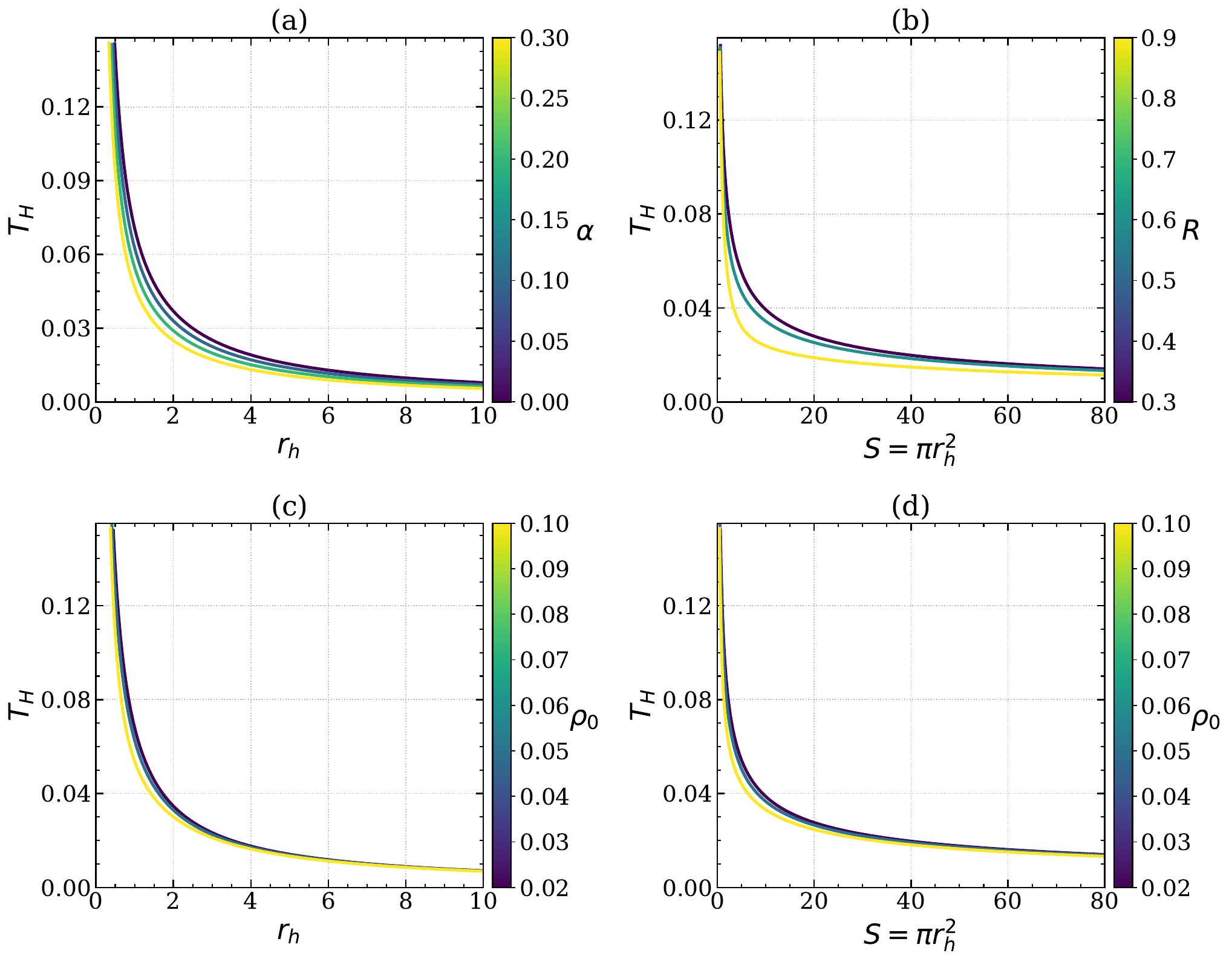}
  \caption{\footnotesize Hawking temperature for the Letelier black hole in a KDM halo given by Eq. (\ref{thermo2}).
  (a) $T_H$ vs.\ $r_h$ at fixed $(\rho_0,R)$ and varying $\alpha$.
  (b) $T_H$ vs.\ $S=\pi r_h^2$ at fixed $(\alpha,\rho_0)$ and varying $R$.
  (c) $T_H$ vs.\ $r_h$ at fixed $(\alpha,R)$ and varying $\rho_0$.
  (d) $T_H$ vs.\ $S$ for the same scan in $\rho_0$.}
  \label{fig:TH-panels}
\end{figure*}

Figure~\ref{fig:CP-panels} summarizes the heat-capacity behavior of the Letelier black hole surrounded by a King (KDM) halo at fixed halo parameters $(\rho_0,R)$ and varying string-cloud strength $\alpha$. The curves implement the thermodynamic relations derived in Sec.~\ref{Sec5}, namely $S=\pi r_h^2$, $T_H(r_h)$ in Eq.~\eqref{thermo2}, and $C_P=T_H\,(dS/dr_h)/(dT_H/dr_h)$ with $dT_H/dr_h$ given by Eq.~\eqref{dTdr}. As anticipated, $C_P$ diverges precisely where $dT_H/dr_h=0$ (second-order phase transition), and changes sign when passing through those radii. The dots along the horizontal axis indicate zeros of $T_H$; $C_P$ also vanishes there because the prefactor $T_H$ in the definition of $C_P$ goes to zero. Figure~\ref{fig:CP-panels}(a) shows the profile of $C_P$ as a function of $r_h$. For small horizons the halo correction $-8\pi\rho_0 R^3 r_h^2 /(r_h^2+R^2)^{3/2}$ in $T_H$ is subdominant and the curves resemble the Letelier limit; as $r_h$ grows, the competition between the $-(1-\alpha)/r_h^2$ and the KDM term in $dT_H/dr_h$ produces a turning point in $T_H$, hence a pole in $C_P$ (vertical dotted lines). Larger $\alpha$ moves this pole to a smaller $r_h$ and widens the negative-$C_P$ sector, indicating that the string cloud disfavors local thermal stability. Beyond the pole, the sign of $C_P$ flips, granting a locally stable branch ($C_P>0$) over a finite interval. At sufficiently large $r_h$ the curves gradually approach the Schwarzschild-like trend, $C_P\to -2\pi r_h^2$ (up to Letelier rescaling), so the stable window is bounded. Figure~\ref{fig:CP-panels}(b) shows the profile of $C_P$ as a function of $S=\pi r_h^2$. Reparametrizing the horizontal axis by $S$ preserves all qualitative features but makes the growth of the unstable sector more apparent as $\alpha$ increases: the divergence of $C_P$ occurs at a smaller entropy, and the negative-$C_P$ domain expands accordingly. The points with $C_P=0$ align with $T_H=0$ as in panel (a), delimiting the physically relevant ($T_H>0$) portions of the curves.

Overall, Fig.~\ref{fig:CP-panels} shows that the KDM halo, in concert with the string cloud, both lowers the Hawking temperature and induces a thermodynamic turning point that splits the equation of state into unstable and stable branches. 
The location of the pole and the width of the stable window are highly sensitive to $\alpha$ (and, implicitly, to $\rho_0$ and $R$), a fact that will be mirrored in the topological classification discussed in Sec.~\ref{Sec6}, where the critical radii correspond to the zero set of the topological vector fields.

\paragraph*{Remark on the non-monotonic pattern of heat-capacity divergences.}

From Eq.~\eqref{C_divergence_condition}, the specific heat diverges whenever
\begin{equation}
\frac{1-\alpha}{r_h^{2}}=\mathcal{R}(r_h)
\equiv \frac{8\pi\rho_0 R^3\,(2r_h^{2}-R^{2})}{(r_h^{2}+R^{2})^{5/2}}.
\label{eq:crit-balance}
\end{equation}
For fixed $(\rho_0,R)$, the right-hand side $\mathcal{R}(r_h)$ is \emph{not} monotonic in $r_h$: it turns on at $r_h=R/\sqrt{2}$, increases to a single maximum, and then decays as $\sim r_h^{-3}$ at large radii. By contrast, the left-hand side $(1-\alpha)/r_h^2$ is a monotonic function of $r_h$ whose overall scale decreases as $\alpha$ increases. As a result, the number and location of intersections of the two curves, hence, the radii $r_c$ where $C_P$ diverges, need \emph{not} vary monotonically with $\alpha$: depending on parameters, one may have 0, 1, or 2 intersections. This explains why in Fig.~\ref{fig:CP-panels} the sequence of divergence positions (and the apparent peak heights in the plots) does not follow a simple ordering with $\alpha$, and why a second divergence can emerge at larger $r_h$. We emphasize that the ``height'' of the rendered peaks is a plotting artifact (true divergences are unbounded); it depends on the chosen vertical range, the local sampling density in $r_h$, and any clipping used to avoid axis blow-ups. The qualitative, physically relevant information is the \emph{existence} and \emph{location} of the critical radii determined by Eq.~\eqref{eq:crit-balance}.

Figure~\ref{fig:TH-panels} displays the Hawking temperature of the Letelier-KDM black hole (Eq. (\ref{thermo2}))
for representative scans of the parameters $(\alpha,\rho_0,R)$.
Figure~\ref{fig:TH-panels}(a) shows $T_H$ as a function of the horizon radius for fixed $(\rho_0,R)$ and increasing string parameter $\alpha$.
Because $\alpha$ enters the square bracket with a minus sign, larger $\alpha$ uniformly suppresses the temperature at all radii.
The curves are monotonic and decay as $T_H\sim (1-\alpha)/(4\pi r_h)$ for large $r_h$, with halo-induced corrections that fall as $1/r_h^2$. Figure~\ref{fig:TH-panels}(b) presents $T_H$ vs. the entropy $S=\pi r_h^2$ at fixed $(\alpha,\rho_0)$ while varying the halo core scale $R$.
The KDM contribution scales as $8\pi\rho_0 R^3 r_h^2/(r_h^2+R^2)^{3/2}$: for $r_h\ll R$ it is $\sim 8\pi\rho_0 r_h^2$ (weakly dependent on $R$),
whereas for $r_h\gg R$ it approaches $\sim 8\pi\rho_0 R^3/r_h$.
Accordingly, increasing $R$ causes a mild reduction of $T_H$ that becomes more visible at intermediate/large $S$, while the small-entropy region is nearly insensitive to $R$. Figures~\ref{fig:TH-panels}(c) and (d) examine the dependence on the central halo density $\rho_0$ at fixed $(\alpha,R)$, plotted against $r_h$ and $S$, respectively.
Since the halo term is linear in $\rho_0$, larger $\rho_0$ systematically lowers the temperature across the whole domain, preserving the ordering of the curves.
In all panels, $T_H$ remains positive within the plotted ranges and decreases monotonically with $r_h$;
The large-$r_h$ tail reproduces the Letelier scaling $\propto 1/r_h$, while the KDM halo introduces controlled, parameter-dependent deviations most pronounced at intermediate radii and vanishing asymptotically.
\begin{figure*}[tbhp]
\centering
\includegraphics[width=0.9\linewidth]{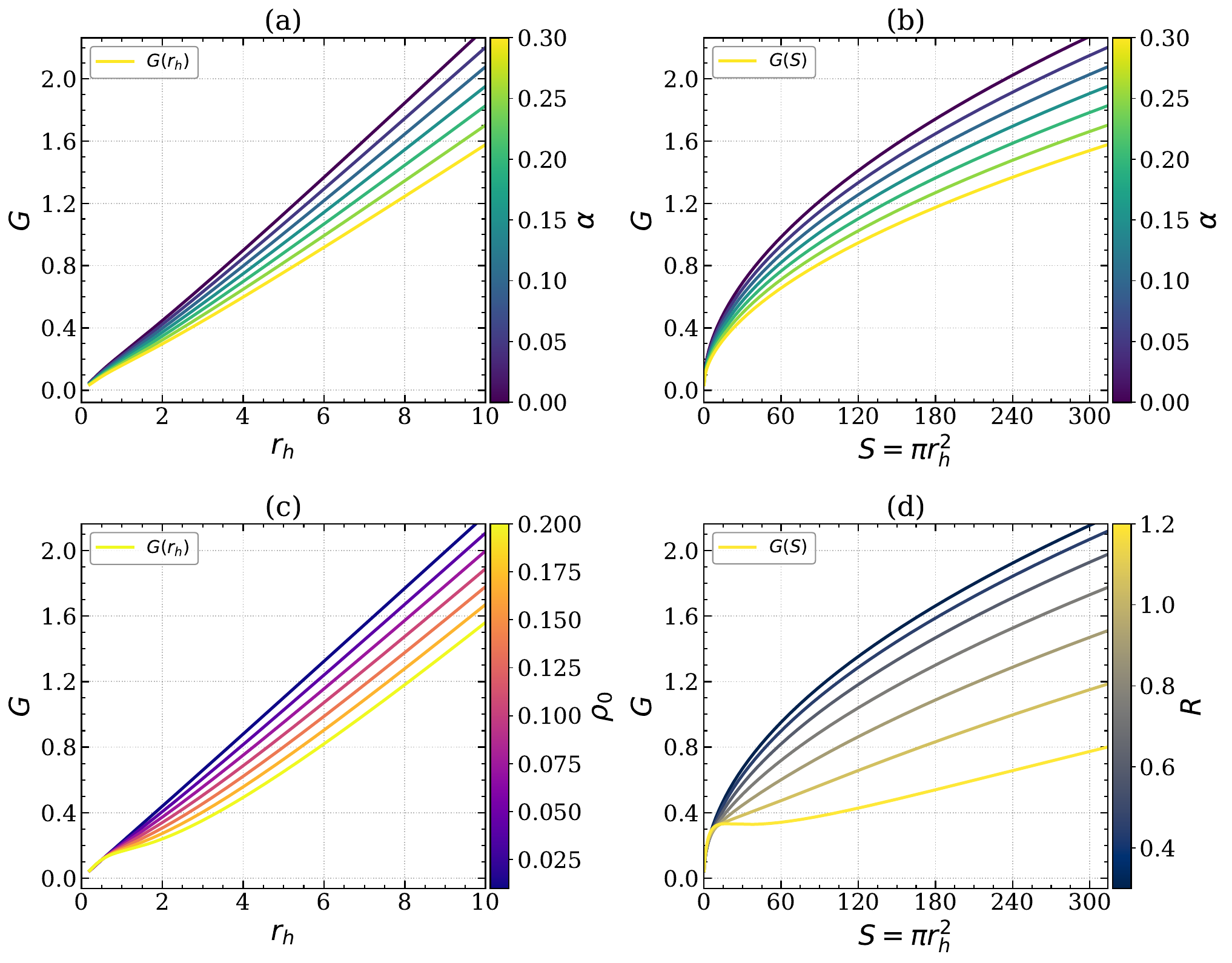}
\caption{\footnotesize
Gibbs free energy $G=M-T_H S$ of the Letelier BH immersed in KDM halo. 
Panels: (a) $G$ vs $r_h$ for fixed $(\rho_0,R)$ while scanning the string-cloud parameter $\alpha$; 
(b) $G$ vs $S=\pi r_h^2$ for the same scan in $\alpha$ at fixed $(\rho_0,R)$; 
(c) $G$ vs $r_h$ for fixed $(\alpha,R)$ while scanning the halo central density $\rho_0$; 
(d) $G$ vs $S$ for fixed $(\alpha,\rho_0)$ while scanning the halo core scale $R$. 
Each panel uses its own colorbar to indicate the scanned parameter. The remaining parameters (when not scanned) are kept fixed as specified in the main text.}
\label{fig:gibbs-panels}
\end{figure*}
\begin{figure}[tbhp]
    \centering
    \includegraphics[width=0.9\linewidth]{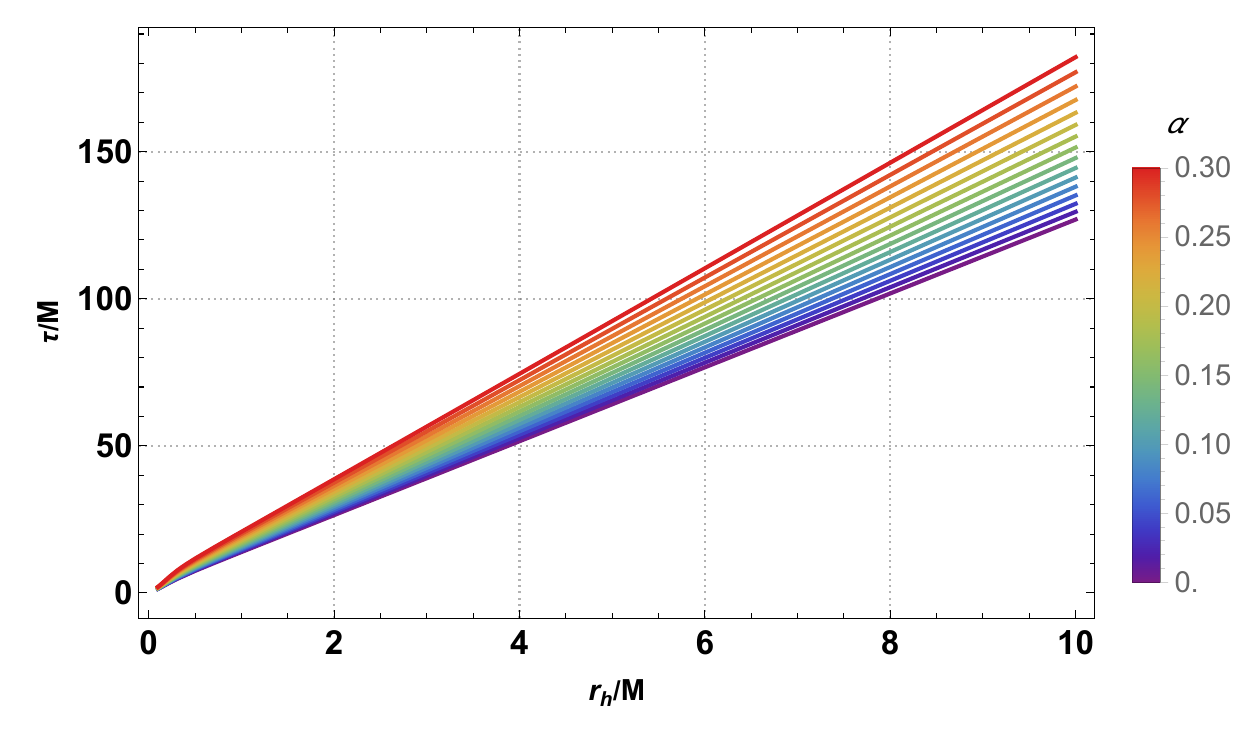}
    \caption{\footnotesize Inverse temperature $\tau/M$ vs. $r_h/M$ for fixed $\rho_0 M^2=0.5$ and $R/M=0.2$. The curves reproduce the on-shell relation $\tau=1/T_H(r_h)$ implied by Eq.~\eqref{ee5}, and their slope changes with $\alpha$ through $T_H$ in Eq.~\eqref{thermo2}.}
    \label{fig:inversion}
\end{figure}

\begin{figure*}[tbhp]
    \centering
    \begin{minipage}{0.46\linewidth}
    \centering
    \includegraphics[width=\linewidth]{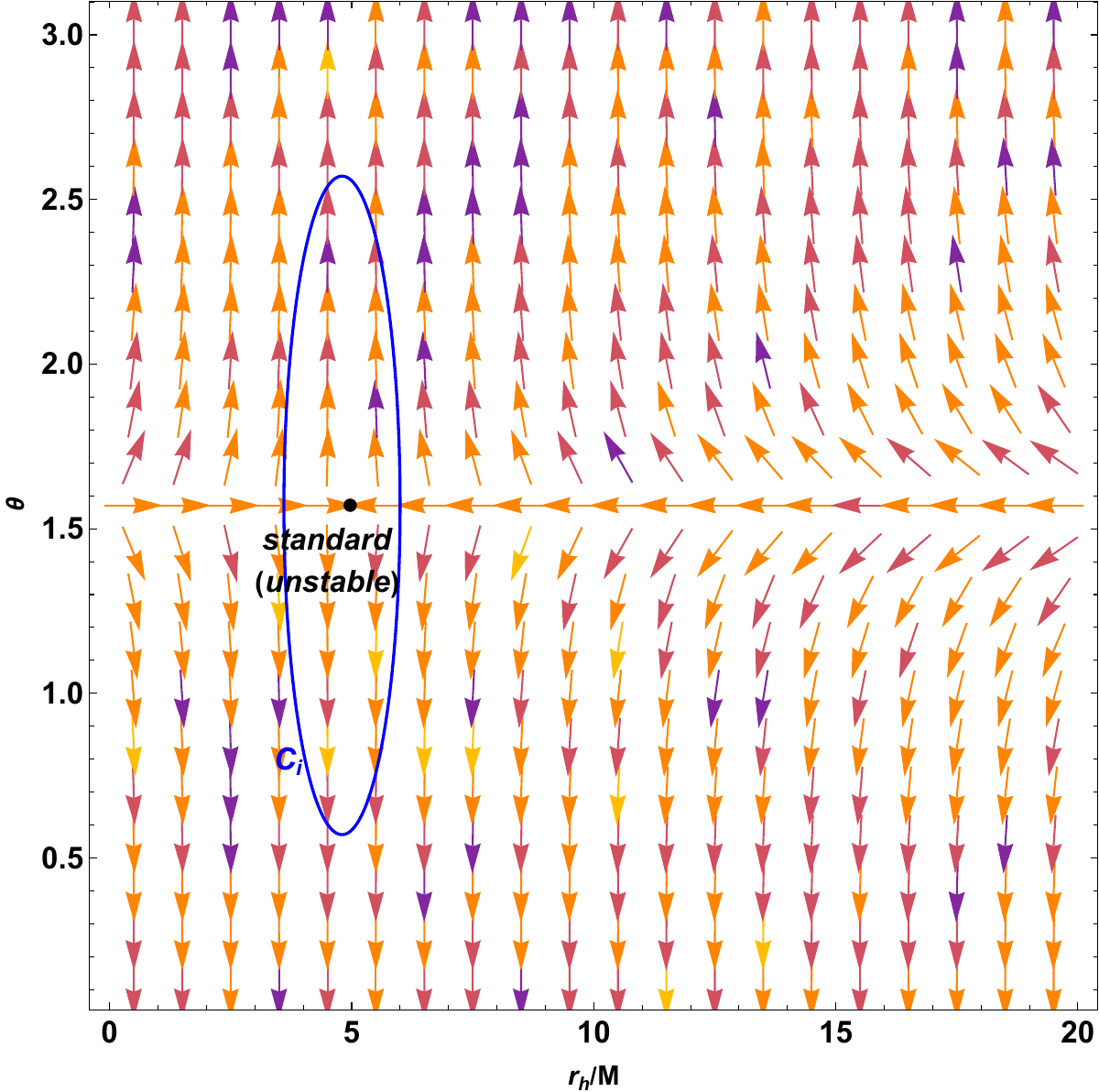}\\[3pt]
    \footnotesize (i) $\alpha=0.05$
    \end{minipage}
    \hfill
    \begin{minipage}{0.46\linewidth}
    \includegraphics[width=\linewidth]{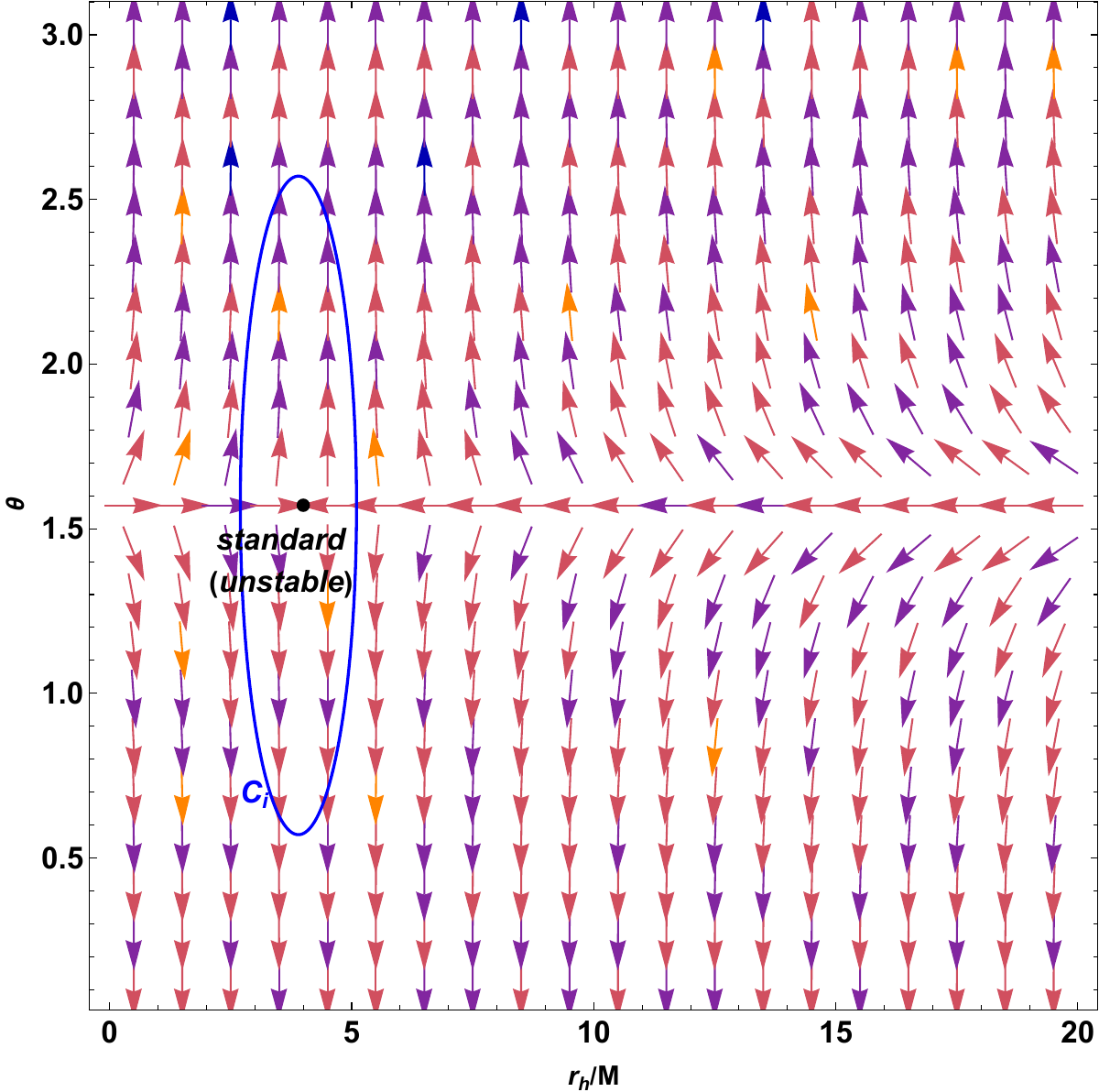}\\[3pt]
    \footnotesize (ii) $\alpha=0.10$
    \end{minipage}
  \caption{\footnotesize The arrows represent the unit vector field $\boldsymbol{n}_{\mathcal{F}}$ of energy $\mathcal{F}$ on a portion of the $r$-$\Theta$ plane for the Letelier black holes surrounded by KDM halo with $R/M = 0.1$, $\rho_0=0.1/M^2$ and $\tau=20\pi M$. The horizon ring (HR), marked with a black dot, is at $(r_h, \theta) =(4.97, \pi/2)$ for $\alpha=0.05$ and at $(r_h, \theta) =(4.0,\pi/2)$ for $\alpha=0.10$. The blue contour $\mathcal{C}_i$ is a closed loop enclosing the horizon ring with a topological charge $Q = -1$.}
  \label{fig:unit-vector-3}
\end{figure*}
Figure~\ref{fig:gibbs-panels} displays the Gibbs free energy $G(r_h)=M(r_h)-T_H(r_h)\,S(r_h)$ for the Letelier-KDM black hole in four complementary sweeps of the model parameters. In all panels, the curves are smooth and monotonic in the horizon radius $r_h$ (or equivalently in the entropy $S=\pi r_h^2$), with no nonconvex segments or swallowtail structures; thus, for the fiducial ranges considered, the plots show no evidence of first-order phase transitions. When varying the string-cloud parameter $\alpha$ at fixed $(\rho_0,R)$ primarily tilts the family of curves (Figs.~\ref{fig:gibbs-panels}(a) and (b)). Since both $M$ and $T_H$ carry the combination $(1-\alpha)$ (Eqs.~\eqref{thermo1}-\eqref{thermo2}), changing $\alpha$ acts as a nearly linear rescaling in $r_h$, shifting the $G$-profiles in a controlled way. For our fiducial values, $G$ is an increasing function of $r_h$ and $S$, and higher $\alpha$ produces a systematic displacement of the curves (as encoded by the colorbar), consistent with the analytical dependence of $M$ and $T_H$ on $\alpha$. On the other hand, scanning the halo central density $\rho_0$ at fixed $(\alpha,R)$ modifies both terms in $G=M-T_H S$ (Figs.~\ref{fig:gibbs-panels}(c)): the mass function acquires the halo contributions in Eq.~\eqref{thermo1}, while the temperature picks up the KDM correction in Eq.~\eqref{thermo2}. The resulting families remain monotonic in $r_h$; in the parameter window shown, increasing $\rho_0$ shifts the curves approximately linearly, reflecting the fact that the KDM corrections enter $G$ with overall factors $\propto \rho_0 R^3$. In Fig.~\ref{fig:gibbs-panels}(d), at fixed $(\alpha,\rho_0)$, increasing the core radius $R$ strengthens the geometric lever arm $R^3/\!(r_h^2+R^2)^{1/2}$ that appears in both $M$ and $T_H$. The family's $G(S)$ remains smooth and strictly increasing with $S$, and the color-coded progression with $R$ indicates a coherent, nearly linear shift over the entropy range displayed.

Overall, the four panels demonstrate that, within the ranges explored, the Gibbs free energy is a smooth, monotonic function of $r_h$ (or $S$) and responds predictably to variations of $\alpha$, $\rho_0$, and $R$. The absence of swallowtail behavior is consistent with the specific-heat analysis in Sec.~\ref{Sec5}, which identifies criticality only through the divergence of $\mathbb{C}$ when $\partial_{r_h}T_H=0$, rather than through competing thermodynamic branches of $G$.

\section{Topological Characteristics of Thermodynamic Potentials}\label{Sec6}

In Sec.~\ref{Sec5} we derived the Hawking temperature $T_H(r_h)$ in Eq.~\eqref{thermo2} and its radial derivative in Eq.~\eqref{dTdr}, and we showed that the specific heat $\mathbb{C}$ diverges at radii where $\partial_{r_h}T_H=0$ (cf. Eq.~\eqref{C_divergence_condition}). We now use a topological approach to examine these critical points and classify the corresponding phase transitions. 
In addition, we delve into the analysis of phase transitions encoded in a generalized (off-shell) free energy, which allows us to determine the thermodynamic topological class of the BH described by Eq.~\eqref{aa1}.

Following Refs.~\cite{barzi2024renyi,chen2024thermal,alipour2023topological,wei2022topology}, we define a potential that depends on the Hawking temperature,
\begin{eqnarray}
\Phi(r_h,\theta)=\frac{T_{\rm H}(r_h)}{\sin \theta},\label{ee1}
\end{eqnarray}
and introduce the associated vector field in the $(r_h,\theta)$ space \cite{yerra2023topology,mehmood2023thermodynamic,zhang2023topology}
\begin{equation}
\phi_{\Phi}^r=\partial_{r_h}\Phi,\qquad
\phi_{\Phi}^\theta=\partial_\theta\Phi.
\end{equation}
The zero points of this field are located at $\theta=\pi/2$ and $\partial_{r_h}T_{\rm H}(r_h)=0$, i.e., precisely at the radii where the heat capacity diverges in Sec.~\ref{Sec5}. Thus, each zero of $\vec\phi_\Phi$ corresponds to a thermodynamic critical point identified by Eq.~\eqref{C_divergence_condition}.

To examine the topological structure in a complementary way, we follow \cite{Wei2022b,wu2023topological} and consider the generalized (off-shell) free energy
\begin{equation}
\mathcal{F}(r_h,\tau)=M(r_h)-\frac{S(r_h)}{\tau},\label{ee2}
\end{equation}
where $\tau$ is an external inverse-temperature parameter and $S$ is the BH entropy. Consistently with Sec.~\ref{Sec5}, we take $S=\pi r_h^2$ (Eq.~\eqref{ee3}).

The corresponding vector field $\phi^a_{\mathcal{F}}$ is defined as \cite{Wei2022a,Wei2022b,wu2025}
\begin{align}
\phi^{r}_{\mathcal{F}}&=\partial_{r_h}\mathcal{F}
=M'(r_h)\;-\;\frac{dS/dr_h}{\tau}\notag \\&
= \frac{1}{2}\Big[1 - \alpha-\frac{8\pi \rho_0 R^3 r_h^{2}}{(r_h^2 + R^2)^{3/2}}\Big]\;-\;\frac{2\pi r_h}{\tau},\\
\phi^{\theta}_{\mathcal{F}}&=-\cot \theta \csc \theta,\label{ee4}
\end{align}
where $a=1,2$, $\phi^{1}_{\mathcal{F}}=\phi^{r}_{\mathcal{F}}$ and $\phi^{2}_{\mathcal{F}}=\phi^{\theta}_{\mathcal{F}}$.
In the last equality for $\phi^{r}_{\mathcal{F}}$ we used the result of Sec.~\ref{Sec5}, $M'(r_h)=2\pi r_h\,T_H(r_h)$ (equivalently, the first law $dM=T_H\,dS$ at fixed $\alpha,\rho_0,R$).

The zero points of $\vec\phi_{\mathcal{F}}$ satisfy $\theta=\pi/2$ and $\partial_{r_h}\mathcal{F}=0$. Using $M'(r_h)=2\pi r_h T_H$ and $dS/dr_h=2\pi r_h$, the condition $\partial_{r_h}\mathcal{F}=0$ gives
\begin{eqnarray}
\tau&=\frac{dS/dr_h}{M'(r_h)}=\frac{2\pi r_h}{M'(r_h)}
=\frac{1}{T_H(r_h)}\notag\\
&=\frac{4\pi r_h}{\,1 - \alpha-\dfrac{8\pi \rho_0 R^3 r_h^{2}}{(r_h^2 + R^2)^{3/2}}\,}.
\label{ee5}
\end{eqnarray}
Therefore, the off-shell zero set of $\vec\phi_{\mathcal{F}}$ coincides with the on-shell equilibrium curve $\tau=1/T_H(r_h)$, ensuring full consistency with the thermodynamics of Sec.~\ref{Sec5}.

We may write $\phi^a_{\mathcal{F}}=\|\phi_{\mathcal{F}}\|\,(\cos\Theta,\sin\Theta)$ and introduce the normalized vector
\begin{equation}
    n^{r}_{\mathcal{F}}=\frac{\phi^{r}_{\mathcal{F}}}{\|\phi_{\mathcal{F}}\|}\,, 
    \qquad 
    n^{\theta}_{\mathcal{F}}=\frac{\phi^{\theta}_{\mathcal{F}}}{\|\phi_{\mathcal{F}}\|},\label{ee6}
\end{equation}
with $\|\phi_{\mathcal{F}}\|=\sqrt{(\phi^{r}_{\mathcal{F}})^2+(\phi^{\theta}_{\mathcal{F}})^2}$.
Using Duan’s $\phi$-mapping topological current theory, we define
\begin{equation}
    j^{\mu}=\frac{1}{2\pi}\,\varepsilon^{\mu\nu\lambda}\,\varepsilon_{ab}\,\partial_{\nu} n^{a}_{\mathcal{F}}\,\partial_{\lambda}\,n^{b}_{\mathcal{F}},
    \qquad \mu,\nu,\lambda=0,1,2.\label{ee7}
\end{equation}
Noether’s theorem implies conservation:
\begin{equation}
\partial_\mu j^\mu = 0.\label{ee8}
\end{equation}
Equivalently,
\begin{equation}
j^\mu = \delta^2(\phi_{\mathcal{F}})\, J^\mu\!\left(\frac{\phi_{\mathcal{F}}}{x}\right),\label{ee9}
\end{equation}
with the Jacobian tensor
\begin{equation}
\varepsilon^{ab} J^\mu\!\left(\frac{\phi_{\mathcal{F}}}{x}\right) 
= \varepsilon^{\mu\nu\rho} \partial_\nu \phi^a_{\mathcal{F}} \partial_\rho \phi^b_{\mathcal{F}}.\label{ee10}
\end{equation}
For $\mu=0$, one recovers the usual Jacobian
\[
J^0\!\left(\frac{\phi_{\mathcal{F}}}{x}\right) 
= \frac{\partial(\phi^1_{\mathcal{F}}, \phi^2_{\mathcal{F}})}{\partial(x^1, x^2)}.
\]
The total topological charge (winding number) is
\begin{equation}
W = \int_\Sigma j^0 \, d^2x = \sum_{i=1}^n \beta_i \eta_i = \sum_{i=1}^n \omega_i.\label{ee12}
\end{equation}
Here $\beta_i>0$ is the Hopf index counting the loops of $\vec\phi_{\mathcal{F}}$ around the zero $z_i$, and $\eta_i=\mathrm{sign}\big[J^0(\phi_{\mathcal{F}}/x)\big]_{z_i}=\pm1$.
A positive (negative) winding number in $(r_h,\theta)$ is associated with a locally stable (unstable) thermodynamic branch, matching the sign of the specific heat discussed in Sec.~\ref{Sec5}.

In summary, the topological characterization based on $\Phi=T_H/\sin\theta$ and on the generalized free energy $\mathcal{F}=M-S/\tau$ is fully consistent with the thermodynamic relations of Sec.~\ref{Sec5}: the equilibrium condition $\partial_{r_h}\mathcal{F}=0$ reproduces $\tau=1/T_H(r_h)$, and the critical set $\partial_{r_h}T_H=0$ coincides with the loci where $\mathbb{C}$ diverges, providing a topological classification of the phase transitions.

The unit vector field $\boldsymbol{n}_{\mathcal{F}}$ of free energy $\mathcal{F}$ on a portion of the $r-\theta$ plane is plotted in Figure \ref{fig:unit-vector-3} with parameters $R/M=0.1$, $\rho_0\,M^2=0.1$, and $\tau=20\pi M$ for two different values of $\alpha$. We observe that there is a horizon ring, marked with a black dot, is located at $(r_h, \theta) =(4.97, \pi/2)$ for $\alpha=0.05$; and (ii) $(r_h, \theta) =(4.0,\pi/2)$ for $\alpha=0.10$. Furthermore, the winding number $W$ associated with the blue contour $C_i$ characterizes the behavior of the vector field analogous to Figs. \ref{fig:unit-vector-1}-\ref{fig:unit-vector-2}. 

\section{Conclusions}\label{Sec8}

We analyzed a Schwarzschild black hole threaded by a spherically symmetric cloud of strings and immersed in a King dark matter halo, showing how the string parameter $\alpha$ and the halo parameters $(\rho_0, R)$ jointly deform the lapse function and impact optical, dynamical, and thermodynamic properties. On the optical side, the photon effective potential and the photon-sphere radius increase with the string-cloud strength and with the halo contribution, which enlarges the shadow predicted from $R_{\rm sh}=r/\sqrt{f(r)}$ at the circular null orbit; the topological current construction based on Duan’s mapping yields a single equatorial zero with winding number $-1$, identifying the unstable photon ring responsible for the shadow. For timelike motion, the effective potential and the derived observables $E_{\rm sp}$, $L_{\rm sp}$, and $\Omega_\phi$ acquire halo- and $\alpha$-dependent corrections, and the innermost stable circular orbit shifts outward; in the halo-free limit one recovers $r_{\rm ISCO}=6M/(1-\alpha)$, continuously connecting to the Letelier and Schwarzschild cases. Thermodynamically, fixing $(\alpha,\rho_0,R)$ lowers the Hawking temperature relative to the halo-free case according to $T_H(r_h)=(4\pi r_h)^{-1}\!\left[1-\alpha-8\pi\rho_0 R^3 r_h^2 (r_h^2+R^2)^{-3/2}\right]$, and the Gibbs free energy is smooth and monotonic throughout the explored ranges, showing no swallowtail structure. The specific heat diverges where $\partial_{r_h} T_H=0$, which separates locally unstable from stable branches; an analytic reduction demonstrates that the response function governing this condition is unimodal, so at most two critical radii occur for fixed parameters. A complementary thermodynamic-topology analysis using the vector fields built from $T_H$ and from the generalized free energy $\mathcal{F}=M-S/\tau$ identifies equilibrium and critical sets that coincide with the zeros and poles inferred from the standard thermodynamics, providing a consistent topological classification of phase transitions. These results suggest observational consequences: the enlargement of the photon sphere and shadow, together with the outward ISCO shift, can affect horizon-scale imaging and thin-disk spectra. Extensions that include rotation, quasinormal-mode spectra, and detailed lensing and accretion modeling in the Letelier-King background are natural next steps toward confronting the framework with gravitational-wave and Event Horizon Telescope data. An important extension of this BH solution involves studying the accretion disk structure, including energy flux, effective temperature, differential luminosity, and emission spectra. These analyses provide key insights into the BH’s observational features and help connect theoretical predictions with X-ray and imaging data from instruments like the Event Horizon Telescope.

The future scope of this BH solution involves a detailed study of perturbations across different spin fields - specifically, spin-0 (scalar), spin-1 (vector/electromagnetic), and spin-1/2 (Dirac) fields. Such perturbative analyses are essential for evaluating the stability of the BH and understanding wave propagation in modified spacetime geometries influenced by dark matter distributions and/or topological structures. A key extension lies in determining quasinormal modes (QNMs) using higher-order WKB approximation methods, which offer precise insights into the dynamical response of the BH to external disturbances. These QNMs have significant implications in gravitational wave astronomy, especially during the ringdown phase of BH mergers.

Another promising avenue is the investigation of the BH’s thermodynamic behavior, particularly through the Joule-Thomson (JT) expansion, which explores temperature variations during adiabatic processes and shows potential phase transitions in extended phase space. Furthermore, incorporating Generalized Uncertainty Principle (GUP) corrections into the thermal analysis allows for the examination of quantum gravitational effects on thermodynamic quantities such as Hawking temperature, entropy, and specific heat. Together, these studies not only enhance theoretical understanding but also connect with observational prospects in high-energy astrophysics and gravitational wave detection.

Beyond the qualitative trends already highlighted, our results admit a compact synthesis in terms of monotonic responses of geometric, orbital, and thermal scales to the parameters $(\alpha,\rho_0,R)$. On the optical side, the photon-sphere and shadow radii grow with either a stronger string cloud ($\alpha\uparrow$) or a denser/larger King core ($\rho_0\uparrow$ or $R\uparrow$), while the effective radial force is reduced in magnitude, consistently weakening light bending at fixed $r/M$. For timelike motion, $r_{\rm ISCO}$ increases in the same directions in parameter space, implying lower characteristic disk temperatures and a systematic outward drift of the inner edge; concomitantly, $\mathrm{E}_{\rm sp}$ and $\mathrm{L}_{\rm sp}$ rise for fixed $r/M$, signaling tighter binding and faster rotation required to sustain circular orbits. Thermodynamically, $T_H(r_h)$ is uniformly suppressed by $\alpha$ and linearly by $\rho_0$, and the specific heat exhibits at most two divergences governed by a unimodal response function, providing a clean separation between locally unstable and stable branches that matches the thermodynamic-topology classification.

These correlated trends suggest a pragmatic route to observational inference. First, the pair (shadow size, ISCO-sensitive spectral proxy) helps break the degeneracy between $\alpha$ and the halo parameters: the shadow reacts to both $\alpha$ and $(\rho_0,R)$, whereas line profiles and QPOs tied to $r_{\rm ISCO}$ respond more steeply to $(\rho_0,R)$; a joint fit therefore tightens constraints. Second, the topological charge $Q=-1$ of the photon ring implies a single unstable ring that controls the eikonal quasinormal sector through the standard Lyapunov-exponent correspondence, enabling consistency checks between shadow-compatible $(\alpha,\rho_0,R)$ and ringdown frequencies. Third, because $T_H$ and $G$ are smooth in the explored ranges, any evidence of multi-branch free-energy structure (e.g., swallowtails) in data would directly falsify parts of the parameter space or point to missing physics (rotation, charge, anisotropy, or non-ideal halo features).

Limitations and extensions are clear. Our background is static and spherically symmetric with an isotropic King halo; allowing rotation (Kerr-like Letelier-King), halo anisotropy, or small departures from spherical symmetry should imprint measurable shifts on the photon-ring Lyapunov exponent, lensing coefficients, and ISCO energetics. On the perturbative side, computing QNMs (scalar, electromagnetic, and Dirac) across the $(\alpha,\rho_0,R)$ space-e.g., via WKB/Leaver methods-will quantify the ringdown-shadow interplay anticipated by our topology analysis. On the thermodynamic side, extending to the $P$-$V$ framework, performing the Joule-Thomson expansion, and adding GUP corrections will probe quantum-thermal effects and possible criticality beyond the non-extended sector studied here. Finally, confronting the model with EHT bounds and X-ray disk spectroscopy through a simple Bayesian pipeline (shadow radius prior $+$ Fe K$\alpha$ line/continuum-fitting likelihood) offers a concrete path to constrain $(\alpha,\rho_0,R)$ with current data and to forecast the discriminating power of next-generation observations.

\section*{Acknowledgments}

F.A. acknowledges the Inter University Centre for Astronomy and Astrophysics (IUCAA), Pune, India, for granting a visiting associateship. E. O. Silva acknowledges the support from grants CNPq/306308/2022-3, FAPEMA/UNIVERSAL-06395/22, FAPEMA/APP-12256/22, and (CAPES) - Brazil (Code 001).

\appendix

\section*{Appendix: Unimodal structure of the response function and the number of heat-capacity divergences}

In Sec. \ref{Sec5}, the divergences of the specific heat are determined by the balance
\begin{equation}
\frac{1-\alpha}{r_h^{2}}=\mathcal{R}(r_h),
\qquad
\mathcal{R}(r_h)\equiv \frac{8\pi\rho_0 R^3\,(2r_h^{2}-R^{2})}{(r_h^{2}+R^{2})^{5/2}}.
\label{eq:AppCritBalance}
\end{equation}
Here we show that $ \mathcal{R}(r_h) $ is \emph{unimodal}: it turns on at $r_h=R/\sqrt{2}$, increases monotonically to a single global maximum, and then decays as $r_h^{-3}$. This immediately implies that Eq.~\eqref{eq:AppCritBalance} admits at most two positive solutions for fixed $0\le\alpha<1$, i.e., at most two $C_P$ divergences.

\paragraph{Dimensionless reduction.}
Set $x\equiv r_h/R\ge0$. Then
\begin{equation}
\mathcal{R}(r_h)=8\pi\rho_0\,\underbrace{\frac{2x^{2}-1}{(1+x^{2})^{5/2}}}_{\displaystyle h(x)},
\label{eq:AppRshape}
\end{equation}
so the shape is fully captured by the dimensionless function $h(x)$.

\paragraph{Stationary points of $h(x)$.}
Differentiate:
\begin{align}
h'(x)
&=\frac{d}{dx}\!\left[(2x^{2}-1)(1+x^{2})^{-5/2}\right]
=\frac{3x\,(3-2x^{2})}{(1+x^{2})^{7/2}}.
\label{eq:Apphprime}
\end{align}
Thus, the stationary points are
\[
x=0, \qquad x_\star=\sqrt{\tfrac{3}{2}}.
\]
Moreover:
\begin{itemize}
\item For $0<x<\sqrt{3/2}$, one has $h'(x)>0$ (strictly increasing).
\item For $x>\sqrt{3/2}$, one has $h'(x)<0$ (strictly decreasing).
\end{itemize}
Since $h(x)\to 0^-$ as $x\to 0^+$ (indeed $h(0)=-1$) and $h(x)\to 0^+$ as $x\to\infty$ with the asymptotics $h(x)\sim 2/x^{3}$, it follows that $x_\star=\sqrt{3/2}$ is the \emph{unique global maximum}. The zero of $h$ occurs at
\[
h(x)=0 \;\Longleftrightarrow\; x=\frac{1}{\sqrt{2}}\,,
\]
so $h(x)<0$ for $0\le x<1/\sqrt{2}$ and $h(x)>0$ for $x>1/\sqrt{2}$.

\paragraph{Maximum value.}
At the unique maximum $x_\star=\sqrt{3/2}$,
\begin{equation}
h_{\max}=h\!\left(\sqrt{\tfrac{3}{2}}\right)
=\frac{2\cdot \tfrac{3}{2}-1}{\bigl(1+\tfrac{3}{2}\bigr)^{5/2}}
=\frac{2}{\left(\tfrac{5}{2}\right)^{5/2}}
=\frac{2^{7/2}}{5^{5/2}}.
\end{equation}
Therefore
\begin{equation}
\mathcal{R}_{\max}=8\pi\rho_0\,h_{\max}
=8\pi\rho_0\,\frac{2^{7/2}}{5^{5/2}}.
\label{eq:AppRmax}
\end{equation}

\paragraph{Asymptotics.}
Near the origin, $h(x)=-1+O(x^{2})$, hence $\mathcal{R}(r_h)\sim -8\pi\rho_0$ (unphysical branch, since the left-hand side of Eq.~\eqref{eq:AppCritBalance} is positive); the curve turns positive for $x>1/\sqrt{2}$. For large $x$,
\begin{align}
h(x)=\frac{2x^{2}-1}{(1+x^{2})^{5/2}}=\frac{2}{x^{3}}+O\!\left(\frac{1}{x^{5}}\right),
\notag \\
\Rightarrow\quad
\mathcal{R}(r_h)=\frac{16\pi\rho_0}{x^{3}}+O\!\left(\frac{1}{x^{5}}\right)
\propto r_h^{-3}.
\end{align}

\paragraph{Consequence for the number of intersections.}
The left-hand side of Eq.~\eqref{eq:AppCritBalance} is the strictly decreasing function
\[
L(r_h)=\frac{1-\alpha}{r_h^{2}}, \qquad 0\le\alpha<1.
\]
The right-hand side $ \mathcal{R}(r_h) $ is unimodal: it vanishes at $r_h=R/\sqrt{2}$, increases monotonically to the unique peak $\mathcal{R}_{\max}$ at $r_h=R\sqrt{3/2}$, and then decreases monotonically to zero. The intersection equation $L(r_h)=\mathcal{R}(r_h)$ can therefore have:
\begin{itemize}
\item no solution (if $L(r_h)>\mathcal{R}_{\max}$ for all $r_h$, i.e. if the horizontal scale set by $1-\alpha$ is too large);
\item exactly one solution (tangency at the peak, a codimension-one case);
\item or exactly two solutions (one on each side of the peak).
\end{itemize}
Hence, the specific heat $C_P$ exhibits at most two divergences (critical radii) for fixed $(\alpha,\rho_0,R)$. This also clarifies the non-monotonic motion of the divergence radii with $\alpha$: increasing $\alpha$ uniformly lowers $L(r_h)$, which can \emph{create}, \emph{annihilate}, or \emph{split} intersections depending on whether the curve slides above, through, or below the unique peak of $\mathcal{R}$.

\bibliographystyle{apsrev4-2}

\begin{thebibliography}{124}%
\makeatletter
\providecommand \@ifxundefined [1]{%
 \@ifx{#1\undefined}
}%
\providecommand \@ifnum [1]{%
 \ifnum #1\expandafter \@firstoftwo
 \else \expandafter \@secondoftwo
 \fi
}%
\providecommand \@ifx [1]{%
 \ifx #1\expandafter \@firstoftwo
 \else \expandafter \@secondoftwo
 \fi
}%
\providecommand \natexlab [1]{#1}%
\providecommand \enquote  [1]{``#1''}%
\providecommand \bibnamefont  [1]{#1}%
\providecommand \bibfnamefont [1]{#1}%
\providecommand \citenamefont [1]{#1}%
\providecommand \href@noop [0]{\@secondoftwo}%
\providecommand \href [0]{\begingroup \@sanitize@url \@href}%
\providecommand \@href[1]{\@@startlink{#1}\@@href}%
\providecommand \@@href[1]{\endgroup#1\@@endlink}%
\providecommand \@sanitize@url [0]{\catcode `\\12\catcode `\$12\catcode
  `\&12\catcode `\#12\catcode `\^12\catcode `\_12\catcode `\%12\relax}%
\providecommand \@@startlink[1]{}%
\providecommand \@@endlink[0]{}%
\providecommand \url  [0]{\begingroup\@sanitize@url \@url }%
\providecommand \@url [1]{\endgroup\@href {#1}{\urlprefix }}%
\providecommand \urlprefix  [0]{URL }%
\providecommand \Eprint [0]{\href }%
\providecommand \doibase [0]{https://doi.org/}%
\providecommand \selectlanguage [0]{\@gobble}%
\providecommand \bibinfo  [0]{\@secondoftwo}%
\providecommand \bibfield  [0]{\@secondoftwo}%
\providecommand \translation [1]{[#1]}%
\providecommand \BibitemOpen [0]{}%
\providecommand \bibitemStop [0]{}%
\providecommand \bibitemNoStop [0]{.\EOS\space}%
\providecommand \EOS [0]{\spacefactor3000\relax}%
\providecommand \BibitemShut  [1]{\csname bibitem#1\endcsname}%
\let\auto@bib@innerbib\@empty
\bibitem [{\citenamefont {Novikov}\ and\ \citenamefont
  {Thorne}(1973)}]{novikov1973astrophysics}%
  \BibitemOpen
  \bibfield  {author} {\bibinfo {author} {\bibfnamefont {I.~D.}\ \bibnamefont
  {Novikov}}\ and\ \bibinfo {author} {\bibfnamefont {K.~S.}\ \bibnamefont
  {Thorne}},\ }in\ \href {https://doi.org/TODO} {\emph {\bibinfo {booktitle}
  {Les Astres Occlus: Les Houches Summer School of Theoretical Physics, August
  1972}}},\ \bibinfo {editor} {edited by\ \bibinfo {editor} {\bibfnamefont
  {C.}~\bibnamefont {DeWitt}} \emph {et~al.}}\ (\bibinfo  {publisher} {Gordon
  and Breach},\ \bibinfo {year} {1973})\ pp.\ \bibinfo {pages}
  {343--550}\BibitemShut {NoStop}%
\bibitem [{\citenamefont {Novikov}\ and\ \citenamefont
  {Frolov}(1989)}]{novikov2013physics}%
  \BibitemOpen
  \bibfield  {author} {\bibinfo {author} {\bibfnamefont {I.~D.}\ \bibnamefont
  {Novikov}}\ and\ \bibinfo {author} {\bibfnamefont {V.}~\bibnamefont
  {Frolov}},\ }\href {https://doi.org/10.1007/978-94-009-2376-2} {\emph
  {\bibinfo {title} {Physics of Black Holes}}}\ (\bibinfo  {publisher} {Kluwer
  Academic Publishers},\ \bibinfo {year} {1989})\BibitemShut {NoStop}%
\bibitem [{\citenamefont {Heckman}\ and\ \citenamefont
  {Best}(2014)}]{heckman2014coevolution}%
  \BibitemOpen
  \bibfield  {author} {\bibinfo {author} {\bibfnamefont {T.~M.}\ \bibnamefont
  {Heckman}}\ and\ \bibinfo {author} {\bibfnamefont {P.~N.}\ \bibnamefont
  {Best}},\ }\href {https://doi.org/10.1146/annurev-astro-081913-035722}
  {\bibfield  {journal} {\bibinfo  {journal} {Annual Review of Astronomy and
  Astrophysics}\ }\textbf {\bibinfo {volume} {52}},\ \bibinfo {pages} {589}
  (\bibinfo {year} {2014})}\BibitemShut {NoStop}%
\bibitem [{\citenamefont {Ruffini}\ and\ \citenamefont
  {Wheeler}(1971)}]{ruffini1971introducing}%
  \BibitemOpen
  \bibfield  {author} {\bibinfo {author} {\bibfnamefont {R.}~\bibnamefont
  {Ruffini}}\ and\ \bibinfo {author} {\bibfnamefont {J.~A.}\ \bibnamefont
  {Wheeler}},\ }\href {https://doi.org/10.1063/1.3022513} {\bibfield  {journal}
  {\bibinfo  {journal} {Physics Today}\ }\textbf {\bibinfo {volume} {24}},\
  \bibinfo {pages} {30} (\bibinfo {year} {1971})}\BibitemShut {NoStop}%
\bibitem [{\citenamefont {Penrose}(1972)}]{penrose1972black}%
  \BibitemOpen
  \bibfield  {author} {\bibinfo {author} {\bibfnamefont {R.}~\bibnamefont
  {Penrose}},\ }\href {https://doi.org/10.1038/scientificamerican0472-38}
  {\bibfield  {journal} {\bibinfo  {journal} {Scientific American}\ }\textbf
  {\bibinfo {volume} {226}},\ \bibinfo {pages} {38} (\bibinfo {year}
  {1972})}\BibitemShut {NoStop}%
\bibitem [{\citenamefont {Rees}(1974)}]{rees1974black}%
  \BibitemOpen
  \bibfield  {author} {\bibinfo {author} {\bibfnamefont {M.~J.}\ \bibnamefont
  {Rees}},\ }\href {https://doi.org/TODO} {\bibfield  {journal} {\bibinfo
  {journal} {The Observatory}\ }\textbf {\bibinfo {volume} {94}},\ \bibinfo
  {pages} {168} (\bibinfo {year} {1974})}\BibitemShut {NoStop}%
\bibitem [{\citenamefont {Novikov}(1995)}]{novikov1995black}%
  \BibitemOpen
  \bibfield  {author} {\bibinfo {author} {\bibfnamefont {I.~D.}\ \bibnamefont
  {Novikov}},\ }\href {https://doi.org/10.1017/CBO9780511564561} {\emph
  {\bibinfo {title} {Black Holes and the Universe}}}\ (\bibinfo  {publisher}
  {Cambridge University Press},\ \bibinfo {year} {1995})\BibitemShut {NoStop}%
\bibitem [{\citenamefont {Chapline}(1975)}]{chapline1975cosmological}%
  \BibitemOpen
  \bibfield  {author} {\bibinfo {author} {\bibfnamefont {G.~F.}\ \bibnamefont
  {Chapline}},\ }\href {https://doi.org/10.1038/253251a0} {\bibfield  {journal}
  {\bibinfo  {journal} {Nature}\ }\textbf {\bibinfo {volume} {253}},\ \bibinfo
  {pages} {251} (\bibinfo {year} {1975})}\BibitemShut {NoStop}%
\bibitem [{\citenamefont {Schneider}\ and\ \citenamefont
  {Ferrara}(2002)}]{schneider2002first}%
  \BibitemOpen
  \bibfield  {author} {\bibinfo {author} {\bibfnamefont {R.}~\bibnamefont
  {Schneider}}\ and\ \bibinfo {author} {\bibfnamefont {A.}~\bibnamefont
  {Ferrara}},\ }\href {https://doi.org/10.1086/339917} {\bibfield  {journal}
  {\bibinfo  {journal} {The Astrophysical Journal}\ }\textbf {\bibinfo {volume}
  {571}},\ \bibinfo {pages} {30} (\bibinfo {year} {2002})}\BibitemShut
  {NoStop}%
\bibitem [{\citenamefont {Cyburt}\ \emph {et~al.}(2003)\citenamefont {Cyburt},
  \citenamefont {Fields},\ and\ \citenamefont {Olive}}]{cyburt2003primordial}%
  \BibitemOpen
  \bibfield  {author} {\bibinfo {author} {\bibfnamefont {R.~H.}\ \bibnamefont
  {Cyburt}}, \bibinfo {author} {\bibfnamefont {B.~D.}\ \bibnamefont {Fields}},\
  and\ \bibinfo {author} {\bibfnamefont {K.~A.}\ \bibnamefont {Olive}},\ }\href
  {https://doi.org/10.1016/S0370-2693(03)00856-7} {\bibfield  {journal}
  {\bibinfo  {journal} {Physics Letters B}\ }\textbf {\bibinfo {volume}
  {567}},\ \bibinfo {pages} {227} (\bibinfo {year} {2003})}\BibitemShut
  {NoStop}%
\bibitem [{\citenamefont {Volonteri}\ and\ \citenamefont
  {Bellovary}(2012)}]{volonteri2012black}%
  \BibitemOpen
  \bibfield  {author} {\bibinfo {author} {\bibfnamefont {M.}~\bibnamefont
  {Volonteri}}\ and\ \bibinfo {author} {\bibfnamefont {J.}~\bibnamefont
  {Bellovary}},\ }\href {https://doi.org/10.1088/0034-4885/75/12/124901}
  {\bibfield  {journal} {\bibinfo  {journal} {Reports on Progress in Physics}\
  }\textbf {\bibinfo {volume} {75}},\ \bibinfo {pages} {124901} (\bibinfo
  {year} {2012})}\BibitemShut {NoStop}%
\bibitem [{\citenamefont {Carr}\ \emph {et~al.}(2016)\citenamefont {Carr},
  \citenamefont {K{\"u}hnel},\ and\ \citenamefont
  {Sandstad}}]{carr2016primordial}%
  \BibitemOpen
  \bibfield  {author} {\bibinfo {author} {\bibfnamefont {B.}~\bibnamefont
  {Carr}}, \bibinfo {author} {\bibfnamefont {F.}~\bibnamefont {K{\"u}hnel}},\
  and\ \bibinfo {author} {\bibfnamefont {M.}~\bibnamefont {Sandstad}},\ }\href
  {https://doi.org/10.1103/PhysRevD.94.083504} {\bibfield  {journal} {\bibinfo
  {journal} {Physical Review D}\ }\textbf {\bibinfo {volume} {94}},\ \bibinfo
  {pages} {083504} (\bibinfo {year} {2016})}\BibitemShut {NoStop}%
\bibitem [{\citenamefont {Lynden-Bell}(1969)}]{lynden1969galactic}%
  \BibitemOpen
  \bibfield  {author} {\bibinfo {author} {\bibfnamefont {D.}~\bibnamefont
  {Lynden-Bell}},\ }\href {https://doi.org/10.1038/223690a0} {\bibfield
  {journal} {\bibinfo  {journal} {Nature}\ }\textbf {\bibinfo {volume} {223}},\
  \bibinfo {pages} {690} (\bibinfo {year} {1969})}\BibitemShut {NoStop}%
\bibitem [{\citenamefont {Shakura}\ and\ \citenamefont
  {Sunyaev}(1973)}]{shakura1973black}%
  \BibitemOpen
  \bibfield  {author} {\bibinfo {author} {\bibfnamefont {N.~I.}\ \bibnamefont
  {Shakura}}\ and\ \bibinfo {author} {\bibfnamefont {R.~A.}\ \bibnamefont
  {Sunyaev}},\ }\href {https://doi.org/TODO} {\emph {\bibinfo {title} {Black
  holes in binary systems: Observational appearances}}}\ (\bibinfo  {publisher}
  {Cambridge University Press},\ \bibinfo {year} {1973})\BibitemShut {NoStop}%
\bibitem [{\citenamefont {Antonucci}(1993)}]{antonucci1993unified}%
  \BibitemOpen
  \bibfield  {author} {\bibinfo {author} {\bibfnamefont {R.}~\bibnamefont
  {Antonucci}},\ }\href {https://doi.org/10.1146/annurev.aa.31.090193.002353}
  {\bibfield  {journal} {\bibinfo  {journal} {Annual Review of Astronomy and
  Astrophysics}\ }\textbf {\bibinfo {volume} {31}},\ \bibinfo {pages} {473}
  (\bibinfo {year} {1993})}\BibitemShut {NoStop}%
\bibitem [{\citenamefont {Urry}\ and\ \citenamefont
  {Padovani}(1995)}]{urry1995unified}%
  \BibitemOpen
  \bibfield  {author} {\bibinfo {author} {\bibfnamefont {C.~M.}\ \bibnamefont
  {Urry}}\ and\ \bibinfo {author} {\bibfnamefont {P.}~\bibnamefont
  {Padovani}},\ }\href {https://doi.org/10.1086/133630} {\bibfield  {journal}
  {\bibinfo  {journal} {Publications of the Astronomical Society of the
  Pacific}\ }\textbf {\bibinfo {volume} {107}},\ \bibinfo {pages} {803}
  (\bibinfo {year} {1995})}\BibitemShut {NoStop}%
\bibitem [{\citenamefont
  {Zwicky}(1933{\natexlab{a}})}]{zwicky1933rotverschiebung}%
  \BibitemOpen
  \bibfield  {author} {\bibinfo {author} {\bibfnamefont {F.}~\bibnamefont
  {Zwicky}},\ }\href {https://doi.org/TODO} {\bibfield  {journal} {\bibinfo
  {journal} {Helvetica Physica Acta}\ }\textbf {\bibinfo {volume} {6}},\
  \bibinfo {pages} {110} (\bibinfo {year} {1933}{\natexlab{a}})}\BibitemShut
  {NoStop}%
\bibitem [{\citenamefont {Rubin}\ \emph
  {et~al.}(1980{\natexlab{a}})\citenamefont {Rubin}, \citenamefont {Ford},\
  and\ \citenamefont {Thonnard}}]{rubin1980rotational}%
  \BibitemOpen
  \bibfield  {author} {\bibinfo {author} {\bibfnamefont {V.~C.}\ \bibnamefont
  {Rubin}}, \bibinfo {author} {\bibfnamefont {J.}~\bibnamefont {Ford},
  \bibfnamefont {W.~K.}},\ and\ \bibinfo {author} {\bibfnamefont
  {N.}~\bibnamefont {Thonnard}},\ }\href {https://doi.org/10.1086/158003}
  {\bibfield  {journal} {\bibinfo  {journal} {The Astrophysical Journal}\
  }\textbf {\bibinfo {volume} {238}},\ \bibinfo {pages} {471} (\bibinfo {year}
  {1980}{\natexlab{a}})}\BibitemShut {NoStop}%
\bibitem [{\citenamefont {Blumenthal}\ \emph {et~al.}(1984)\citenamefont
  {Blumenthal}, \citenamefont {Faber}, \citenamefont {Primack},\ and\
  \citenamefont {Rees}}]{blumenthal1984formation}%
  \BibitemOpen
  \bibfield  {author} {\bibinfo {author} {\bibfnamefont {G.~R.}\ \bibnamefont
  {Blumenthal}}, \bibinfo {author} {\bibfnamefont {S.~M.}\ \bibnamefont
  {Faber}}, \bibinfo {author} {\bibfnamefont {J.~R.}\ \bibnamefont {Primack}},\
  and\ \bibinfo {author} {\bibfnamefont {M.~J.}\ \bibnamefont {Rees}},\ }\href
  {https://doi.org/10.1038/311517a0} {\bibfield  {journal} {\bibinfo  {journal}
  {Nature}\ }\textbf {\bibinfo {volume} {311}},\ \bibinfo {pages} {517}
  (\bibinfo {year} {1984})}\BibitemShut {NoStop}%
\bibitem [{\citenamefont {Trimble}(1987)}]{trimble1987existence}%
  \BibitemOpen
  \bibfield  {author} {\bibinfo {author} {\bibfnamefont {V.}~\bibnamefont
  {Trimble}},\ }\href {https://doi.org/10.1146/annurev.aa.25.090187.002233}
  {\bibfield  {journal} {\bibinfo  {journal} {Annual Review of Astronomy and
  Astrophysics}\ }\textbf {\bibinfo {volume} {25}},\ \bibinfo {pages} {425}
  (\bibinfo {year} {1987})}\BibitemShut {NoStop}%
\bibitem [{\citenamefont {Rubin}\ \emph
  {et~al.}(1980{\natexlab{b}})\citenamefont {Rubin}, \citenamefont {Ford},\
  and\ \citenamefont {Thonnard}}]{DM1980}%
  \BibitemOpen
  \bibfield  {author} {\bibinfo {author} {\bibfnamefont {V.~C.}\ \bibnamefont
  {Rubin}}, \bibinfo {author} {\bibfnamefont {J.}~\bibnamefont {Ford},
  \bibfnamefont {W.~K.}},\ and\ \bibinfo {author} {\bibfnamefont
  {N.}~\bibnamefont {Thonnard}},\ }\href {https://doi.org/10.1086/158003}
  {\bibfield  {journal} {\bibinfo  {journal} {The Astrophysical Journal}\
  }\textbf {\bibinfo {volume} {238}},\ \bibinfo {pages} {471} (\bibinfo {year}
  {1980}{\natexlab{b}})}\BibitemShut {NoStop}%
\bibitem [{\citenamefont {Persic}\ \emph
  {et~al.}(1996{\natexlab{a}})\citenamefont {Persic}, \citenamefont {Salucci},\
  and\ \citenamefont {Stel}}]{DM1996}%
  \BibitemOpen
  \bibfield  {author} {\bibinfo {author} {\bibfnamefont {M.}~\bibnamefont
  {Persic}}, \bibinfo {author} {\bibfnamefont {P.}~\bibnamefont {Salucci}},\
  and\ \bibinfo {author} {\bibfnamefont {F.}~\bibnamefont {Stel}},\ }\href
  {https://doi.org/10.1093/mnras/281.1.27} {\bibfield  {journal} {\bibinfo
  {journal} {Monthly Notices of the Royal Astronomical Society}\ }\textbf
  {\bibinfo {volume} {281}},\ \bibinfo {pages} {27} (\bibinfo {year}
  {1996}{\natexlab{a}})}\BibitemShut {NoStop}%
\bibitem [{\citenamefont {Tyson}\ \emph {et~al.}(1998)\citenamefont {Tyson},
  \citenamefont {Kochanski},\ and\ \citenamefont {Dell’Antonio}}]{DM1998}%
  \BibitemOpen
  \bibfield  {author} {\bibinfo {author} {\bibfnamefont {J.~A.}\ \bibnamefont
  {Tyson}}, \bibinfo {author} {\bibfnamefont {G.~P.}\ \bibnamefont
  {Kochanski}},\ and\ \bibinfo {author} {\bibfnamefont {I.~P.}\ \bibnamefont
  {Dell’Antonio}},\ }\href {https://doi.org/10.1086/311310} {\bibfield
  {journal} {\bibinfo  {journal} {The Astrophysical Journal Letters}\ }\textbf
  {\bibinfo {volume} {498}},\ \bibinfo {pages} {L107} (\bibinfo {year}
  {1998})}\BibitemShut {NoStop}%
\bibitem [{\citenamefont {Zwicky}(1933{\natexlab{b}})}]{DM9}%
  \BibitemOpen
  \bibfield  {author} {\bibinfo {author} {\bibfnamefont {F.}~\bibnamefont
  {Zwicky}},\ }\href {https://doi.org/TODO} {\bibfield  {journal} {\bibinfo
  {journal} {Helvetica Physica Acta}\ }\textbf {\bibinfo {volume} {6}},\
  \bibinfo {pages} {110} (\bibinfo {year} {1933}{\natexlab{b}})}\BibitemShut
  {NoStop}%
\bibitem [{\citenamefont {Rubin}\ and\ \citenamefont {Ford}(1970)}]{DM10}%
  \BibitemOpen
  \bibfield  {author} {\bibinfo {author} {\bibfnamefont {V.~C.}\ \bibnamefont
  {Rubin}}\ and\ \bibinfo {author} {\bibfnamefont {J.}~\bibnamefont {Ford},
  \bibfnamefont {W.~K.}},\ }\href {https://doi.org/10.1086/150317} {\bibfield
  {journal} {\bibinfo  {journal} {The Astrophysical Journal}\ }\textbf
  {\bibinfo {volume} {159}},\ \bibinfo {pages} {379} (\bibinfo {year}
  {1970})}\BibitemShut {NoStop}%
\bibitem [{\citenamefont {Persic}\ \emph
  {et~al.}(1996{\natexlab{b}})\citenamefont {Persic}, \citenamefont {Salucci},\
  and\ \citenamefont {Stel}}]{DM11}%
  \BibitemOpen
  \bibfield  {author} {\bibinfo {author} {\bibfnamefont {M.}~\bibnamefont
  {Persic}}, \bibinfo {author} {\bibfnamefont {P.}~\bibnamefont {Salucci}},\
  and\ \bibinfo {author} {\bibfnamefont {F.}~\bibnamefont {Stel}},\ }\href
  {https://doi.org/10.1093/mnras/281.1.27} {\bibfield  {journal} {\bibinfo
  {journal} {Monthly Notices of the Royal Astronomical Society}\ }\textbf
  {\bibinfo {volume} {281}},\ \bibinfo {pages} {27} (\bibinfo {year}
  {1996}{\natexlab{b}})}\BibitemShut {NoStop}%
\bibitem [{\citenamefont {Bertone}\ and\ \citenamefont
  {Hooper}(2018{\natexlab{a}})}]{DM12}%
  \BibitemOpen
  \bibfield  {author} {\bibinfo {author} {\bibfnamefont {G.}~\bibnamefont
  {Bertone}}\ and\ \bibinfo {author} {\bibfnamefont {D.}~\bibnamefont
  {Hooper}},\ }\href {https://doi.org/10.1103/RevModPhys.90.045002} {\bibfield
  {journal} {\bibinfo  {journal} {Reviews of Modern Physics}\ }\textbf
  {\bibinfo {volume} {90}},\ \bibinfo {pages} {045002} (\bibinfo {year}
  {2018}{\natexlab{a}})}\BibitemShut {NoStop}%
\bibitem [{\citenamefont {Marsh}\ \emph {et~al.}(2024)\citenamefont {Marsh},
  \citenamefont {Ellis},\ and\ \citenamefont {Mehta}}]{DM15}%
  \BibitemOpen
  \bibfield  {author} {\bibinfo {author} {\bibfnamefont {D.~J.~E.}\
  \bibnamefont {Marsh}}, \bibinfo {author} {\bibfnamefont {D.}~\bibnamefont
  {Ellis}},\ and\ \bibinfo {author} {\bibfnamefont {V.~M.}\ \bibnamefont
  {Mehta}},\ }\href {https://doi.org/TODO} {\emph {\bibinfo {title} {Dark
  Matter: Evidence, Theory, and Constraints}}},\ Princeton Series in
  Astrophysics\ (\bibinfo  {publisher} {Princeton University Press},\ \bibinfo
  {year} {2024})\BibitemShut {NoStop}%
\bibitem [{\citenamefont {Tsai}\ \emph {et~al.}(2023)\citenamefont {Tsai},
  \citenamefont {Eby},\ and\ \citenamefont {Safronova}}]{DM16}%
  \BibitemOpen
  \bibfield  {author} {\bibinfo {author} {\bibfnamefont {Y.-D.}\ \bibnamefont
  {Tsai}}, \bibinfo {author} {\bibfnamefont {J.}~\bibnamefont {Eby}},\ and\
  \bibinfo {author} {\bibfnamefont {M.~S.}\ \bibnamefont {Safronova}},\ }\href
  {https://doi.org/10.1038/s41550-022-01875-7} {\bibfield  {journal} {\bibinfo
  {journal} {Nature Astronomy}\ }\textbf {\bibinfo {volume} {7}},\ \bibinfo
  {pages} {113} (\bibinfo {year} {2023})}\BibitemShut {NoStop}%
\bibitem [{\citenamefont {Souza}\ \emph {et~al.}(2025)\citenamefont {Souza},
  \citenamefont {Muniz}, \citenamefont {Neves},\ and\ \citenamefont
  {Cruz}}]{DM17}%
  \BibitemOpen
  \bibfield  {author} {\bibinfo {author} {\bibfnamefont {A.~D.~S.}\
  \bibnamefont {Souza}}, \bibinfo {author} {\bibfnamefont {C.~R.}\ \bibnamefont
  {Muniz}}, \bibinfo {author} {\bibfnamefont {R.~M.~P.}\ \bibnamefont
  {Neves}},\ and\ \bibinfo {author} {\bibfnamefont {M.~B.}\ \bibnamefont
  {Cruz}},\ }\href {https://doi.org/10.1016/j.aop.2024.169859} {\bibfield
  {journal} {\bibinfo  {journal} {Annals of Physics}\ }\textbf {\bibinfo
  {volume} {472}},\ \bibinfo {pages} {169859} (\bibinfo {year}
  {2025})}\BibitemShut {NoStop}%
\bibitem [{\citenamefont {Navarro}\ \emph {et~al.}(1996)\citenamefont
  {Navarro}, \citenamefont {Frenk},\ and\ \citenamefont {White}}]{DM26}%
  \BibitemOpen
  \bibfield  {author} {\bibinfo {author} {\bibfnamefont {J.~F.}\ \bibnamefont
  {Navarro}}, \bibinfo {author} {\bibfnamefont {C.~S.}\ \bibnamefont {Frenk}},\
  and\ \bibinfo {author} {\bibfnamefont {S.~D.~M.}\ \bibnamefont {White}},\
  }\href {https://doi.org/10.1086/177173} {\bibfield  {journal} {\bibinfo
  {journal} {The Astrophysical Journal}\ }\textbf {\bibinfo {volume} {462}},\
  \bibinfo {pages} {563} (\bibinfo {year} {1996})}\BibitemShut {NoStop}%
\bibitem [{\citenamefont {Begeman}\ \emph {et~al.}(1991)\citenamefont
  {Begeman}, \citenamefont {Broeils},\ and\ \citenamefont {Sanders}}]{DM27}%
  \BibitemOpen
  \bibfield  {author} {\bibinfo {author} {\bibfnamefont {K.~G.}\ \bibnamefont
  {Begeman}}, \bibinfo {author} {\bibfnamefont {A.~H.}\ \bibnamefont
  {Broeils}},\ and\ \bibinfo {author} {\bibfnamefont {R.~H.}\ \bibnamefont
  {Sanders}},\ }\href {https://doi.org/10.1093/mnras/249.3.523} {\bibfield
  {journal} {\bibinfo  {journal} {Monthly Notices of the Royal Astronomical
  Society}\ }\textbf {\bibinfo {volume} {249}},\ \bibinfo {pages} {523}
  (\bibinfo {year} {1991})}\BibitemShut {NoStop}%
\bibitem [{\citenamefont {Li}\ and\ \citenamefont {Yang}(2012)}]{LiYang2012}%
  \BibitemOpen
  \bibfield  {author} {\bibinfo {author} {\bibfnamefont {M.-H.}\ \bibnamefont
  {Li}}\ and\ \bibinfo {author} {\bibfnamefont {K.-C.}\ \bibnamefont {Yang}},\
  }\href {https://doi.org/10.1103/PhysRevD.86.123015} {\bibfield  {journal}
  {\bibinfo  {journal} {Physical Review D}\ }\textbf {\bibinfo {volume} {86}},\
  \bibinfo {pages} {123015} (\bibinfo {year} {2012})}\BibitemShut {NoStop}%
\bibitem [{\citenamefont {Hendi}\ \emph {et~al.}(2020)\citenamefont {Hendi},
  \citenamefont {Nemati}, \citenamefont {Lin},\ and\ \citenamefont
  {Jamil}}]{Hendi2020}%
  \BibitemOpen
  \bibfield  {author} {\bibinfo {author} {\bibfnamefont {S.~H.}\ \bibnamefont
  {Hendi}}, \bibinfo {author} {\bibfnamefont {A.}~\bibnamefont {Nemati}},
  \bibinfo {author} {\bibfnamefont {K.}~\bibnamefont {Lin}},\ and\ \bibinfo
  {author} {\bibfnamefont {M.}~\bibnamefont {Jamil}},\ }\href
  {https://doi.org/10.1140/epjc/s10052-020-7834-2} {\bibfield  {journal}
  {\bibinfo  {journal} {European Physical Journal C}\ }\textbf {\bibinfo
  {volume} {80}},\ \bibinfo {pages} {296} (\bibinfo {year} {2020})}\BibitemShut
  {NoStop}%
\bibitem [{\citenamefont {Rizwan}\ \emph {et~al.}(2019)\citenamefont {Rizwan},
  \citenamefont {Jamil},\ and\ \citenamefont {Jusufi}}]{Rizwan2019}%
  \BibitemOpen
  \bibfield  {author} {\bibinfo {author} {\bibfnamefont {M.}~\bibnamefont
  {Rizwan}}, \bibinfo {author} {\bibfnamefont {M.}~\bibnamefont {Jamil}},\ and\
  \bibinfo {author} {\bibfnamefont {K.}~\bibnamefont {Jusufi}},\ }\href
  {https://doi.org/10.1103/PhysRevD.99.024050} {\bibfield  {journal} {\bibinfo
  {journal} {Physical Review D}\ }\textbf {\bibinfo {volume} {99}},\ \bibinfo
  {pages} {024050} (\bibinfo {year} {2019})}\BibitemShut {NoStop}%
\bibitem [{\citenamefont {Narzilloev}\ \emph {et~al.}(2020)\citenamefont
  {Narzilloev}, \citenamefont {Rayimbaev}, \citenamefont {Shaymatov},
  \citenamefont {Abdujabbarov}, \citenamefont {Ahmedov},\ and\ \citenamefont
  {Bambi}}]{Narzilloev2020}%
  \BibitemOpen
  \bibfield  {author} {\bibinfo {author} {\bibfnamefont {B.}~\bibnamefont
  {Narzilloev}}, \bibinfo {author} {\bibfnamefont {J.}~\bibnamefont
  {Rayimbaev}}, \bibinfo {author} {\bibfnamefont {S.}~\bibnamefont
  {Shaymatov}}, \bibinfo {author} {\bibfnamefont {A.}~\bibnamefont
  {Abdujabbarov}}, \bibinfo {author} {\bibfnamefont {B.}~\bibnamefont
  {Ahmedov}},\ and\ \bibinfo {author} {\bibfnamefont {C.}~\bibnamefont
  {Bambi}},\ }\href {https://doi.org/10.1103/PhysRevD.102.104062} {\bibfield
  {journal} {\bibinfo  {journal} {Physical Review D}\ }\textbf {\bibinfo
  {volume} {102}},\ \bibinfo {pages} {104062} (\bibinfo {year}
  {2020})}\BibitemShut {NoStop}%
\bibitem [{\citenamefont {Shaymatov}\ \emph
  {et~al.}(2021{\natexlab{a}})\citenamefont {Shaymatov}, \citenamefont
  {Ahmedov},\ and\ \citenamefont {Jamil}}]{Shaymatov2021a}%
  \BibitemOpen
  \bibfield  {author} {\bibinfo {author} {\bibfnamefont {S.}~\bibnamefont
  {Shaymatov}}, \bibinfo {author} {\bibfnamefont {B.}~\bibnamefont {Ahmedov}},\
  and\ \bibinfo {author} {\bibfnamefont {M.}~\bibnamefont {Jamil}},\ }\href
  {https://doi.org/10.1140/epjc/s10052-021-09366-3} {\bibfield  {journal}
  {\bibinfo  {journal} {European Physical Journal C}\ }\textbf {\bibinfo
  {volume} {81}},\ \bibinfo {pages} {588} (\bibinfo {year}
  {2021}{\natexlab{a}})}\BibitemShut {NoStop}%
\bibitem [{\citenamefont {Rayimbaev}\ \emph {et~al.}(2021)\citenamefont
  {Rayimbaev}, \citenamefont {Shaymatov},\ and\ \citenamefont
  {Jamil}}]{Rayimbaev2021}%
  \BibitemOpen
  \bibfield  {author} {\bibinfo {author} {\bibfnamefont {J.}~\bibnamefont
  {Rayimbaev}}, \bibinfo {author} {\bibfnamefont {S.}~\bibnamefont
  {Shaymatov}},\ and\ \bibinfo {author} {\bibfnamefont {M.}~\bibnamefont
  {Jamil}},\ }\href {https://doi.org/10.1140/epjc/s10052-021-09465-1}
  {\bibfield  {journal} {\bibinfo  {journal} {European Physical Journal C}\
  }\textbf {\bibinfo {volume} {81}},\ \bibinfo {pages} {699} (\bibinfo {year}
  {2021})}\BibitemShut {NoStop}%
\bibitem [{\citenamefont {Shaymatov}\ \emph
  {et~al.}(2021{\natexlab{b}})\citenamefont {Shaymatov}, \citenamefont
  {Malafarina},\ and\ \citenamefont {Ahmedov}}]{ShaymatovMalafarina2021}%
  \BibitemOpen
  \bibfield  {author} {\bibinfo {author} {\bibfnamefont {S.}~\bibnamefont
  {Shaymatov}}, \bibinfo {author} {\bibfnamefont {D.}~\bibnamefont
  {Malafarina}},\ and\ \bibinfo {author} {\bibfnamefont {B.}~\bibnamefont
  {Ahmedov}},\ }\href {https://doi.org/10.1016/j.dark.2021.100891} {\bibfield
  {journal} {\bibinfo  {journal} {Physics of the Dark Universe}\ }\textbf
  {\bibinfo {volume} {34}},\ \bibinfo {pages} {100891} (\bibinfo {year}
  {2021}{\natexlab{b}})}\BibitemShut {NoStop}%
\bibitem [{\citenamefont {Shaymatov}\ \emph {et~al.}(2022)\citenamefont
  {Shaymatov}, \citenamefont {Sheoran},\ and\ \citenamefont
  {Siwach}}]{ShaymatovSheoran2022}%
  \BibitemOpen
  \bibfield  {author} {\bibinfo {author} {\bibfnamefont {S.}~\bibnamefont
  {Shaymatov}}, \bibinfo {author} {\bibfnamefont {P.}~\bibnamefont {Sheoran}},\
  and\ \bibinfo {author} {\bibfnamefont {S.}~\bibnamefont {Siwach}},\ }\href
  {https://doi.org/10.1103/PhysRevD.105.104059} {\bibfield  {journal} {\bibinfo
   {journal} {Physical Review D}\ }\textbf {\bibinfo {volume} {105}},\ \bibinfo
  {pages} {104059} (\bibinfo {year} {2022})}\BibitemShut {NoStop}%
\bibitem [{\citenamefont {Cardoso}\ \emph {et~al.}(2022)\citenamefont
  {Cardoso}, \citenamefont {Destounis}, \citenamefont {Duque}, \citenamefont
  {Macedo},\ and\ \citenamefont {Maselli}}]{Cardoso2022}%
  \BibitemOpen
  \bibfield  {author} {\bibinfo {author} {\bibfnamefont {V.}~\bibnamefont
  {Cardoso}}, \bibinfo {author} {\bibfnamefont {K.}~\bibnamefont {Destounis}},
  \bibinfo {author} {\bibfnamefont {F.}~\bibnamefont {Duque}}, \bibinfo
  {author} {\bibfnamefont {R.~P.}\ \bibnamefont {Macedo}},\ and\ \bibinfo
  {author} {\bibfnamefont {A.}~\bibnamefont {Maselli}},\ }\href
  {https://doi.org/10.1103/PhysRevD.105.L061501} {\bibfield  {journal}
  {\bibinfo  {journal} {Physical Review D}\ }\textbf {\bibinfo {volume}
  {105}},\ \bibinfo {pages} {L061501} (\bibinfo {year} {2022})}\BibitemShut
  {NoStop}%
\bibitem [{\citenamefont {Shen}\ \emph
  {et~al.}(2024{\natexlab{a}})\citenamefont {Shen}, \citenamefont {Wang},
  \citenamefont {Gong},\ and\ \citenamefont {Yin}}]{Shen2024}%
  \BibitemOpen
  \bibfield  {author} {\bibinfo {author} {\bibfnamefont {Z.}~\bibnamefont
  {Shen}}, \bibinfo {author} {\bibfnamefont {A.}~\bibnamefont {Wang}}, \bibinfo
  {author} {\bibfnamefont {Y.}~\bibnamefont {Gong}},\ and\ \bibinfo {author}
  {\bibfnamefont {S.}~\bibnamefont {Yin}},\ }\href
  {https://doi.org/10.1016/j.physletb.2024.138797} {\bibfield  {journal}
  {\bibinfo  {journal} {Physics Letters B}\ }\textbf {\bibinfo {volume}
  {855}},\ \bibinfo {pages} {138797} (\bibinfo {year}
  {2024}{\natexlab{a}})}\BibitemShut {NoStop}%
\bibitem [{\citenamefont {Hou}\ \emph {et~al.}(2018)\citenamefont {Hou},
  \citenamefont {Xu}, \citenamefont {Zhou},\ and\ \citenamefont
  {Wang}}]{Hou2018}%
  \BibitemOpen
  \bibfield  {author} {\bibinfo {author} {\bibfnamefont {X.}~\bibnamefont
  {Hou}}, \bibinfo {author} {\bibfnamefont {Z.}~\bibnamefont {Xu}}, \bibinfo
  {author} {\bibfnamefont {M.}~\bibnamefont {Zhou}},\ and\ \bibinfo {author}
  {\bibfnamefont {J.}~\bibnamefont {Wang}},\ }\href
  {https://doi.org/10.1088/1475-7516/2018/12/015} {\bibfield  {journal}
  {\bibinfo  {journal} {JCAP}\ }\textbf {\bibinfo {volume} {2018}}\bibinfo
  {number} { (12)},\ \bibinfo {pages} {015}}\BibitemShut {NoStop}%
\bibitem [{\citenamefont {Dehnen}(1993{\natexlab{a}})}]{Dehnen1993}%
  \BibitemOpen
\bibfield  {number} {  }\bibfield  {author} {\bibinfo {author} {\bibfnamefont
  {W.}~\bibnamefont {Dehnen}},\ }\href
  {https://doi.org/10.1093/mnras/265.1.250} {\bibfield  {journal} {\bibinfo
  {journal} {Monthly Notices of the Royal Astronomical Society}\ }\textbf
  {\bibinfo {volume} {265}},\ \bibinfo {pages} {250} (\bibinfo {year}
  {1993}{\natexlab{a}})}\BibitemShut {NoStop}%
\bibitem [{\citenamefont {Gohain}\ \emph
  {et~al.}(2024{\natexlab{a}})\citenamefont {Gohain}, \citenamefont {Phukon},\
  and\ \citenamefont {Bhuyan}}]{Gohain2024}%
  \BibitemOpen
  \bibfield  {author} {\bibinfo {author} {\bibfnamefont {M.~M.}\ \bibnamefont
  {Gohain}}, \bibinfo {author} {\bibfnamefont {P.}~\bibnamefont {Phukon}},\
  and\ \bibinfo {author} {\bibfnamefont {K.}~\bibnamefont {Bhuyan}},\ }\href
  {https://doi.org/TODO} {\bibinfo {title} {Title unavailable}} (\bibinfo
  {year} {2024}{\natexlab{a}}),\ \Eprint {https://arxiv.org/abs/2407.02872}
  {arXiv:2407.02872 [gr-qc]} \BibitemShut {NoStop}%
\bibitem [{\citenamefont {Al-Badawi}\ and\ \citenamefont
  {Shaymatov}(2025)}]{B1}%
  \BibitemOpen
  \bibfield  {author} {\bibinfo {author} {\bibfnamefont {A.}~\bibnamefont
  {Al-Badawi}}\ and\ \bibinfo {author} {\bibfnamefont {S.}~\bibnamefont
  {Shaymatov}},\ }\href {https://doi.org/10.1088/1674-1137/ad7b2a} {\bibfield
  {journal} {\bibinfo  {journal} {Chinese Physics C}\ }\textbf {\bibinfo
  {volume} {49}},\ \bibinfo {pages} {055101} (\bibinfo {year}
  {2025})}\BibitemShut {NoStop}%
\bibitem [{\citenamefont {Xamidov}\ \emph {et~al.}(2025)\citenamefont
  {Xamidov}, \citenamefont {Uktamov}, \citenamefont {Shaymatov},\ and\
  \citenamefont {Ahmedov}}]{B2}%
  \BibitemOpen
  \bibfield  {author} {\bibinfo {author} {\bibfnamefont {T.}~\bibnamefont
  {Xamidov}}, \bibinfo {author} {\bibfnamefont {U.}~\bibnamefont {Uktamov}},
  \bibinfo {author} {\bibfnamefont {S.}~\bibnamefont {Shaymatov}},\ and\
  \bibinfo {author} {\bibfnamefont {B.}~\bibnamefont {Ahmedov}},\ }\href
  {https://doi.org/10.1016/j.dark.2024.101805} {\bibfield  {journal} {\bibinfo
  {journal} {Physics of the Dark Universe}\ }\textbf {\bibinfo {volume} {47}},\
  \bibinfo {pages} {101805} (\bibinfo {year} {2025})}\BibitemShut {NoStop}%
\bibitem [{\citenamefont {Ashraf}\ \emph {et~al.}(2025)\citenamefont {Ashraf}
  \emph {et~al.}}]{B6}%
  \BibitemOpen
  \bibfield  {author} {\bibinfo {author} {\bibfnamefont {A.}~\bibnamefont
  {Ashraf}} \emph {et~al.},\ }\href
  {https://doi.org/10.1016/j.dark.2024.101823} {\bibfield  {journal} {\bibinfo
  {journal} {Physics of the Dark Universe}\ }\textbf {\bibinfo {volume} {47}},\
  \bibinfo {pages} {101823} (\bibinfo {year} {2025})}\BibitemShut {NoStop}%
\bibitem [{\citenamefont {Hawley}\ and\ \citenamefont
  {Holcomb}(2005)}]{hawley2005foundations}%
  \BibitemOpen
  \bibfield  {author} {\bibinfo {author} {\bibfnamefont {J.~F.}\ \bibnamefont
  {Hawley}}\ and\ \bibinfo {author} {\bibfnamefont {K.~A.}\ \bibnamefont
  {Holcomb}},\ }\href {https://doi.org/10.1093/oso/9780198530961.001.0001}
  {\emph {\bibinfo {title} {Foundations of Modern Cosmology}}}\ (\bibinfo
  {publisher} {Oxford University Press},\ \bibinfo {year} {2005})\BibitemShut
  {NoStop}%
\bibitem [{\citenamefont {Turner}(1991)}]{turner1991dark}%
  \BibitemOpen
  \bibfield  {author} {\bibinfo {author} {\bibfnamefont {M.~S.}\ \bibnamefont
  {Turner}},\ }\href {https://doi.org/10.1088/0031-8949/1991/T36/020}
  {\bibfield  {journal} {\bibinfo  {journal} {Physica Scripta}\ }\textbf
  {\bibinfo {volume} {T36}},\ \bibinfo {pages} {167} (\bibinfo {year}
  {1991})}\BibitemShut {NoStop}%
\bibitem [{\citenamefont {Overduin}\ and\ \citenamefont
  {Wesson}(2004)}]{overduin2004dark}%
  \BibitemOpen
  \bibfield  {author} {\bibinfo {author} {\bibfnamefont {J.~M.}\ \bibnamefont
  {Overduin}}\ and\ \bibinfo {author} {\bibfnamefont {P.~S.}\ \bibnamefont
  {Wesson}},\ }\href {https://doi.org/10.1016/j.physrep.2004.07.006} {\bibfield
   {journal} {\bibinfo  {journal} {Physics Reports}\ }\textbf {\bibinfo
  {volume} {402}},\ \bibinfo {pages} {267} (\bibinfo {year}
  {2004})}\BibitemShut {NoStop}%
\bibitem [{\citenamefont {Spergel}(2015)}]{spergel2015dark}%
  \BibitemOpen
  \bibfield  {author} {\bibinfo {author} {\bibfnamefont {D.~N.}\ \bibnamefont
  {Spergel}},\ }\href {https://doi.org/10.1126/science.aaa0980} {\bibfield
  {journal} {\bibinfo  {journal} {Science}\ }\textbf {\bibinfo {volume}
  {347}},\ \bibinfo {pages} {1100} (\bibinfo {year} {2015})}\BibitemShut
  {NoStop}%
\bibitem [{\citenamefont {Bertone}\ and\ \citenamefont
  {Hooper}(2018{\natexlab{b}})}]{bertone2018history}%
  \BibitemOpen
  \bibfield  {author} {\bibinfo {author} {\bibfnamefont {G.}~\bibnamefont
  {Bertone}}\ and\ \bibinfo {author} {\bibfnamefont {D.}~\bibnamefont
  {Hooper}},\ }\href {https://doi.org/10.1103/RevModPhys.90.045002} {\bibfield
  {journal} {\bibinfo  {journal} {Reviews of Modern Physics}\ }\textbf
  {\bibinfo {volume} {90}},\ \bibinfo {pages} {045002} (\bibinfo {year}
  {2018}{\natexlab{b}})}\BibitemShut {NoStop}%
\bibitem [{\citenamefont {Wang}\ \emph {et~al.}(2016)\citenamefont {Wang},
  \citenamefont {Abdalla}, \citenamefont {Atrio-Barandela},\ and\ \citenamefont
  {Pavon}}]{wang2016dark}%
  \BibitemOpen
  \bibfield  {author} {\bibinfo {author} {\bibfnamefont {B.}~\bibnamefont
  {Wang}}, \bibinfo {author} {\bibfnamefont {E.}~\bibnamefont {Abdalla}},
  \bibinfo {author} {\bibfnamefont {F.}~\bibnamefont {Atrio-Barandela}},\ and\
  \bibinfo {author} {\bibfnamefont {D.}~\bibnamefont {Pavon}},\ }\href
  {https://doi.org/10.1088/0034-4885/79/9/096901} {\bibfield  {journal}
  {\bibinfo  {journal} {Reports on Progress in Physics}\ }\textbf {\bibinfo
  {volume} {79}},\ \bibinfo {pages} {096901} (\bibinfo {year}
  {2016})}\BibitemShut {NoStop}%
\bibitem [{\citenamefont {Oks}(2021)}]{oks2021brief}%
  \BibitemOpen
  \bibfield  {author} {\bibinfo {author} {\bibfnamefont {E.}~\bibnamefont
  {Oks}},\ }\href {https://doi.org/10.1016/j.newar.2021.101632} {\bibfield
  {journal} {\bibinfo  {journal} {New Astronomy Reviews}\ }\textbf {\bibinfo
  {volume} {93}},\ \bibinfo {pages} {101632} (\bibinfo {year}
  {2021})}\BibitemShut {NoStop}%
\bibitem [{\citenamefont {Shen}\ \emph
  {et~al.}(2024{\natexlab{b}})\citenamefont {Shen}, \citenamefont {Wang},
  \citenamefont {Gong},\ and\ \citenamefont {Yin}}]{shen2024analytical}%
  \BibitemOpen
  \bibfield  {author} {\bibinfo {author} {\bibfnamefont {Z.}~\bibnamefont
  {Shen}}, \bibinfo {author} {\bibfnamefont {A.}~\bibnamefont {Wang}}, \bibinfo
  {author} {\bibfnamefont {Y.}~\bibnamefont {Gong}},\ and\ \bibinfo {author}
  {\bibfnamefont {S.}~\bibnamefont {Yin}},\ }\href
  {https://doi.org/10.1016/j.physletb.2024.138797} {\bibfield  {journal}
  {\bibinfo  {journal} {Physics Letters B}\ }\textbf {\bibinfo {volume}
  {855}},\ \bibinfo {pages} {138797} (\bibinfo {year}
  {2024}{\natexlab{b}})}\BibitemShut {NoStop}%
\bibitem [{\citenamefont {Navarro}(1996)}]{navarro1996structure}%
  \BibitemOpen
  \bibfield  {author} {\bibinfo {author} {\bibfnamefont {J.~F.}\ \bibnamefont
  {Navarro}},\ }\href {https://doi.org/TODO} {\emph {\bibinfo {title} {The
  Structure of Cold Dark Matter Halos}}}\ (\bibinfo  {publisher} {Cambridge
  University Press},\ \bibinfo {year} {1996})\BibitemShut {NoStop}%
\bibitem [{\citenamefont {Hernquist}(1990)}]{hernquist1990analytical}%
  \BibitemOpen
  \bibfield  {author} {\bibinfo {author} {\bibfnamefont {L.}~\bibnamefont
  {Hernquist}},\ }\href {https://doi.org/10.1086/168845} {\bibfield  {journal}
  {\bibinfo  {journal} {The Astrophysical Journal}\ }\textbf {\bibinfo {volume}
  {356}},\ \bibinfo {pages} {359} (\bibinfo {year} {1990})}\BibitemShut
  {NoStop}%
\bibitem [{\citenamefont {Burkert}(1995)}]{burkert1995structure}%
  \BibitemOpen
  \bibfield  {author} {\bibinfo {author} {\bibfnamefont {A.}~\bibnamefont
  {Burkert}},\ }\href {https://doi.org/10.1086/309560} {\bibfield  {journal}
  {\bibinfo  {journal} {The Astrophysical Journal}\ }\textbf {\bibinfo {volume}
  {447}},\ \bibinfo {pages} {L25} (\bibinfo {year} {1995})}\BibitemShut
  {NoStop}%
\bibitem [{\citenamefont {Moore}(1994)}]{moore1994evidence}%
  \BibitemOpen
  \bibfield  {author} {\bibinfo {author} {\bibfnamefont {B.}~\bibnamefont
  {Moore}},\ }\href {https://doi.org/10.1038/370629a0} {\bibfield  {journal}
  {\bibinfo  {journal} {Nature}\ }\textbf {\bibinfo {volume} {370}},\ \bibinfo
  {pages} {629} (\bibinfo {year} {1994})}\BibitemShut {NoStop}%
\bibitem [{\citenamefont {Dehnen}(1993{\natexlab{b}})}]{dehnen1993family}%
  \BibitemOpen
  \bibfield  {author} {\bibinfo {author} {\bibfnamefont {W.}~\bibnamefont
  {Dehnen}},\ }\href {https://doi.org/10.1093/mnras/265.1.250} {\bibfield
  {journal} {\bibinfo  {journal} {Monthly Notices of the Royal Astronomical
  Society}\ }\textbf {\bibinfo {volume} {265}},\ \bibinfo {pages} {250}
  (\bibinfo {year} {1993}{\natexlab{b}})}\BibitemShut {NoStop}%
\bibitem [{\citenamefont {Xu}\ \emph {et~al.}(2019)\citenamefont {Xu},
  \citenamefont {Hou}, \citenamefont {Wang},\ and\ \citenamefont
  {Liao}}]{xu2019perfect}%
  \BibitemOpen
  \bibfield  {author} {\bibinfo {author} {\bibfnamefont {Z.}~\bibnamefont
  {Xu}}, \bibinfo {author} {\bibfnamefont {X.}~\bibnamefont {Hou}}, \bibinfo
  {author} {\bibfnamefont {J.}~\bibnamefont {Wang}},\ and\ \bibinfo {author}
  {\bibfnamefont {Y.}~\bibnamefont {Liao}},\ }\href
  {https://doi.org/10.1155/2019/2434390} {\bibfield  {journal} {\bibinfo
  {journal} {Advances in High Energy Physics}\ }\textbf {\bibinfo {volume}
  {2019}},\ \bibinfo {pages} {2434390} (\bibinfo {year} {2019})}\BibitemShut
  {NoStop}%
\bibitem [{\citenamefont {Singh}\ \emph {et~al.}(2021)\citenamefont {Singh},
  \citenamefont {Ghosh},\ and\ \citenamefont
  {Bhamidipati}}]{singh2021thermodynamic}%
  \BibitemOpen
  \bibfield  {author} {\bibinfo {author} {\bibfnamefont {A.}~\bibnamefont
  {Singh}}, \bibinfo {author} {\bibfnamefont {A.}~\bibnamefont {Ghosh}},\ and\
  \bibinfo {author} {\bibfnamefont {C.}~\bibnamefont {Bhamidipati}},\ }\href
  {https://doi.org/10.3389/fphy.2021.631471} {\bibfield  {journal} {\bibinfo
  {journal} {Frontiers in Physics}\ }\textbf {\bibinfo {volume} {9}},\ \bibinfo
  {pages} {631471} (\bibinfo {year} {2021})}\BibitemShut {NoStop}%
\bibitem [{\citenamefont {Pantig}\ and\ \citenamefont
  {{\"O}vg{\"u}n}(2023)}]{pantig2023black}%
  \BibitemOpen
  \bibfield  {author} {\bibinfo {author} {\bibfnamefont {R.~C.}\ \bibnamefont
  {Pantig}}\ and\ \bibinfo {author} {\bibfnamefont {A.}~\bibnamefont
  {{\"O}vg{\"u}n}},\ }\href {https://doi.org/10.1002/prop.202200164} {\bibfield
   {journal} {\bibinfo  {journal} {Fortschritte der Physik}\ }\textbf {\bibinfo
  {volume} {71}},\ \bibinfo {pages} {2200164} (\bibinfo {year}
  {2023})}\BibitemShut {NoStop}%
\bibitem [{\citenamefont {Carvalho}\ \emph {et~al.}(2023)\citenamefont
  {Carvalho}, \citenamefont {Alencar},\ and\ \citenamefont
  {Muniz}}]{carvalho2023thermodynamics}%
  \BibitemOpen
  \bibfield  {author} {\bibinfo {author} {\bibfnamefont {{\'I}.~D.~D.}\
  \bibnamefont {Carvalho}}, \bibinfo {author} {\bibfnamefont {G.}~\bibnamefont
  {Alencar}},\ and\ \bibinfo {author} {\bibfnamefont {C.~R.}\ \bibnamefont
  {Muniz}},\ }\href {https://doi.org/10.1016/j.dark.2023.101290} {\bibfield
  {journal} {\bibinfo  {journal} {Physics of the Dark Universe}\ }\textbf
  {\bibinfo {volume} {42}},\ \bibinfo {pages} {101290} (\bibinfo {year}
  {2023})}\BibitemShut {NoStop}%
\bibitem [{\citenamefont {Gohain}\ \emph
  {et~al.}(2024{\natexlab{b}})\citenamefont {Gohain}, \citenamefont {Phukon},\
  and\ \citenamefont {Bhuyan}}]{gohain2024thermodynamics}%
  \BibitemOpen
  \bibfield  {author} {\bibinfo {author} {\bibfnamefont {M.~M.}\ \bibnamefont
  {Gohain}}, \bibinfo {author} {\bibfnamefont {P.}~\bibnamefont {Phukon}},\
  and\ \bibinfo {author} {\bibfnamefont {K.}~\bibnamefont {Bhuyan}},\ }\href
  {https://doi.org/10.1016/j.dark.2024.101683} {\bibfield  {journal} {\bibinfo
  {journal} {Physics of the Dark Universe}\ }\textbf {\bibinfo {volume} {46}},\
  \bibinfo {pages} {101683} (\bibinfo {year} {2024}{\natexlab{b}})}\BibitemShut
  {NoStop}%
\bibitem [{\citenamefont {Cardoso}\ \emph {et~al.}(2016)\citenamefont
  {Cardoso}, \citenamefont {Macedo}, \citenamefont {Pani},\ and\ \citenamefont
  {Ferrari}}]{cardoso2016black}%
  \BibitemOpen
  \bibfield  {author} {\bibinfo {author} {\bibfnamefont {V.}~\bibnamefont
  {Cardoso}}, \bibinfo {author} {\bibfnamefont {C.~F.~B.}\ \bibnamefont
  {Macedo}}, \bibinfo {author} {\bibfnamefont {P.}~\bibnamefont {Pani}},\ and\
  \bibinfo {author} {\bibfnamefont {V.}~\bibnamefont {Ferrari}},\ }\href
  {https://doi.org/10.1088/1475-7516/2016/05/054} {\bibfield  {journal}
  {\bibinfo  {journal} {Journal of Cosmology and Astroparticle Physics}\
  }\textbf {\bibinfo {volume} {2016}}\bibinfo  {number} { (05)},\ \bibinfo
  {pages} {054}}\BibitemShut {NoStop}%
\bibitem [{\citenamefont {Jusufi}(2020)}]{jusufi2020quasinormal}%
  \BibitemOpen
\bibfield  {number} {  }\bibfield  {author} {\bibinfo {author} {\bibfnamefont
  {K.}~\bibnamefont {Jusufi}},\ }\href
  {https://doi.org/10.1103/PhysRevD.101.084055} {\bibfield  {journal} {\bibinfo
   {journal} {Physical Review D}\ }\textbf {\bibinfo {volume} {101}},\ \bibinfo
  {pages} {084055} (\bibinfo {year} {2020})}\BibitemShut {NoStop}%
\bibitem [{\citenamefont {Bamber}\ \emph {et~al.}(2021)\citenamefont {Bamber},
  \citenamefont {Tattersall}, \citenamefont {Clough},\ and\ \citenamefont
  {Ferreira}}]{bamber2021quasinormal}%
  \BibitemOpen
  \bibfield  {author} {\bibinfo {author} {\bibfnamefont {J.}~\bibnamefont
  {Bamber}}, \bibinfo {author} {\bibfnamefont {O.~J.}\ \bibnamefont
  {Tattersall}}, \bibinfo {author} {\bibfnamefont {K.}~\bibnamefont {Clough}},\
  and\ \bibinfo {author} {\bibfnamefont {P.~G.}\ \bibnamefont {Ferreira}},\
  }\href {https://doi.org/10.1103/PhysRevD.103.124013} {\bibfield  {journal}
  {\bibinfo  {journal} {Physical Review D}\ }\textbf {\bibinfo {volume}
  {103}},\ \bibinfo {pages} {124013} (\bibinfo {year} {2021})}\BibitemShut
  {NoStop}%
\bibitem [{\citenamefont {Konoplya}(2021)}]{konoplya2021black}%
  \BibitemOpen
  \bibfield  {author} {\bibinfo {author} {\bibfnamefont {R.~A.}\ \bibnamefont
  {Konoplya}},\ }\href {https://doi.org/10.1016/j.physletb.2021.136734}
  {\bibfield  {journal} {\bibinfo  {journal} {Physics Letters B}\ }\textbf
  {\bibinfo {volume} {823}},\ \bibinfo {pages} {136734} (\bibinfo {year}
  {2021})}\BibitemShut {NoStop}%
\bibitem [{\citenamefont {Das}\ \emph {et~al.}(2023)\citenamefont {Das},
  \citenamefont {Chowdhury},\ and\ \citenamefont
  {Gangopadhyay}}]{das2023stability}%
  \BibitemOpen
  \bibfield  {author} {\bibinfo {author} {\bibfnamefont {A.}~\bibnamefont
  {Das}}, \bibinfo {author} {\bibfnamefont {A.~R.}\ \bibnamefont {Chowdhury}},\
  and\ \bibinfo {author} {\bibfnamefont {S.}~\bibnamefont {Gangopadhyay}},\
  }\href {https://doi.org/10.1088/1361-6382/ad07a6} {\bibfield  {journal}
  {\bibinfo  {journal} {Classical and Quantum Gravity}\ }\textbf {\bibinfo
  {volume} {41}},\ \bibinfo {pages} {015018} (\bibinfo {year}
  {2023})}\BibitemShut {NoStop}%
\bibitem [{\citenamefont {Konoplya}\ \emph {et~al.}(2025)\citenamefont
  {Konoplya}, \citenamefont {Khrabustovskyi}, \citenamefont {K{\v r}{\'\i}{\v
  z}},\ and\ \citenamefont {Zhidenko}}]{konoplya2025quasinormal}%
  \BibitemOpen
  \bibfield  {author} {\bibinfo {author} {\bibfnamefont {R.~A.}\ \bibnamefont
  {Konoplya}}, \bibinfo {author} {\bibfnamefont {A.}~\bibnamefont
  {Khrabustovskyi}}, \bibinfo {author} {\bibfnamefont {J.}~\bibnamefont {K{\v
  r}{\'\i}{\v z}}},\ and\ \bibinfo {author} {\bibfnamefont {A.}~\bibnamefont
  {Zhidenko}},\ }\href {https://doi.org/10.1088/1475-7516/2025/04/062}
  {\bibfield  {journal} {\bibinfo  {journal} {JCAP}\ }\textbf {\bibinfo
  {volume} {2025}}\bibinfo  {number} { (04)},\ \bibinfo {pages}
  {062}}\BibitemShut {NoStop}%
\bibitem [{\citenamefont {Konoplya}(2019)}]{konoplya2019shadow}%
  \BibitemOpen
\bibfield  {number} {  }\bibfield  {author} {\bibinfo {author} {\bibfnamefont
  {R.~A.}\ \bibnamefont {Konoplya}},\ }\href
  {https://doi.org/10.1016/j.physletb.2019.05.043} {\bibfield  {journal}
  {\bibinfo  {journal} {Physics Letters B}\ }\textbf {\bibinfo {volume}
  {795}},\ \bibinfo {pages} {1} (\bibinfo {year} {2019})}\BibitemShut {NoStop}%
\bibitem [{\citenamefont {Jusufi}\ \emph {et~al.}(2019)\citenamefont {Jusufi},
  \citenamefont {Jamil}, \citenamefont {Salucci}, \citenamefont {Zhu},\ and\
  \citenamefont {Haroon}}]{jusufi2019black}%
  \BibitemOpen
  \bibfield  {author} {\bibinfo {author} {\bibfnamefont {K.}~\bibnamefont
  {Jusufi}}, \bibinfo {author} {\bibfnamefont {M.}~\bibnamefont {Jamil}},
  \bibinfo {author} {\bibfnamefont {P.}~\bibnamefont {Salucci}}, \bibinfo
  {author} {\bibfnamefont {T.}~\bibnamefont {Zhu}},\ and\ \bibinfo {author}
  {\bibfnamefont {S.}~\bibnamefont {Haroon}},\ }\href
  {https://doi.org/10.1103/PhysRevD.100.044012} {\bibfield  {journal} {\bibinfo
   {journal} {Physical Review D}\ }\textbf {\bibinfo {volume} {100}},\ \bibinfo
  {pages} {044012} (\bibinfo {year} {2019})}\BibitemShut {NoStop}%
\bibitem [{\citenamefont {Saurabh}\ and\ \citenamefont
  {Jusufi}(2021)}]{saurabh2021imprints}%
  \BibitemOpen
  \bibfield  {author} {\bibinfo {author} {\bibfnamefont {K.}~\bibnamefont
  {Saurabh}}\ and\ \bibinfo {author} {\bibfnamefont {K.}~\bibnamefont
  {Jusufi}},\ }\href {https://doi.org/10.1140/epjc/s10052-021-09273-7}
  {\bibfield  {journal} {\bibinfo  {journal} {European Physical Journal C}\
  }\textbf {\bibinfo {volume} {81}},\ \bibinfo {pages} {490} (\bibinfo {year}
  {2021})}\BibitemShut {NoStop}%
\bibitem [{\citenamefont {Figueiredo}\ \emph {et~al.}(2023)\citenamefont
  {Figueiredo}, \citenamefont {Maselli},\ and\ \citenamefont
  {Cardoso}}]{figueiredo2023black}%
  \BibitemOpen
  \bibfield  {author} {\bibinfo {author} {\bibfnamefont {E.}~\bibnamefont
  {Figueiredo}}, \bibinfo {author} {\bibfnamefont {A.}~\bibnamefont
  {Maselli}},\ and\ \bibinfo {author} {\bibfnamefont {V.}~\bibnamefont
  {Cardoso}},\ }\href {https://doi.org/10.1103/PhysRevD.107.104033} {\bibfield
  {journal} {\bibinfo  {journal} {Physical Review D}\ }\textbf {\bibinfo
  {volume} {107}},\ \bibinfo {pages} {104033} (\bibinfo {year}
  {2023})}\BibitemShut {NoStop}%
\bibitem [{\citenamefont {Capozziello}\ \emph {et~al.}(2023)\citenamefont
  {Capozziello}, \citenamefont {Zare}, \citenamefont {Mota},\ and\
  \citenamefont {Hassanabadi}}]{capozziello2023dark}%
  \BibitemOpen
  \bibfield  {author} {\bibinfo {author} {\bibfnamefont {S.}~\bibnamefont
  {Capozziello}}, \bibinfo {author} {\bibfnamefont {S.}~\bibnamefont {Zare}},
  \bibinfo {author} {\bibfnamefont {D.~F.}\ \bibnamefont {Mota}},\ and\
  \bibinfo {author} {\bibfnamefont {H.}~\bibnamefont {Hassanabadi}},\ }\href
  {https://doi.org/10.1088/1475-7516/2023/10/027} {\bibfield  {journal}
  {\bibinfo  {journal} {JCAP}\ }\textbf {\bibinfo {volume} {2023}}\bibinfo
  {number} { (10)},\ \bibinfo {pages} {027}}\BibitemShut {NoStop}%
\bibitem [{\citenamefont {Chowdhury}\ \emph {et~al.}(2025)\citenamefont
  {Chowdhury}, \citenamefont {Sen}, \citenamefont {Chakrabarti},\ and\
  \citenamefont {Das}}]{chowdhury2025effect}%
  \BibitemOpen
\bibfield  {number} {  }\bibfield  {author} {\bibinfo {author} {\bibfnamefont
  {A.}~\bibnamefont {Chowdhury}}, \bibinfo {author} {\bibfnamefont
  {G.}~\bibnamefont {Sen}}, \bibinfo {author} {\bibfnamefont {S.}~\bibnamefont
  {Chakrabarti}},\ and\ \bibinfo {author} {\bibfnamefont {S.}~\bibnamefont
  {Das}},\ }\href {https://doi.org/10.1103/PhysRevD.112.064041} {\bibfield
  {journal} {\bibinfo  {journal} {Physical Review D}\ }\textbf {\bibinfo
  {volume} {112}},\ \bibinfo {pages} {064041} (\bibinfo {year}
  {2025})}\BibitemShut {NoStop}%
\bibitem [{\citenamefont {Xu}\ \emph {et~al.}(2018)\citenamefont {Xu},
  \citenamefont {Hou}, \citenamefont {Gong},\ and\ \citenamefont
  {Wang}}]{xu2018black}%
  \BibitemOpen
  \bibfield  {author} {\bibinfo {author} {\bibfnamefont {Z.}~\bibnamefont
  {Xu}}, \bibinfo {author} {\bibfnamefont {X.}~\bibnamefont {Hou}}, \bibinfo
  {author} {\bibfnamefont {X.}~\bibnamefont {Gong}},\ and\ \bibinfo {author}
  {\bibfnamefont {J.}~\bibnamefont {Wang}},\ }\href
  {https://doi.org/10.1088/1475-7516/2018/09/038} {\bibfield  {journal}
  {\bibinfo  {journal} {Journal of Cosmology and Astroparticle Physics}\
  }\textbf {\bibinfo {volume} {2018}}\bibinfo  {number} { (09)},\ \bibinfo
  {pages} {038}}\BibitemShut {NoStop}%
\bibitem [{\citenamefont {Kavanagh}\ \emph {et~al.}(2020)\citenamefont
  {Kavanagh}, \citenamefont {Nichols}, \citenamefont {Bertone},\ and\
  \citenamefont {Gaggero}}]{kavanagh2020detecting}%
  \BibitemOpen
\bibfield  {number} {  }\bibfield  {author} {\bibinfo {author} {\bibfnamefont
  {B.~J.}\ \bibnamefont {Kavanagh}}, \bibinfo {author} {\bibfnamefont {D.~A.}\
  \bibnamefont {Nichols}}, \bibinfo {author} {\bibfnamefont {G.}~\bibnamefont
  {Bertone}},\ and\ \bibinfo {author} {\bibfnamefont {D.}~\bibnamefont
  {Gaggero}},\ }\href {https://doi.org/10.1103/PhysRevD.102.083006} {\bibfield
  {journal} {\bibinfo  {journal} {Physical Review D}\ }\textbf {\bibinfo
  {volume} {102}},\ \bibinfo {pages} {083006} (\bibinfo {year}
  {2020})}\BibitemShut {NoStop}%
\bibitem [{\citenamefont {Konoplya}\ and\ \citenamefont
  {Zinhailo}(2019)}]{konoplya2019hawking}%
  \BibitemOpen
  \bibfield  {author} {\bibinfo {author} {\bibfnamefont {R.~A.}\ \bibnamefont
  {Konoplya}}\ and\ \bibinfo {author} {\bibfnamefont {A.~F.}\ \bibnamefont
  {Zinhailo}},\ }\href {https://doi.org/10.1103/PhysRevD.99.104060} {\bibfield
  {journal} {\bibinfo  {journal} {Physical Review D}\ }\textbf {\bibinfo
  {volume} {99}},\ \bibinfo {pages} {104060} (\bibinfo {year}
  {2019})}\BibitemShut {NoStop}%
\bibitem [{\citenamefont {King}(1962)}]{king1962structure}%
  \BibitemOpen
  \bibfield  {author} {\bibinfo {author} {\bibfnamefont {I.}~\bibnamefont
  {King}},\ }\href {https://doi.org/10.1086/108756} {\bibfield  {journal}
  {\bibinfo  {journal} {The Astronomical Journal}\ }\textbf {\bibinfo {volume}
  {67}},\ \bibinfo {pages} {471} (\bibinfo {year} {1962})}\BibitemShut
  {NoStop}%
\bibitem [{\citenamefont {Kar}\ and\ \citenamefont
  {Kar}(2025)}]{kar2025diverse}%
  \BibitemOpen
  \bibfield  {author} {\bibinfo {author} {\bibfnamefont {A.}~\bibnamefont
  {Kar}}\ and\ \bibinfo {author} {\bibfnamefont {S.}~\bibnamefont {Kar}},\
  }\href {https://doi.org/10.1140/epjc/s10052-025-13122-8} {\bibfield
  {journal} {\bibinfo  {journal} {European Physical Journal C}\ }\textbf
  {\bibinfo {volume} {85}},\ \bibinfo {pages} {773} (\bibinfo {year}
  {2025})}\BibitemShut {NoStop}%
\bibitem [{\citenamefont {Letelier}(1979)}]{Letelier1979}%
  \BibitemOpen
  \bibfield  {author} {\bibinfo {author} {\bibfnamefont {P.~S.}\ \bibnamefont
  {Letelier}},\ }\href {https://doi.org/10.1103/PhysRevD.20.1294} {\bibfield
  {journal} {\bibinfo  {journal} {Physical Review D}\ }\textbf {\bibinfo
  {volume} {20}},\ \bibinfo {pages} {1294} (\bibinfo {year}
  {1979})}\BibitemShut {NoStop}%
\bibitem [{\citenamefont {Konoplya}\ and\ \citenamefont
  {Zhidenko}(2022)}]{konoplya2022solutions}%
  \BibitemOpen
  \bibfield  {author} {\bibinfo {author} {\bibfnamefont {R.~A.}\ \bibnamefont
  {Konoplya}}\ and\ \bibinfo {author} {\bibfnamefont {A.}~\bibnamefont
  {Zhidenko}},\ }\href {https://doi.org/10.3847/1538-4357/ac6ff9} {\bibfield
  {journal} {\bibinfo  {journal} {The Astrophysical Journal}\ }\textbf
  {\bibinfo {volume} {933}},\ \bibinfo {pages} {166} (\bibinfo {year}
  {2022})}\BibitemShut {NoStop}%
\bibitem [{\citenamefont {Zare}\ \emph {et~al.}(2025)\citenamefont {Zare},
  \citenamefont {Hosseinifar}, \citenamefont {Nieto}, \citenamefont {Gogoi},
  \citenamefont {Boshkayev}, \citenamefont {Urazalina},\ and\ \citenamefont
  {Hassanabadi}}]{SZ2025}%
  \BibitemOpen
  \bibfield  {author} {\bibinfo {author} {\bibfnamefont {S.}~\bibnamefont
  {Zare}}, \bibinfo {author} {\bibfnamefont {F.}~\bibnamefont {Hosseinifar}},
  \bibinfo {author} {\bibfnamefont {L.~M.}\ \bibnamefont {Nieto}}, \bibinfo
  {author} {\bibfnamefont {D.~J.}\ \bibnamefont {Gogoi}}, \bibinfo {author}
  {\bibfnamefont {K.}~\bibnamefont {Boshkayev}}, \bibinfo {author}
  {\bibfnamefont {A.}~\bibnamefont {Urazalina}},\ and\ \bibinfo {author}
  {\bibfnamefont {H.}~\bibnamefont {Hassanabadi}},\ }\href
  {https://doi.org/10.48550/arXiv.2510.03925} {\bibinfo {title} {Title
  unavailable}} (\bibinfo {year} {2025}),\ \Eprint
  {https://arxiv.org/abs/2510.03925} {arXiv:2510.03925 [gr-qc]} \BibitemShut
  {NoStop}%
\bibitem [{\citenamefont {Zhao}(1996)}]{zhao1996analytical}%
  \BibitemOpen
  \bibfield  {author} {\bibinfo {author} {\bibfnamefont {H.}~\bibnamefont
  {Zhao}},\ }\href {https://doi.org/10.1093/mnras/278.2.488} {\bibfield
  {journal} {\bibinfo  {journal} {Monthly Notices of the Royal Astronomical
  Society}\ }\textbf {\bibinfo {volume} {278}},\ \bibinfo {pages} {488}
  (\bibinfo {year} {1996})}\BibitemShut {NoStop}%
\bibitem [{\citenamefont {Sofue}(2020)}]{sofue2020rotation}%
  \BibitemOpen
  \bibfield  {author} {\bibinfo {author} {\bibfnamefont {Y.}~\bibnamefont
  {Sofue}},\ }\href {https://doi.org/10.3390/galaxies8020037} {\bibfield
  {journal} {\bibinfo  {journal} {Galaxies}\ }\textbf {\bibinfo {volume} {8}},\
  \bibinfo {pages} {37} (\bibinfo {year} {2020})}\BibitemShut {NoStop}%
\bibitem [{\citenamefont {Akiyama}\ and\ \citenamefont {et~al. (Event Horizon
  Telescope~Collaboration)}(2019)}]{akiyama2019first}%
  \BibitemOpen
  \bibfield  {author} {\bibinfo {author} {\bibfnamefont {K.}~\bibnamefont
  {Akiyama}}\ and\ \bibinfo {author} {\bibnamefont {et~al. (Event Horizon
  Telescope~Collaboration)}},\ }\href
  {https://doi.org/10.3847/2041-8213/ab0ec7} {\bibfield  {journal} {\bibinfo
  {journal} {The Astrophysical Journal Letters}\ }\textbf {\bibinfo {volume}
  {875}},\ \bibinfo {pages} {L1} (\bibinfo {year} {2019})}\BibitemShut
  {NoStop}%
\bibitem [{\citenamefont {Chael}\ \emph {et~al.}(2021)\citenamefont {Chael},
  \citenamefont {Johnson},\ and\ \citenamefont
  {Lupsasca}}]{chael2021observing}%
  \BibitemOpen
  \bibfield  {author} {\bibinfo {author} {\bibfnamefont {A.}~\bibnamefont
  {Chael}}, \bibinfo {author} {\bibfnamefont {M.~D.}\ \bibnamefont {Johnson}},\
  and\ \bibinfo {author} {\bibfnamefont {A.}~\bibnamefont {Lupsasca}},\ }\href
  {https://doi.org/10.3847/1538-4357/ac0cca} {\bibfield  {journal} {\bibinfo
  {journal} {The Astrophysical Journal}\ }\textbf {\bibinfo {volume} {918}},\
  \bibinfo {pages} {6} (\bibinfo {year} {2021})}\BibitemShut {NoStop}%
\bibitem [{\citenamefont {Johnson}\ \emph {et~al.}(2023)\citenamefont
  {Johnson}, \citenamefont {Akiyama} \emph {et~al.}}]{johnson2023key}%
  \BibitemOpen
  \bibfield  {author} {\bibinfo {author} {\bibfnamefont {M.~D.}\ \bibnamefont
  {Johnson}}, \bibinfo {author} {\bibfnamefont {K.}~\bibnamefont {Akiyama}},
  \emph {et~al.},\ }\href {https://doi.org/10.3390/galaxies11030061} {\bibfield
   {journal} {\bibinfo  {journal} {Galaxies}\ }\textbf {\bibinfo {volume}
  {11}},\ \bibinfo {pages} {61} (\bibinfo {year} {2023})}\BibitemShut {NoStop}%
\bibitem [{\citenamefont {Perlick}\ and\ \citenamefont
  {Tsupko}(2022)}]{perlick2022calculating}%
  \BibitemOpen
  \bibfield  {author} {\bibinfo {author} {\bibfnamefont {V.}~\bibnamefont
  {Perlick}}\ and\ \bibinfo {author} {\bibfnamefont {O.~Y.}\ \bibnamefont
  {Tsupko}},\ }\href {https://doi.org/10.1016/j.physrep.2021.10.004} {\bibfield
   {journal} {\bibinfo  {journal} {Physics Reports}\ }\textbf {\bibinfo
  {volume} {947}},\ \bibinfo {pages} {1} (\bibinfo {year} {2022})}\BibitemShut
  {NoStop}%
\bibitem [{\citenamefont {Solanki}\ and\ \citenamefont
  {Perlick}(2022)}]{solanki2022photon}%
  \BibitemOpen
  \bibfield  {author} {\bibinfo {author} {\bibfnamefont {J.}~\bibnamefont
  {Solanki}}\ and\ \bibinfo {author} {\bibfnamefont {V.}~\bibnamefont
  {Perlick}},\ }\href {https://doi.org/10.1103/PhysRevD.105.064056} {\bibfield
  {journal} {\bibinfo  {journal} {Physical Review D}\ }\textbf {\bibinfo
  {volume} {105}},\ \bibinfo {pages} {064056} (\bibinfo {year}
  {2022})}\BibitemShut {NoStop}%
\bibitem [{\citenamefont
  {Chandrasekhar}(1998)}]{chandrasekhar1998mathematical}%
  \BibitemOpen
  \bibfield  {author} {\bibinfo {author} {\bibfnamefont {S.}~\bibnamefont
  {Chandrasekhar}},\ }\href
  {https://doi.org/10.1093/oso/9780198503705.001.0001} {\emph {\bibinfo {title}
  {The Mathematical Theory of Black Holes}}}\ (\bibinfo  {publisher} {Oxford
  University Press},\ \bibinfo {address} {Oxford},\ \bibinfo {year}
  {1998})\BibitemShut {NoStop}%
\bibitem [{\citenamefont {Virbhadra}\ and\ \citenamefont
  {Ellis}(2000)}]{virbhadra2000schwarzschild}%
  \BibitemOpen
  \bibfield  {author} {\bibinfo {author} {\bibfnamefont {K.~S.}\ \bibnamefont
  {Virbhadra}}\ and\ \bibinfo {author} {\bibfnamefont {G.~F.~R.}\ \bibnamefont
  {Ellis}},\ }\href {https://doi.org/10.1103/PhysRevD.62.084003} {\bibfield
  {journal} {\bibinfo  {journal} {Physical Review D}\ }\textbf {\bibinfo
  {volume} {62}},\ \bibinfo {pages} {084003} (\bibinfo {year}
  {2000})}\BibitemShut {NoStop}%
\bibitem [{\citenamefont {Vagnozzi}\ \emph {et~al.}(2023)\citenamefont
  {Vagnozzi}, \citenamefont {Roy} \emph {et~al.}}]{vagnozzi2023horizon}%
  \BibitemOpen
  \bibfield  {author} {\bibinfo {author} {\bibfnamefont {S.}~\bibnamefont
  {Vagnozzi}}, \bibinfo {author} {\bibfnamefont {R.}~\bibnamefont {Roy}}, \emph
  {et~al.},\ }\href {https://doi.org/10.1088/1361-6382/acd97b} {\bibfield
  {journal} {\bibinfo  {journal} {Classical and Quantum Gravity}\ }\textbf
  {\bibinfo {volume} {40}},\ \bibinfo {pages} {165007} (\bibinfo {year}
  {2023})}\BibitemShut {NoStop}%
\bibitem [{\citenamefont {Qiao}(2025)}]{qiao2025existence}%
  \BibitemOpen
  \bibfield  {author} {\bibinfo {author} {\bibfnamefont {C.~K.}\ \bibnamefont
  {Qiao}},\ }\href {https://doi.org/10.1140/epjc/s10052-025-13062-3} {\bibfield
   {journal} {\bibinfo  {journal} {European Physical Journal C}\ }\textbf
  {\bibinfo {volume} {85}},\ \bibinfo {pages} {191} (\bibinfo {year}
  {2025})}\BibitemShut {NoStop}%
\bibitem [{\citenamefont {Qiao}(2022)}]{qiao2022curvatures}%
  \BibitemOpen
  \bibfield  {author} {\bibinfo {author} {\bibfnamefont {C.~K.}\ \bibnamefont
  {Qiao}},\ }\href {https://doi.org/10.1103/PhysRevD.106.084060} {\bibfield
  {journal} {\bibinfo  {journal} {Physical Review D}\ }\textbf {\bibinfo
  {volume} {106}},\ \bibinfo {pages} {084060} (\bibinfo {year}
  {2022})}\BibitemShut {NoStop}%
\bibitem [{\citenamefont {Koga}\ and\ \citenamefont
  {Harada}(2019)}]{koga2019stability}%
  \BibitemOpen
  \bibfield  {author} {\bibinfo {author} {\bibfnamefont {Y.}~\bibnamefont
  {Koga}}\ and\ \bibinfo {author} {\bibfnamefont {T.}~\bibnamefont {Harada}},\
  }\href {https://doi.org/10.1103/PhysRevD.100.064040} {\bibfield  {journal}
  {\bibinfo  {journal} {Physical Review D}\ }\textbf {\bibinfo {volume}
  {100}},\ \bibinfo {pages} {064040} (\bibinfo {year} {2019})}\BibitemShut
  {NoStop}%
\bibitem [{\citenamefont {Shoom}(2017)}]{shoom2017metamorphoses}%
  \BibitemOpen
  \bibfield  {author} {\bibinfo {author} {\bibfnamefont {A.~A.}\ \bibnamefont
  {Shoom}},\ }\href {https://doi.org/10.1103/PhysRevD.96.084056} {\bibfield
  {journal} {\bibinfo  {journal} {Physical Review D}\ }\textbf {\bibinfo
  {volume} {96}},\ \bibinfo {pages} {084056} (\bibinfo {year}
  {2017})}\BibitemShut {NoStop}%
\bibitem [{\citenamefont {Cveti{\v c}}\ \emph {et~al.}(2016)\citenamefont
  {Cveti{\v c}}, \citenamefont {Gibbons},\ and\ \citenamefont
  {Pope}}]{cvetivc2016photon}%
  \BibitemOpen
  \bibfield  {author} {\bibinfo {author} {\bibfnamefont {M.}~\bibnamefont
  {Cveti{\v c}}}, \bibinfo {author} {\bibfnamefont {G.~W.}\ \bibnamefont
  {Gibbons}},\ and\ \bibinfo {author} {\bibfnamefont {C.~N.}\ \bibnamefont
  {Pope}},\ }\href {https://doi.org/10.1103/PhysRevD.94.106005} {\bibfield
  {journal} {\bibinfo  {journal} {Physical Review D}\ }\textbf {\bibinfo
  {volume} {94}},\ \bibinfo {pages} {106005} (\bibinfo {year}
  {2016})}\BibitemShut {NoStop}%
\bibitem [{\citenamefont {Cunha}\ \emph {et~al.}(2017)\citenamefont {Cunha},
  \citenamefont {Herdeiro},\ and\ \citenamefont {Radu}}]{cunha2017fundamental}%
  \BibitemOpen
  \bibfield  {author} {\bibinfo {author} {\bibfnamefont {P.~V.~P.}\
  \bibnamefont {Cunha}}, \bibinfo {author} {\bibfnamefont {C.~A.~R.}\
  \bibnamefont {Herdeiro}},\ and\ \bibinfo {author} {\bibfnamefont
  {E.}~\bibnamefont {Radu}},\ }\href
  {https://doi.org/10.1103/PhysRevD.96.024039} {\bibfield  {journal} {\bibinfo
  {journal} {Physical Review D}\ }\textbf {\bibinfo {volume} {96}},\ \bibinfo
  {pages} {024039} (\bibinfo {year} {2017})}\BibitemShut {NoStop}%
\bibitem [{\citenamefont {Wei}(2020{\natexlab{a}})}]{wei2020topological}%
  \BibitemOpen
  \bibfield  {author} {\bibinfo {author} {\bibfnamefont {S.~W.}\ \bibnamefont
  {Wei}},\ }\href {https://doi.org/10.1103/PhysRevD.102.064039} {\bibfield
  {journal} {\bibinfo  {journal} {Physical Review D}\ }\textbf {\bibinfo
  {volume} {102}},\ \bibinfo {pages} {064039} (\bibinfo {year}
  {2020}{\natexlab{a}})}\BibitemShut {NoStop}%
\bibitem [{\citenamefont {Sadeghi}\ and\ \citenamefont
  {Afshar}(2024)}]{sadeghi2024role}%
  \BibitemOpen
  \bibfield  {author} {\bibinfo {author} {\bibfnamefont {J.}~\bibnamefont
  {Sadeghi}}\ and\ \bibinfo {author} {\bibfnamefont {M.~A.~S.}\ \bibnamefont
  {Afshar}},\ }\href {https://doi.org/10.1016/j.astropartphys.2023.102994}
  {\bibfield  {journal} {\bibinfo  {journal} {Astroparticle Physics}\ }\textbf
  {\bibinfo {volume} {162}},\ \bibinfo {pages} {102994} (\bibinfo {year}
  {2024})}\BibitemShut {NoStop}%
\bibitem [{\citenamefont {Sadeghi}\ \emph {et~al.}(2024)\citenamefont
  {Sadeghi}, \citenamefont {Afshar}, \citenamefont {Gashti},\ and\
  \citenamefont {Alipour}}]{sadeghi2024thermodynamic}%
  \BibitemOpen
  \bibfield  {author} {\bibinfo {author} {\bibfnamefont {J.}~\bibnamefont
  {Sadeghi}}, \bibinfo {author} {\bibfnamefont {M.~A.~S.}\ \bibnamefont
  {Afshar}}, \bibinfo {author} {\bibfnamefont {S.~N.}\ \bibnamefont {Gashti}},\
  and\ \bibinfo {author} {\bibfnamefont {M.~R.}\ \bibnamefont {Alipour}},\
  }\href {https://doi.org/10.1016/j.astropartphys.2023.102920} {\bibfield
  {journal} {\bibinfo  {journal} {Astroparticle Physics}\ }\textbf {\bibinfo
  {volume} {156}},\ \bibinfo {pages} {102920} (\bibinfo {year}
  {2024})}\BibitemShut {NoStop}%
\bibitem [{\citenamefont {Shahzad}\ \emph {et~al.}(2025)\citenamefont
  {Shahzad}, \citenamefont {Alessa}, \citenamefont {Mehmood},\ and\
  \citenamefont {Mamedov}}]{shahzad2025topological}%
  \BibitemOpen
  \bibfield  {author} {\bibinfo {author} {\bibfnamefont {M.~U.}\ \bibnamefont
  {Shahzad}}, \bibinfo {author} {\bibfnamefont {N.}~\bibnamefont {Alessa}},
  \bibinfo {author} {\bibfnamefont {A.}~\bibnamefont {Mehmood}},\ and\ \bibinfo
  {author} {\bibfnamefont {S.}~\bibnamefont {Mamedov}},\ }\href
  {https://doi.org/10.1007/s10773-024-05752-6} {\bibfield  {journal} {\bibinfo
  {journal} {International Journal of Theoretical Physics}\ }\textbf {\bibinfo
  {volume} {64}},\ \bibinfo {pages} {1} (\bibinfo {year} {2025})}\BibitemShut
  {NoStop}%
\bibitem [{\citenamefont {Zhu}\ \emph {et~al.}(2025)\citenamefont {Zhu},
  \citenamefont {Liu},\ and\ \citenamefont {Wu}}]{zhu2025universal}%
  \BibitemOpen
  \bibfield  {author} {\bibinfo {author} {\bibfnamefont {X.~D.}\ \bibnamefont
  {Zhu}}, \bibinfo {author} {\bibfnamefont {W.}~\bibnamefont {Liu}},\ and\
  \bibinfo {author} {\bibfnamefont {D.}~\bibnamefont {Wu}},\ }\href
  {https://doi.org/10.1016/j.physletb.2024.139163} {\bibfield  {journal}
  {\bibinfo  {journal} {Physics Letters B}\ }\textbf {\bibinfo {volume}
  {860}},\ \bibinfo {pages} {139163} (\bibinfo {year} {2025})}\BibitemShut
  {NoStop}%
\bibitem [{\citenamefont {Dong}\ \emph {et~al.}(2025)\citenamefont {Dong},
  \citenamefont {Hosseinifar}, \citenamefont {Studni{\v c}ka}, \citenamefont
  {Chung},\ and\ \citenamefont {Hassanabadi}}]{dong2025thermodynamic}%
  \BibitemOpen
  \bibfield  {author} {\bibinfo {author} {\bibfnamefont {S.~H.}\ \bibnamefont
  {Dong}}, \bibinfo {author} {\bibfnamefont {F.}~\bibnamefont {Hosseinifar}},
  \bibinfo {author} {\bibfnamefont {F.}~\bibnamefont {Studni{\v c}ka}},
  \bibinfo {author} {\bibfnamefont {W.~S.}\ \bibnamefont {Chung}},\ and\
  \bibinfo {author} {\bibfnamefont {H.}~\bibnamefont {Hassanabadi}},\ }\href
  {https://doi.org/10.1016/j.dark.2025.101962} {\bibfield  {journal} {\bibinfo
  {journal} {Physics of the Dark Universe}\ }\textbf {\bibinfo {volume} {48}},\
  \bibinfo {pages} {101962} (\bibinfo {year} {2025})}\BibitemShut {NoStop}%
\bibitem [{\citenamefont {Cunha}\ and\ \citenamefont
  {Herdeiro}(2020)}]{Cunha2020}%
  \BibitemOpen
  \bibfield  {author} {\bibinfo {author} {\bibfnamefont {P.~V.~P.}\
  \bibnamefont {Cunha}}\ and\ \bibinfo {author} {\bibfnamefont {C.~A.~R.}\
  \bibnamefont {Herdeiro}},\ }\href
  {https://doi.org/10.1103/PhysRevLett.124.181101} {\bibfield  {journal}
  {\bibinfo  {journal} {Physical Review Letters}\ }\textbf {\bibinfo {volume}
  {124}},\ \bibinfo {pages} {181101} (\bibinfo {year} {2020})}\BibitemShut
  {NoStop}%
\bibitem [{\citenamefont {Wei}(2020{\natexlab{b}})}]{Wei2020}%
  \BibitemOpen
  \bibfield  {author} {\bibinfo {author} {\bibfnamefont {S.-W.}\ \bibnamefont
  {Wei}},\ }\href {https://doi.org/10.1103/PhysRevD.102.064039} {\bibfield
  {journal} {\bibinfo  {journal} {Physical Review D}\ }\textbf {\bibinfo
  {volume} {102}},\ \bibinfo {pages} {064039} (\bibinfo {year}
  {2020}{\natexlab{b}})}\BibitemShut {NoStop}%
\bibitem [{\citenamefont {Bekenstein}(1973)}]{Bekenstein1973}%
  \BibitemOpen
  \bibfield  {author} {\bibinfo {author} {\bibfnamefont {J.~D.}\ \bibnamefont
  {Bekenstein}},\ }\href {https://doi.org/10.1103/PhysRevD.7.2333} {\bibfield
  {journal} {\bibinfo  {journal} {Physical Review D}\ }\textbf {\bibinfo
  {volume} {7}},\ \bibinfo {pages} {2333} (\bibinfo {year} {1973})}\BibitemShut
  {NoStop}%
\bibitem [{\citenamefont {Hawking}(1975)}]{Hawking1975}%
  \BibitemOpen
  \bibfield  {author} {\bibinfo {author} {\bibfnamefont {S.~W.}\ \bibnamefont
  {Hawking}},\ }\href {https://doi.org/10.1007/BF02345020} {\bibfield
  {journal} {\bibinfo  {journal} {Communications in Mathematical Physics}\
  }\textbf {\bibinfo {volume} {43}},\ \bibinfo {pages} {199} (\bibinfo {year}
  {1975})},\ \bibinfo {note} {erratum: Commun. Math. Phys. 46, 206
  (1976)}\BibitemShut {NoStop}%
\bibitem [{\citenamefont {Wald}(2001)}]{Wald2001}%
  \BibitemOpen
  \bibfield  {author} {\bibinfo {author} {\bibfnamefont {R.~M.}\ \bibnamefont
  {Wald}},\ }\href {https://doi.org/10.12942/lrr-2001-6} {\bibfield  {journal}
  {\bibinfo  {journal} {Living Reviews in Relativity}\ }\textbf {\bibinfo
  {volume} {4}},\ \bibinfo {pages} {6} (\bibinfo {year} {2001})}\BibitemShut
  {NoStop}%
\bibitem [{\citenamefont {Wei}\ and\ \citenamefont
  {Liu}(2022{\natexlab{a}})}]{Wei2022a}%
  \BibitemOpen
  \bibfield  {author} {\bibinfo {author} {\bibfnamefont {S.-W.}\ \bibnamefont
  {Wei}}\ and\ \bibinfo {author} {\bibfnamefont {Y.-X.}\ \bibnamefont {Liu}},\
  }\href {https://doi.org/10.1103/PhysRevD.105.104003} {\bibfield  {journal}
  {\bibinfo  {journal} {Physical Review D}\ }\textbf {\bibinfo {volume}
  {105}},\ \bibinfo {pages} {104003} (\bibinfo {year}
  {2022}{\natexlab{a}})}\BibitemShut {NoStop}%
\bibitem [{\citenamefont {Wei}\ \emph {et~al.}(2022)\citenamefont {Wei},
  \citenamefont {Liu},\ and\ \citenamefont {Mann}}]{Wei2022b}%
  \BibitemOpen
  \bibfield  {author} {\bibinfo {author} {\bibfnamefont {S.-W.}\ \bibnamefont
  {Wei}}, \bibinfo {author} {\bibfnamefont {Y.-X.}\ \bibnamefont {Liu}},\ and\
  \bibinfo {author} {\bibfnamefont {R.~B.}\ \bibnamefont {Mann}},\ }\href
  {https://doi.org/10.1103/PhysRevLett.129.191101} {\bibfield  {journal}
  {\bibinfo  {journal} {Physical Review Letters}\ }\textbf {\bibinfo {volume}
  {129}},\ \bibinfo {pages} {191101} (\bibinfo {year} {2022})}\BibitemShut
  {NoStop}%
\bibitem [{\citenamefont {Wu}(2023)}]{wu2023topological}%
  \BibitemOpen
  \bibfield  {author} {\bibinfo {author} {\bibfnamefont {D.}~\bibnamefont
  {Wu}},\ }\href {https://doi.org/10.1103/PhysRevD.107.024024} {\bibfield
  {journal} {\bibinfo  {journal} {Physical Review D}\ }\textbf {\bibinfo
  {volume} {107}},\ \bibinfo {pages} {024024} (\bibinfo {year}
  {2023})}\BibitemShut {NoStop}%
\bibitem [{\citenamefont {Barzi}\ \emph {et~al.}(2024)\citenamefont {Barzi},
  \citenamefont {El~Moumni},\ and\ \citenamefont {Masmar}}]{barzi2024renyi}%
  \BibitemOpen
  \bibfield  {author} {\bibinfo {author} {\bibfnamefont {F.}~\bibnamefont
  {Barzi}}, \bibinfo {author} {\bibfnamefont {H.}~\bibnamefont {El~Moumni}},\
  and\ \bibinfo {author} {\bibfnamefont {K.}~\bibnamefont {Masmar}},\ }\href
  {https://doi.org/10.1016/j.jheap.2024.03.001} {\bibfield  {journal} {\bibinfo
   {journal} {Journal of High Energy Astrophysics}\ }\textbf {\bibinfo {volume}
  {42}},\ \bibinfo {pages} {63} (\bibinfo {year} {2024})}\BibitemShut {NoStop}%
\bibitem [{\citenamefont {Chen}\ \emph {et~al.}(2024)\citenamefont {Chen},
  \citenamefont {Zhang}, \citenamefont {Ara{\'u}jo~Filho}, \citenamefont
  {Hosseinifar},\ and\ \citenamefont {Hassanabadi}}]{chen2024thermal}%
  \BibitemOpen
  \bibfield  {author} {\bibinfo {author} {\bibfnamefont {H.}~\bibnamefont
  {Chen}}, \bibinfo {author} {\bibfnamefont {M.~Y.}\ \bibnamefont {Zhang}},
  \bibinfo {author} {\bibfnamefont {A.~A.}\ \bibnamefont {Ara{\'u}jo~Filho}},
  \bibinfo {author} {\bibfnamefont {F.}~\bibnamefont {Hosseinifar}},\ and\
  \bibinfo {author} {\bibfnamefont {H.}~\bibnamefont {Hassanabadi}},\ }\href
  {https://doi.org/10.48550/arXiv.2408.03090} {\bibinfo {title} {Thermal
  topology of black holes}} (\bibinfo {year} {2024}),\ \Eprint
  {https://arxiv.org/abs/2408.03090} {arXiv:2408.03090 [gr-qc]} \BibitemShut
  {NoStop}%
\bibitem [{\citenamefont {Alipour}\ \emph {et~al.}(2023)\citenamefont
  {Alipour}, \citenamefont {Afshar}, \citenamefont {Gashti},\ and\
  \citenamefont {Sadeghi}}]{alipour2023topological}%
  \BibitemOpen
  \bibfield  {author} {\bibinfo {author} {\bibfnamefont {M.~R.}\ \bibnamefont
  {Alipour}}, \bibinfo {author} {\bibfnamefont {M.~A.~S.}\ \bibnamefont
  {Afshar}}, \bibinfo {author} {\bibfnamefont {S.~N.}\ \bibnamefont {Gashti}},\
  and\ \bibinfo {author} {\bibfnamefont {J.}~\bibnamefont {Sadeghi}},\ }\href
  {https://doi.org/10.1016/j.dark.2023.101361} {\bibfield  {journal} {\bibinfo
  {journal} {Physics of the Dark Universe}\ }\textbf {\bibinfo {volume} {42}},\
  \bibinfo {pages} {101361} (\bibinfo {year} {2023})}\BibitemShut {NoStop}%
\bibitem [{\citenamefont {Wei}\ and\ \citenamefont
  {Liu}(2022{\natexlab{b}})}]{wei2022topology}%
  \BibitemOpen
  \bibfield  {author} {\bibinfo {author} {\bibfnamefont {S.-W.}\ \bibnamefont
  {Wei}}\ and\ \bibinfo {author} {\bibfnamefont {Y.-X.}\ \bibnamefont {Liu}},\
  }\href {https://doi.org/10.1103/PhysRevD.105.104003} {\bibfield  {journal}
  {\bibinfo  {journal} {Physical Review D}\ }\textbf {\bibinfo {volume}
  {105}},\ \bibinfo {pages} {104003} (\bibinfo {year}
  {2022}{\natexlab{b}})}\BibitemShut {NoStop}%
\bibitem [{\citenamefont {Yerra}\ \emph {et~al.}(2023)\citenamefont {Yerra},
  \citenamefont {Bhamidipati},\ and\ \citenamefont
  {Mukherji}}]{yerra2023topology}%
  \BibitemOpen
  \bibfield  {author} {\bibinfo {author} {\bibfnamefont {P.~K.}\ \bibnamefont
  {Yerra}}, \bibinfo {author} {\bibfnamefont {C.}~\bibnamefont {Bhamidipati}},\
  and\ \bibinfo {author} {\bibfnamefont {S.}~\bibnamefont {Mukherji}},\ }\href
  {https://doi.org/10.1088/1742-6596/2667/1/012031} {\bibfield  {journal}
  {\bibinfo  {journal} {Journal of Physics: Conference Series}\ }\textbf
  {\bibinfo {volume} {2667}},\ \bibinfo {pages} {012031} (\bibinfo {year}
  {2023})}\BibitemShut {NoStop}%
\bibitem [{\citenamefont {Mehmood}\ and\ \citenamefont
  {Shahzad}(2023)}]{mehmood2023thermodynamic}%
  \BibitemOpen
  \bibfield  {author} {\bibinfo {author} {\bibfnamefont {A.}~\bibnamefont
  {Mehmood}}\ and\ \bibinfo {author} {\bibfnamefont {M.~U.}\ \bibnamefont
  {Shahzad}},\ }\href {https://doi.org/10.48550/arXiv.2310.09907} {\bibinfo
  {title} {Thermodynamic topology of ads black holes}} (\bibinfo {year}
  {2023}),\ \Eprint {https://arxiv.org/abs/2310.09907} {arXiv:2310.09907
  [gr-qc]} \BibitemShut {NoStop}%
\bibitem [{\citenamefont {Zhang}\ \emph {et~al.}(2023)\citenamefont {Zhang},
  \citenamefont {Chen}, \citenamefont {Hassanabadi}, \citenamefont {Long},\
  and\ \citenamefont {Yang}}]{zhang2023topology}%
  \BibitemOpen
  \bibfield  {author} {\bibinfo {author} {\bibfnamefont {M.~Y.}\ \bibnamefont
  {Zhang}}, \bibinfo {author} {\bibfnamefont {H.}~\bibnamefont {Chen}},
  \bibinfo {author} {\bibfnamefont {H.}~\bibnamefont {Hassanabadi}}, \bibinfo
  {author} {\bibfnamefont {Z.~W.}\ \bibnamefont {Long}},\ and\ \bibinfo
  {author} {\bibfnamefont {H.}~\bibnamefont {Yang}},\ }\href
  {https://doi.org/10.1140/epjc/s10052-023-11825-5} {\bibfield  {journal}
  {\bibinfo  {journal} {European Physical Journal C}\ }\textbf {\bibinfo
  {volume} {83}},\ \bibinfo {pages} {773} (\bibinfo {year} {2023})}\BibitemShut
  {NoStop}%
\bibitem [{\citenamefont {Wu}\ \emph {et~al.}(2025)\citenamefont {Wu},
  \citenamefont {Liu}, \citenamefont {Wu},\ and\ \citenamefont
  {Mann}}]{wu2025}%
  \BibitemOpen
  \bibfield  {author} {\bibinfo {author} {\bibfnamefont {D.}~\bibnamefont
  {Wu}}, \bibinfo {author} {\bibfnamefont {W.}~\bibnamefont {Liu}}, \bibinfo
  {author} {\bibfnamefont {S.~Q.}\ \bibnamefont {Wu}},\ and\ \bibinfo {author}
  {\bibfnamefont {R.~B.}\ \bibnamefont {Mann}},\ }\href
  {https://doi.org/10.1103/PhysRevD.111.L061501} {\bibfield  {journal}
  {\bibinfo  {journal} {Physical Review D}\ }\textbf {\bibinfo {volume}
  {111}},\ \bibinfo {pages} {L061501} (\bibinfo {year} {2025})}\BibitemShut
  {NoStop}%
\end{thebibliography}
%

\end{document}